\documentclass[letterpaper,twocolumn,10pt]{article}
\usepackage{usenix-2020-09}

\usepackage{tikz}
\usepackage{amsmath}

\usepackage{filecontents}
\usepackage{amsmath , amssymb , amsthm}
\usepackage{thmtools}
\usepackage{thm-restate}

\usepackage{scalerel}
\usepackage{cite}

\usepackage[utf8]{inputenc}

\usepackage[ruled, lined, linesnumbered, commentsnumbered, longend]{algorithm2e}

\usepackage[noend]{algpseudocode}

\usepackage{graphicx}

\usepackage{microtype}
\DisableLigatures{encoding = *, family = * }

\usepackage[english]{babel}
\usepackage{blindtext}

\usepackage{url}

\usepackage{subcaption}

\usepackage{mathtools}

\usepackage{xcolor}

\newsavebox\thmbox

\newcommand{\name}{\textsc{PowerTCP}\xspace}
\newcommand{\nameapprox}{\ensuremath{\theta}-\textsc{PowerTCP}\xspace}

\newcommand{\myitem}[1]{\vspace*{0.02in}\noindent\textbf{#1}}

\newcommand{\first}{\emph{(i)}\xspace}
\newcommand{\second}{\emph{(ii)}\xspace}
\newcommand{\third}{\emph{(iii)}\xspace}
\newcommand{\fourth}{\emph{(iv)}\xspace}
\newcommand{\fifth}{\emph{(v)}\xspace}

\newcommand{\ie}{i.e., \@}
\newcommand{\eg}{e.g., \@}
\newcommand{\etc}{etc., \@}

\theoremstyle{plain}

\newtheorem{property}{Property}

\newtheorem{Reviewers-Comment}{reveiwcomment}

  {\begin{lrbox}{\thmbox}%
   \begin{minipage}{\dimexpr\linewidth-2\fboxsep}
   \myitem{Takeaway.}}%
  {\end{minipage}%
   \end{lrbox}%
   \begin{trivlist}
     \item[]\colorbox{lightgray}{\usebox\thmbox}
   \end{trivlist}}

\usepackage{titlesec}
\titlespacing*{\section}{0ex}{2ex plus .2ex minus .2ex}{1ex plus .2ex minus .2ex}
\titlespacing*{\subsection}{0ex}{1ex plus .2ex minus .2ex}{1ex plus .2ex minus .2ex}

\begin{document}

\title{\name: Pushing the Performance Limits of Datacenter Networks\thanks{Authors version. Final version of the paper to appear in Usenix NSDI 2022. This work is part of a project that has received funding from the European Research Council (ERC) under the European Union’s Horizon 2020 research and innovation programme, consolidator project Self-Adjusting Networks (AdjustNet), grant agreement No. 864228, Horizon 2020, 2020-2025.
}}

\author{
{\rm Vamsi Addanki}\\
University of Vienna \\TU Berlin
\and
{\rm Oliver Michel}\\
University of Vienna \\ Princeton University
\and
{\rm Stefan Schmid}\\
University of Vienna \\ TU Berlin
} 

\maketitle

\begin{abstract}
Increasingly stringent throughput and latency requirements in datacenter networks demand fast and accurate congestion control.
We observe that the reaction time and accuracy of existing datacenter congestion control schemes are inherently limited.
They either rely only on explicit feedback about the network state (e.g., queue lengths in DCTCP) or only on variations of state (e.g., RTT gradient in TIMELY).
To overcome these limitations, we propose a novel congestion control algorithm, \name, which achieves much more fine-grained congestion control by adapting to the bandwidth-window product (henceforth called power).
\name leverages in-band network telemetry to react to changes in the network instantaneously without loss of throughput and while keeping queues short.
Due to its fast reaction time, our algorithm is particularly well-suited for dynamic network environments and bursty traffic patterns.
We show analytically and empirically that \name can significantly outperform the state-of-the-art in both traditional datacenter topologies and emerging reconfigurable datacenters where frequent bandwidth changes make congestion control challenging.
In traditional datacenter networks, \name reduces tail flow completion times of short flows by 80\% compared to DCQCN and TIMELY, and by 33\% compared to HPCC even at 60\% network load.
In reconfigurable datacenters, \name achieves $85\%$ circuit utilization without incurring additional latency and cuts tail latency by at least 2x compared to existing approaches.

\end{abstract}

\section{Introduction}
\label{sec:intro}

The performance of more and more cloud-based applications critically depends on the underlying network, requiring datacenter networks (DCNs) to provide extremely low latency and high bandwidth.
For example, in distributed machine learning applications that periodically require large data transfers, the network is increasingly becoming a bottleneck \cite{li2019hpcc}.
Similarly, stringent performance requirements are introduced by today's trend of resource disaggregation in datacenters where fast access to remote resources (e.g., GPUs or memory) is pivotal for the overall system performance \cite{li2019hpcc}.
Building systems with strict performance requirements is especially challenging under bursty traffic patterns as they are commonly observed in datacenter networks \cite{sigmetrics20complexity,woodruff2019measuring,zhang2017high,chen2009understanding,phanishayee2008measurement}.

These requirements introduce the need for fast and accurate network resource management algorithms that optimally utilize the available bandwidth while minimizing packet latencies and flow completion times.
Congestion control (CC) plays an important role in this context being ``a key enabler (or limiter) of system performance in the datacenter'' \cite{kumar2020swift}.
In fact, fast reacting congestion control is not only essential to efficiently adapt to bursty traffic \cite{kandula2009nature,roy2015inside}, but is also becoming increasingly important in the context of emerging reconfigurable datacenter networks (RDCNs) \cite{projector,sirius,rotornet,opera,ton15splay,apocs21renets,spaa21rdcn}.
In these networks, a congestion control algorithm must be able to quickly ramp up its sending rate when high-bandwidth circuits become available \cite{mukerjee2020adapting}.

Traditional congestion control in datacenters revolves around a bottleneck link model: the control action is related to the state \ie queue length at the bottleneck link. A common goal is to efficiently control queue buildup while achieving high throughput. Existing algorithms can be broadly classified into two types based on the feedback that they react to. In the following, we will use an analogy to electrical circuits\footnote{This analogy is inspired from S.~Keshav's lecture series based on mathematical foundations of computer networking~\cite{keshav2012mathematical}. We emphasize that our power analogy is meant for our networking context and it should not be applied to other domains of science.} to describe these two types. The first category of algorithms react to the absolute network state, such as the queue length or the RTT: a function of network ``effort'' or \textbf{voltage} defined as the sum of the bandwidth-delay product and in-network queuing.
The second category of algorithms rather react to variations, such as the change of RTT.
Since these changes are related to the network ``flow'', we say that these approaches depend on the \textbf{current} defined as the total transmission rate.
We tabulate our analogy and corresponding network quantities in Table~\ref{table:analogy}.
According to this classification, we call congestion control protocols such as CUBIC~\cite{ha2008cubic}, DCTCP~\cite{alizadeh2010data}, or Vegas~\cite{brakmo1994tcp} \textbf{voltage-based CC} algorithms as they react to absolute properties such as the bottleneck queue length, delay, Explicit Congestion Notification (ECN), or loss.
Recent proposals such as TIMELY~\cite{mittal2015timely} are \textbf{current-based CC} algorithms as they react to the variations, such as the RTT-gradient.
In conclusion, we find that existing congestion control algorithms are fundamentally limited to one of the two dimensions (voltage or current) in the way they update the congestion window.

\begin{table}[t]
\begin{center}
\begin{tabular}{|c|c|c|c|}
\hline
\textbf{Quantity} & \textbf{Analogy}\\
\hline
Total transmission rate (network flow) & Current ($\lambda$) \\
BDP + buffered bytes (network effort) &  Voltage ($\nu$) \\ 
Current $\times$ Voltage &  Power ($\Gamma$) \\
\hline
\end{tabular}
\end{center}
\vspace{-5mm}
\caption{Analogy between metrics in networks and in electrical circuits. Note that the network here is the ``pipe'' seen by a flow and not the whole network.}
\vspace{-5mm}
\label{table:analogy}
\end{table}
\begin{figure}[t]
\centering
\includegraphics[width=1\linewidth]{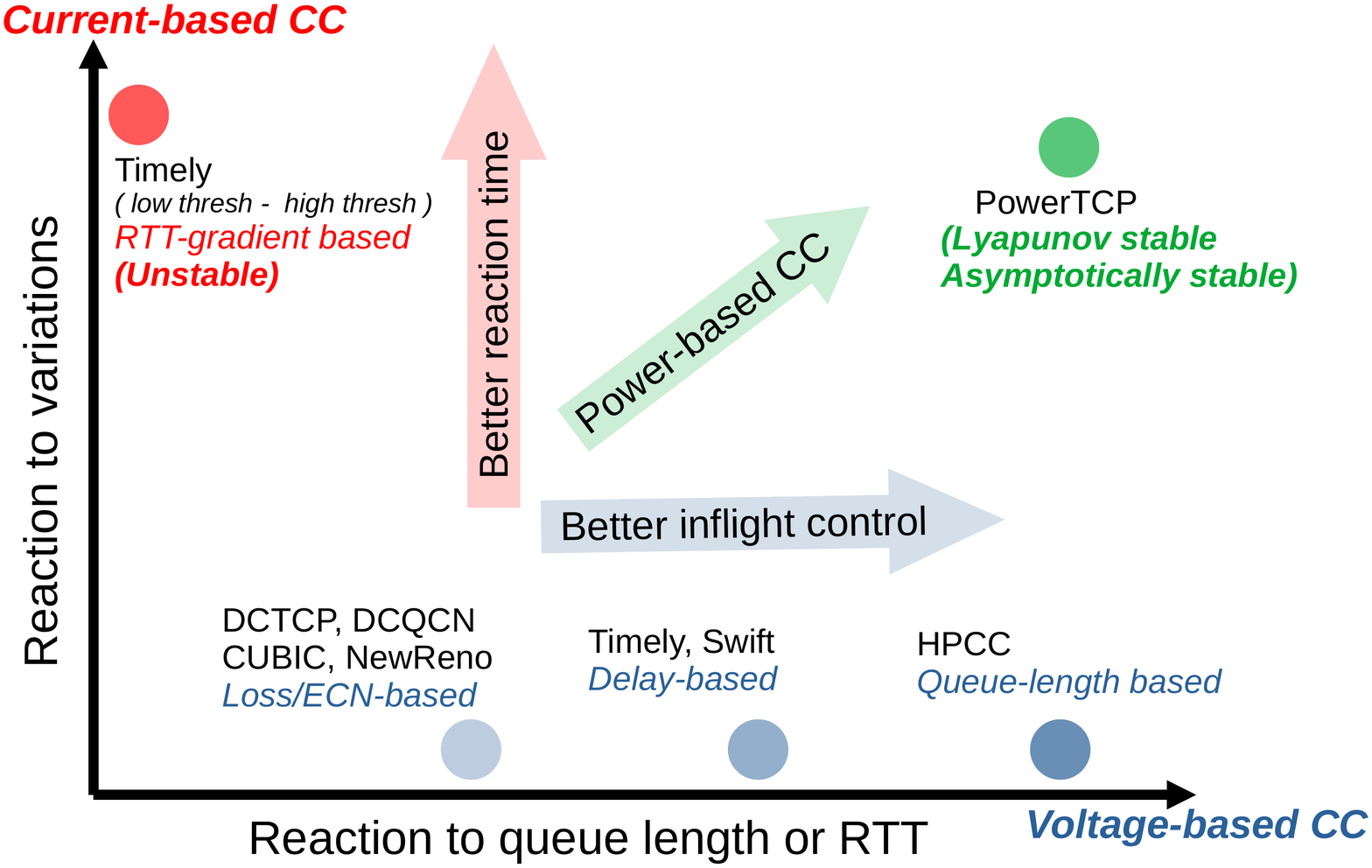}
\vspace{-8mm}
\caption{Existing congestion control algorithms are fundamentally limited to a single dimension in their window (or rate) update decisions and are unable to distinguish between two scenarios across multiple dimensions. }
\vspace{-5mm}
\label{fig:dimensions}
\end{figure}

We argue that the input to a congestion control algorithm should rather be a function of the two-dimensional state of the network (\ie both voltage and current) to allow for more informed and accurate reaction, improving performance and stability.
In our work, we show that there exists an accurate relationship between the optimal adjustment of the congestion window, the network voltage and the network current.
We analytically show that the optimal window adjustment depends on the product of network voltage and network current. We call this product \textbf{network power}: current $\times$ voltage, a function of both queue lengths and queue dynamics.

Figure~\ref{fig:dimensions} illustrates our classification. Existing congestion control protocols depend on a single dimension, voltage or current. This can result in imprecise congestion control as the protocol is unable to distinguish between fundamentally different scenarios, and, as a result, either reacts too slowly or overreacts, both impeding performance. 
Accounting for both voltage and current, \ie power, balances accurate inflight control and fast reaction, effectively providing the best of both worlds.

In this paper we present \name, a novel \emph{power}-based congestion control algorithm that accurately captures both \emph{voltage} and \emph{current} dimensions for every control action using measurements taken within the network and propagated through in-band network telemetry (INT).
\name is able to utilize available bandwidth within one or two RTTs while being stable, maintaining low queue lengths, and resolving congestion rapidly.
Furthermore, we show that \name is Lyapunov-stable, as well as asymptotically stable and has a convergence time as low as five update intervals (Appendix~\ref{sec:AppendixAnalysis}).
This makes \name highly suitable for today's datacenter networks and dynamic network environments such as in reconfigurable datacenters.

\name leverages in-network measurements at programmable switches to accurately obtain the bottleneck link state. 
Our switch component is lightweight and the required INT header fields are standard in the literature~\cite{li2019hpcc}.
We also discuss an approximation of \name for use with non-programmable, legacy switches.

To evaluate \name, we focus on a deployment scenario in the context of RDMA networks where the CC algorithm is implemented on a NIC. Our results from large-scale simulations show that \name reduces the 99.9-percentile short flow completion times by $80\%$ compared to DCQCN~\cite{zhu2015congestion} and by $33\%$ compared to the state-of-the-art low-latency protocol HPCC~\cite{li2019hpcc}.
We show that \name  maintains near-zero queue lengths without affecting throughput or incurring long flow completion times even at $80\%$ load.
As a case study, we explore the benefits of \name in reconfigurable datacenter networks where it achieves $80-85\%$ circuit utilization and reduces tail latency by at least $2\times$ compared to the state-of-the-art~\cite{mukerjee2020adapting}.
Finally, as a proof-of-concept, we implemented \name in the Linux kernel and the telemetry component on an Intel Tofino programmable line-rate switch using P4 \cite{tofino}.

\begin{figure*}
\centering
\begin{subfigure}{0.25\linewidth}
\centering
\includegraphics[trim=0 18 0 18,clip,width=0.9\linewidth]{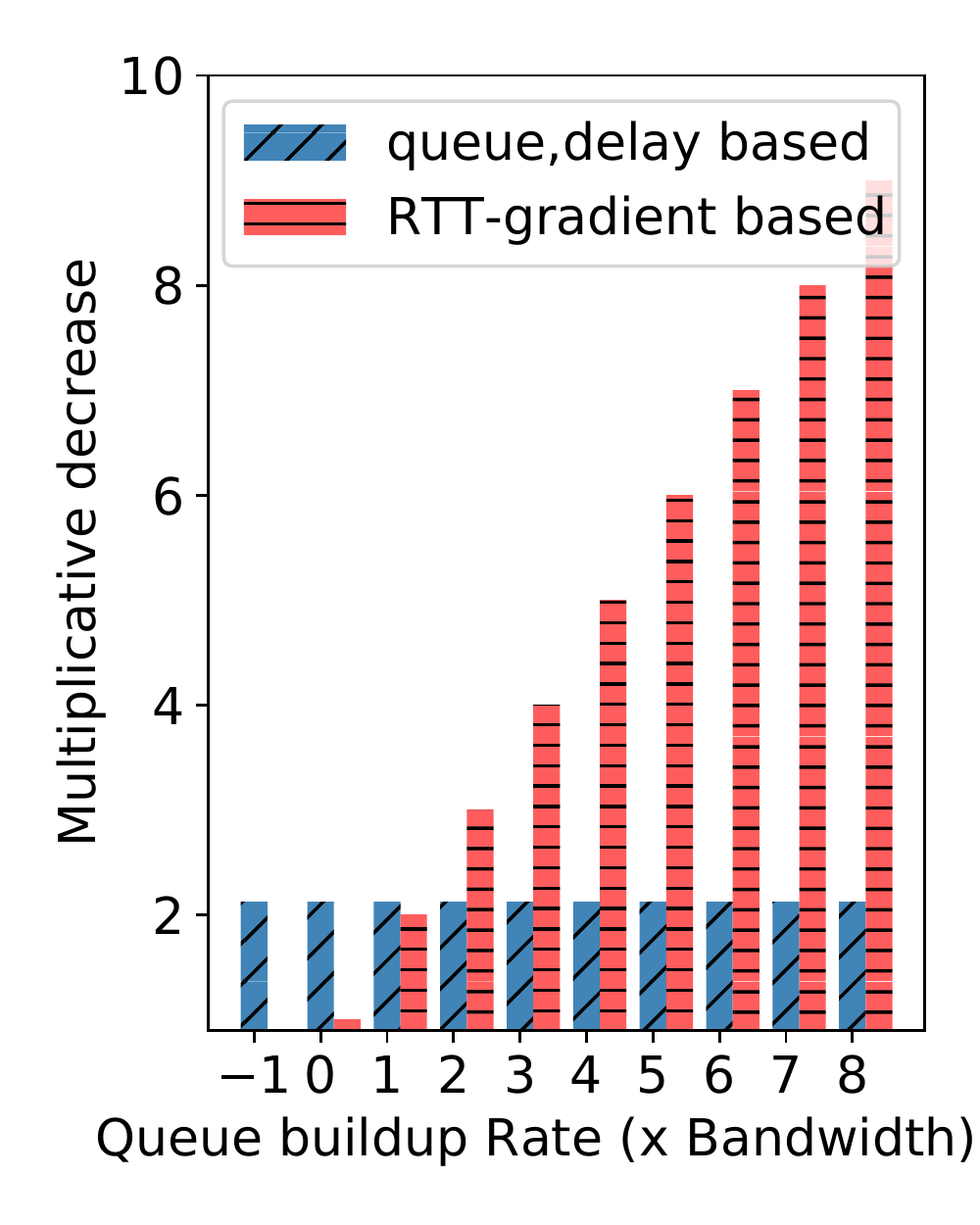}
\caption{Voltage-based CC is oblivious to queue buildup rate.}
\label{fig:currentDrawbacks}
\end{subfigure}\hfill
\begin{subfigure}{0.25\linewidth}
\centering
\includegraphics[trim=0 18 0 18,clip, width=0.9\linewidth]{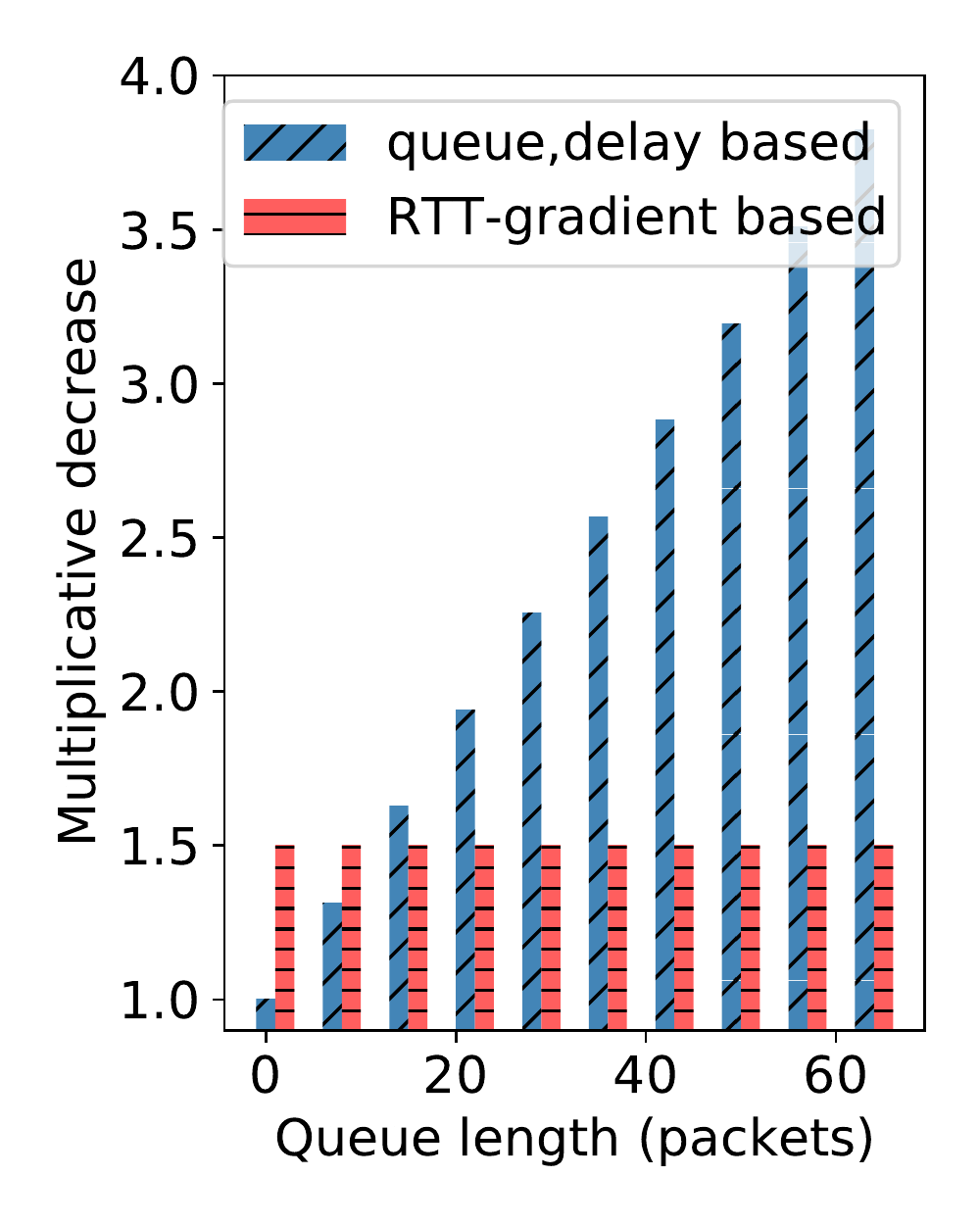}
\caption{Current-based CC is oblivious to queue lengths.}
\label{fig:trendDrawbacks}
\end{subfigure}\hfill
\begin{subfigure}{0.45\linewidth}
\centering
\includegraphics[width=0.9\linewidth]{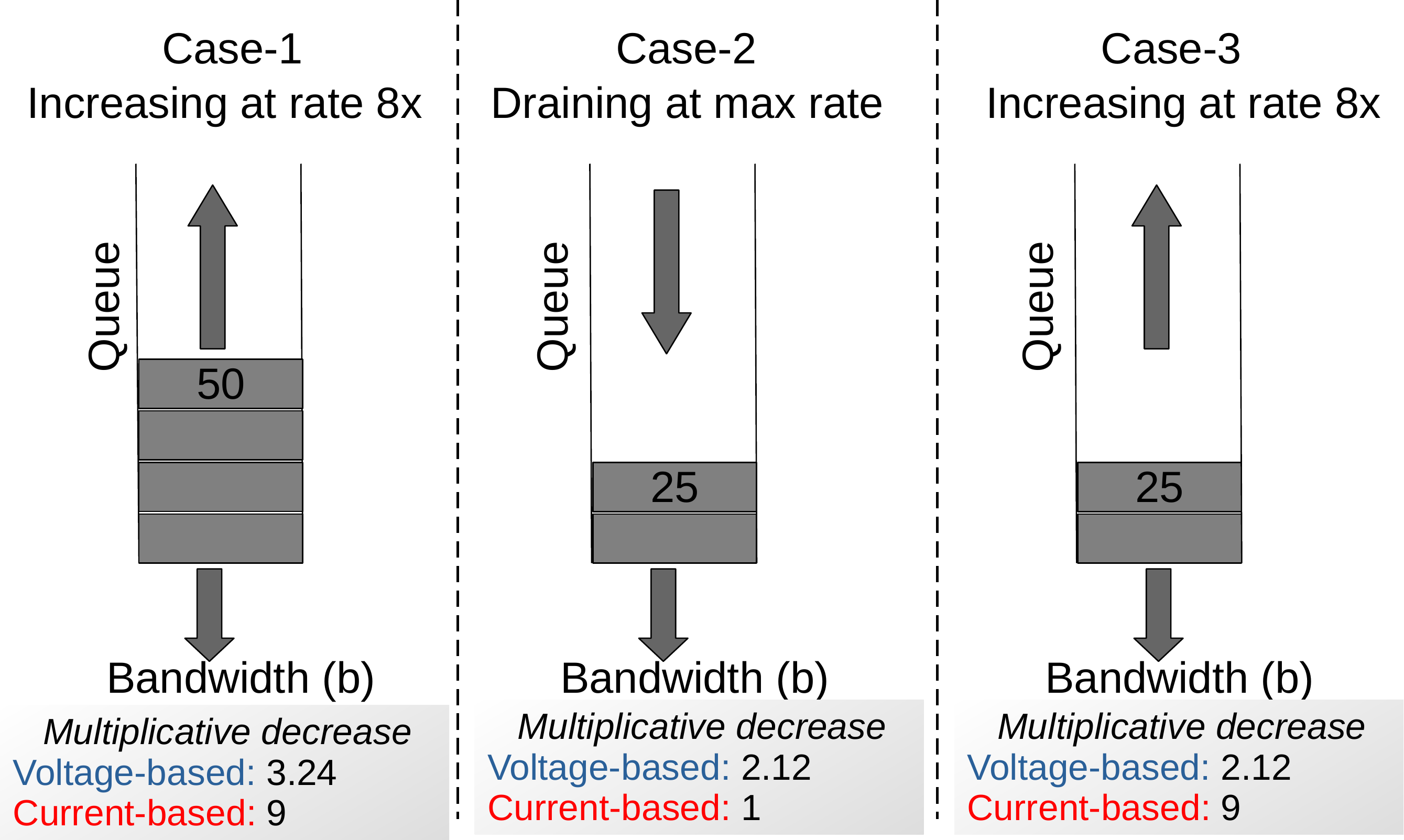}
\caption{Voltage-based CC cannot differentiate case-2 vs case-3; whereas current-based CC cannot differentiate case-1 vs case-3. }
\label{fig:drawbackExamples}
\end{subfigure}
\caption{Existing CC schemes, classified as voltage and current-based, are orthogonal in their response to queue length and queue buildup rate.}
\label{fig:drawbacks}
\vspace{-5mm}
\end{figure*}

In summary, our key contributions in this paper are:
\begin{itemize}
    \item We reveal the shortcomings of existing congestion control approaches 
    which either only react to the current state or the dynamics of the network,
    and introduce the notion of \emph{power} to account for both.
    \item \name, a power-based approach to congestion control at the end-host which reacts faster to changes in the network such as an arrival of burst, fluctuations in available bandwidth \etc
    \item An evaluation of the benefits of \name in traditional DCNs and RDCNs. 
    \item As a contribution to the research community and in order to ensure reproducibility and facilitate future work, we will make all our artefacts (source code, results) publicly available together with the final version of this paper.
\end{itemize}

\section{ Motivation }
\label{sec:motivation}

We first provide a more detailed motivation of our work by highlighting the benefits and drawbacks of existing congestion control approaches. In the following, \textbf{voltage-based CC} refers to the class of end-host congestion control algorithms that react to the state of the network in absolute values related to the bandwidth-delay product, such as bottleneck queue length, delay, loss, or ECN; \textbf{current-based CC} refers to the class of algorithms that react to changes in the state, such as the RTT-gradient. 
Voltage-based CC algorithms are likely to exhibit better stability but are fundamentally limited in their reaction time.
Current-based CC algorithms detect congestion faster but ensuring stability may be more challenging.
Indeed, TIMELY~\cite{mittal2015timely}, a current-based CC, deployed at Google datacenters, turned out to be unstable~\cite{zhu2016ecn} and evolved to SWIFT~\cite{kumar2020swift}, a voltage-based CC. 

Orthogonal to our approach, receiver-driven transport protocols~\cite{10.1145/3098822.3098825,10.1145/3387514.3405878,10.1145/3230543.3230564} have been proposed which show significant performance improvements.
A receiver-driven transport approach relies on the assumption that datacenter networks are well-provisioned and claims that congestion control is unnecessary; for example ``NDP performs no congestion control whatsoever in a Clos topology''~\cite{10.1145/3098822.3098825}.
The key difference is that receiver-driven approaches take feedback from the ToR downlink at the receiver whereas sender-based approaches rely on a variety of feedback signals to identify congestion anywhere along the path. 
In this paper, we focus on the sender-based congestion control approach which can in principle handle congestion anywhere along the round-trip path between a sender and a receiver, even in oversubscribed datacenters. 

To take a leap forward and design fine-grained datacenter congestion control algorithms, we present an analytical approach and study the fundamental problems faced by existing algorithms.
We first formally express the desirable properties of a datacenter congestion control law (\S\ref{sec:theSearch}) and then analytically identify the limitations and drawbacks of existing control laws (\S\ref{sec:problems}).
Finally, we discuss the lessons learned and formulate our design goals (\S\ref{sec:lessonsGoals}).

\begin{figure*}
\centering
\begin{subfigure}{0.30\linewidth}
\centering
\includegraphics[width=1\linewidth]{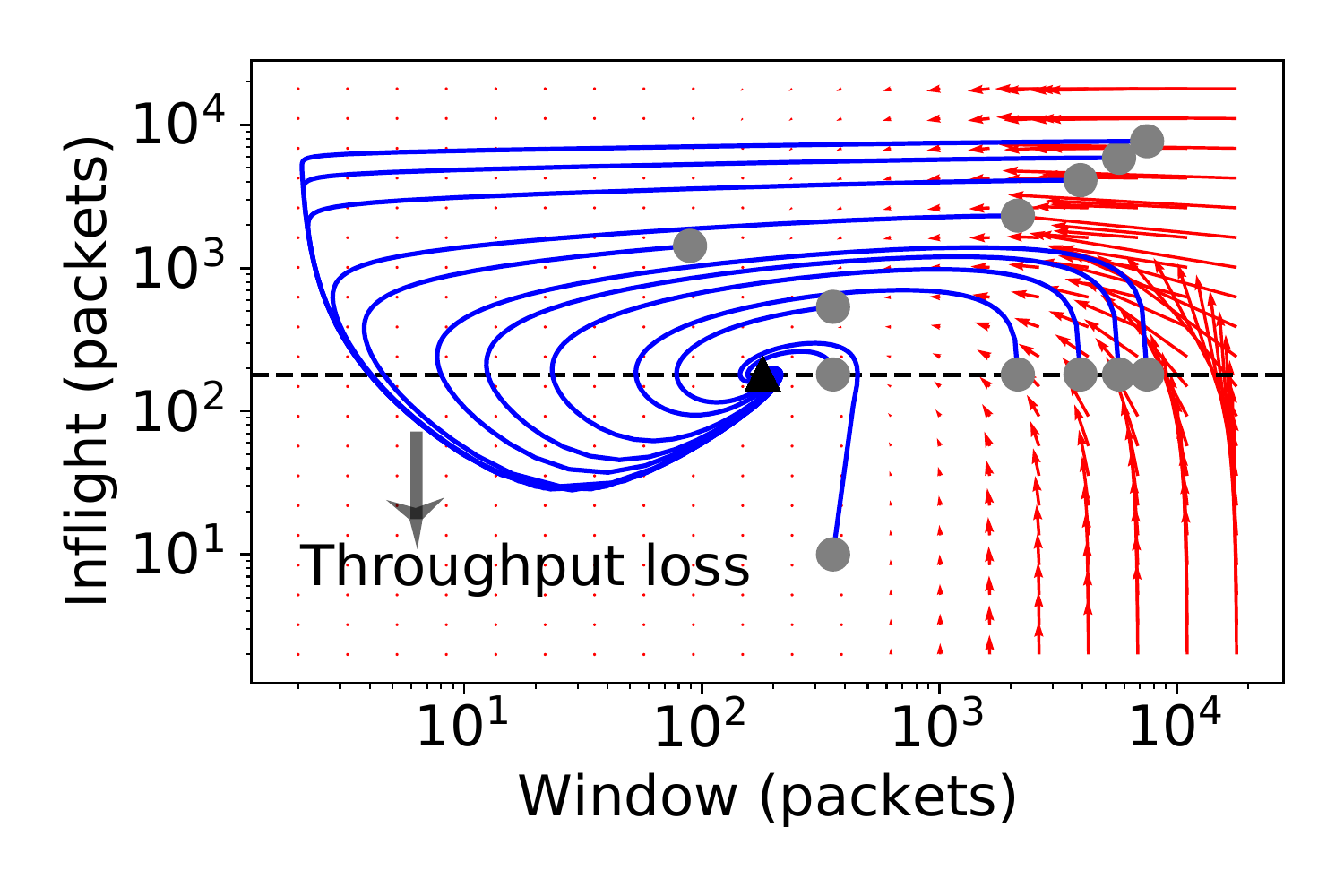}
\caption{Voltage-based CC (RTT or queue length) exhibits equilibrium properties but has an imprecise reaction leading to throughput loss.}
\label{fig:voltagePhaseplot}
\end{subfigure}
\hfill
\begin{subfigure}{0.30\linewidth}
\centering
\includegraphics[width=1\linewidth]{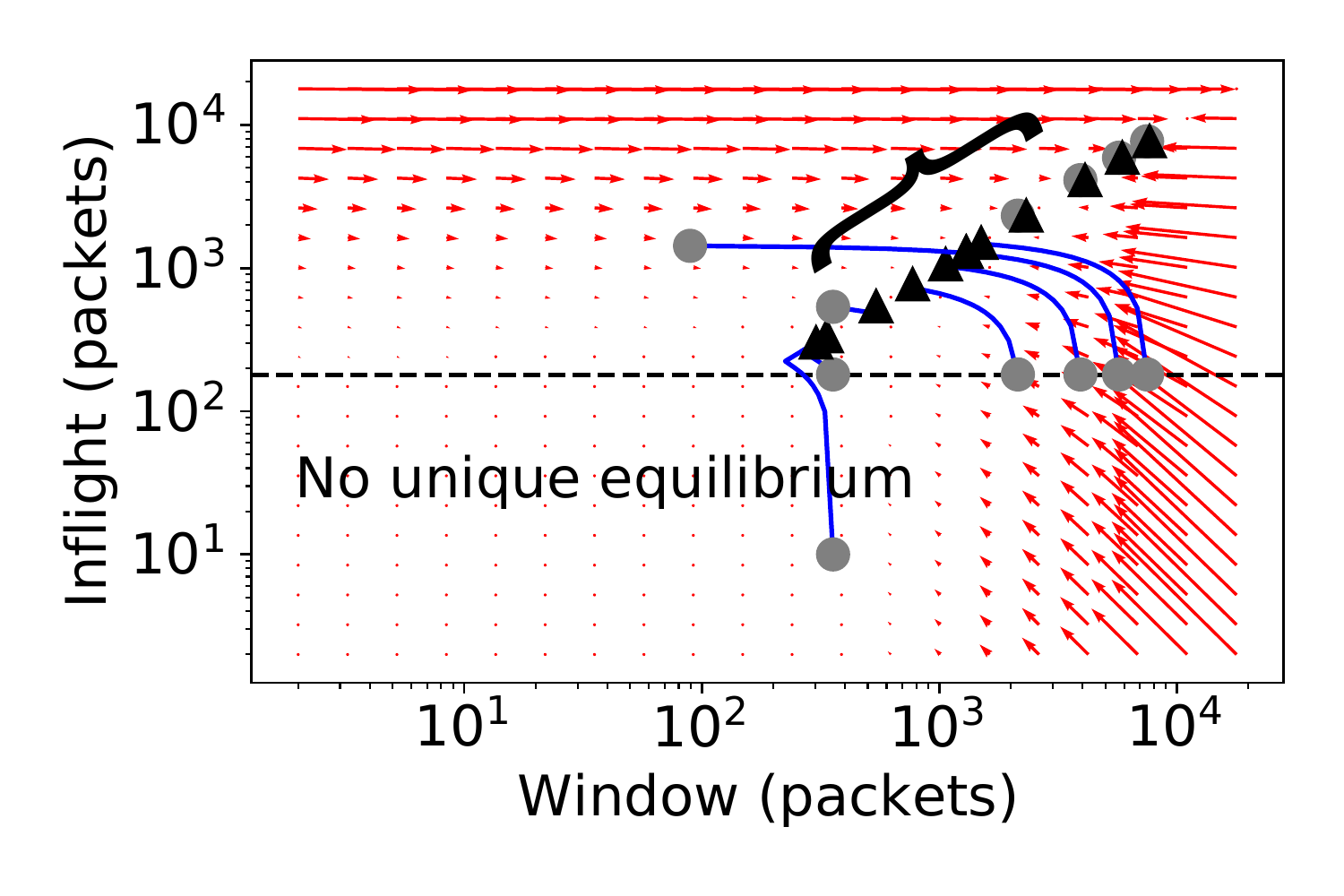}
\caption{Current-based CC (RTT-gradient) reacts faster but has no unique equilibrium point, and is thereby unable to stabilize queue lengths.}
\label{fig:currentPhaseplot}
\end{subfigure}
\hfill
\begin{subfigure}{0.30\linewidth}
\centering
\includegraphics[width=1\linewidth]{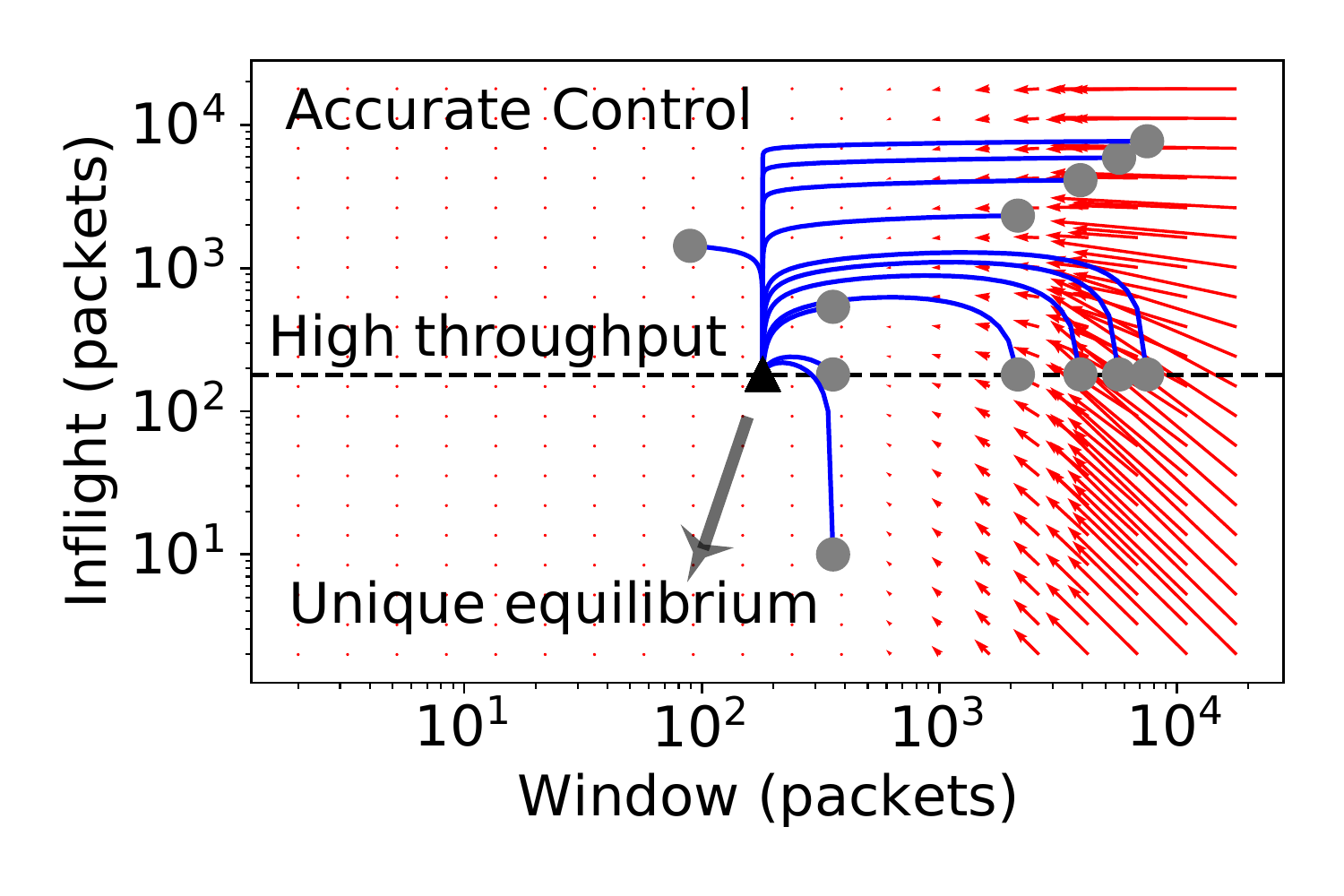}
\caption{\name, a power-based CC, exhibits equilibrium properties and has a precise reaction to perturbations.\\}
\label{fig:gradientTcpPhaseplot}
\end{subfigure}
\caption{Phase plots showing the trajectories of existing schemes and our approach \name from different initial states (circles) to equilibrium (triangles). At each point on the plane, arrows show the direction in which the system moves. An example is depicted with bottleneck link bandwidth $100$Gbps and a base RTT of $20\mu s$. BDP is shown by a horizontal dotted line and any trajectory going below this line indicates throughput loss.}
\label{fig:phaseplots}
\vspace{-2mm}
\end{figure*}

\subsection{Desirable Control Law Properties}
\label{sec:theSearch}

Among various desired properties of datacenter congestion control, high throughput and low tail latency are most important \cite{mittal2015timely, alizadeh2010data, li2019hpcc} with fairness and stability being essential as well \cite{zarchy2019axiomatizing, zhu2016ecn}.
Achieving these properties simultaneously can be challenging. For example, to realize high throughput, we may aim to keep the queue length at the bottleneck link large; however, this may increase latency.
Thus, an ideal CC algorithm must be capable of maintaining near-zero queue lengths, achieving both high throughput and low latency. It must further minimize throughput loss and latency penalty caused by perturbations, such as bursty traffic.

In order to formalize our requirements, we consider a single-bottleneck link model widely used in the literature~\cite{zarchy2019axiomatizing,zhu2016ecn,hollot2001control,misra2000fluid}. Specifically, we assume that all senders use the same protocol, transmit long flows sharing a common bottleneck link with bandwidth $b$, and have a base round trip time $\tau$ (excluding queuing delays).
In this model, equilibrium is a state reached when the window size and bottleneck queue length stabilize.
We now formally express the desired equilibrium state that captures our performance requirements in terms of the sum of window sizes of all flows (aggregate window size) $w(t)$, bandwidth delay product $b\cdot \tau$, and bottleneck queue length $q(t)$:
\begin{equation}
\label{eq:Desire}
0 < q(t) < \epsilon
\end{equation}\vspace{-5mm}
\[
b\cdot\tau  \le w(t) < b\cdot\tau + \epsilon
\]
\[
\dot{q}(t)=0; \ \dot{w}(t)=0
\]

\noindent where $\epsilon$ is a positive integer. First, this captures the requirement for high throughput \ie when $w(t) > b\cdot \tau $ and $ q(t) >0$, the number of inflight bytes are greater than the bandwidth-delay product (BDP) and the queue length is greater than zero. 
Second, from $w(t) < b\cdot\tau + \epsilon$ and $q(t)< \epsilon$, the queue length is at most $\epsilon$, thereby achieving low latency. Finally, for the system to stabilize, we need that $\dot{q}(t)=0$ and $\dot{w}(t)=0$. 

As simple as these requirements are, it is challenging to control the aggregate window size $w(t)$ while CC operates per flow. 
In addition to the equilibrium state requirement, we need fast response to perturbations. 
The response must minimize the distance from the equilibrium \ie minimize the latency or throughput penalty caused by a perturbation (\eg incast or changes in available bandwidth).

In this work, we ask two fundamental questions:

\myitem{\textit{(Q1)} Equilibrium point:} Do existing algorithms satisfy the equilibrium state in Eq.~\ref{eq:Desire} for the aggregate window size?

\noindent In addition to the equilibrium behavior, we are also interested in the reaction to a perturbation.

\myitem{\textit{(Q2)} Response to perturbation:} What is the trajectory followed after a perturbation, \ie the dynamics of the bottleneck queue as well as the TCP window sizes, from an initial point to the equilibrium point?

\subsection{Drawbacks of Existing Control Laws}
\label{sec:problems}

We now aim to analytically answer our questions above and shed light on the inefficiencies of existing protocols, both voltage-based and current-based. 
We begin by simplifying the congestion avoidance model of existing CC approaches we are interested in, specifically delay, queue length, and RTT-gradient based CC approaches as follows:
\begin{equation}
\label{eq:simplifiedModel}
w_i(t+\delta t) = \gamma\cdot \left( w_i(t)\cdot \frac{e}{f(t)} + \beta \right) + (1-\gamma)\cdot w_i(t)
\end{equation}

Here $w_i$ is the window of a flow $i$, $\beta$ is the additive increase term, $e$ is the equilibrium point that the algorithm is expected to reach, $f(t)$ is the measured feedback and $\gamma$ is the exponential moving average parameter.
A queue length-based CC~\cite{li2019hpcc} sets the desired equilibrium point $e$ as $b\cdot\tau$ (BDP) and the feedback $f(t)$ as the sum of bottleneck queue length and BDP \ie $\text{voltage}\ (\nu)$.
A delay-based CC~\cite{kumar2020swift} sets $e$ to $\tau$ (base RTT) and the feedback $f(t)$ as RTT which is the sum of queuing delay and base RTT \ie $\frac{\text{voltage}}{\text{bandwidth}}\ (\frac{\nu}{b})$.
Similarly, the RTT-gradient approach~\cite{mittal2015timely}
sets $e$ to 1 and the feedback $f(t)$ as one plus RTT-gradient \ie $\frac{\text{current}}{\text{bandwidth}}\ (\frac{\lambda}{b})$.
In Appendix~\ref{sec:justification}, we further justify how Eq.~\ref{eq:simplifiedModel} captures existing control laws\footnote{TIMELY, for example, is rate-based while our simplification is window-based. However, window and rate are interchangeable for update calculations.}.
Note that our simplified model does not capture loss/ECN-based CC algorithms; however, there exists rich literature on the analysis of loss/ECN-based CC algorithms~\cite{hollot2001control,980245} including DCTCP~\cite{alizadeh2010data,10.1145/2007116.2007125}.
We now use Euler's first order approximation to obtain the window dynamics as follows:
\begin{equation}
\label{eq:simplifiedModelDynamics}
\dot{w_i}(t) = \frac{\gamma}{\delta t}\cdot \left( w_i(t)\cdot \frac{e}{f(t)} -w_i(t) + \beta \right)
\end{equation}

\noindent Each flow $i$ has a sending rate $\lambda_i$ and hence the bottleneck queue experiences an aggregate arrival rate of $\lambda$.
In our analogy, $\lambda$ is the network current.
We additionally use the traditional model of queue length dynamics which is independent of the control law~\cite{hollot2001control,misra2000fluid}:

\begin{equation}
\label{eq:queueRate}
\dot{q}(t) =
\begin{cases}
\lambda(t-t^f) - \mu(t) & q(t)>0\\
0           & \textit{otherwise}\\
\end{cases}
\end{equation}

\noindent where $\lambda(t)=\frac{w(t)}{\theta(t)}$. An equilibrium point is the window size $w_e$ and queue length $q_e$ that satisfies $\dot{w}(t)=0$ and $\dot{q}(t)=0$.

We are now ready to answer the questions raised.

\myitem{Equilibrium point:}
It is well-known from literature that loss/ECN-based schemes operate by maintaining a standing queue~\cite{hollot2001control,10.1145/2007116.2007125,10.1145/52325.52356}. For example, TCP NewReno flows fill the queue to maximum (say $q_{max}$) and then react by reducing windows by half. Consequently, the bottleneck queue-length oscillates between $q_{max}$ and $q_{max}-b\cdot\tau$ or zero if $q_{max}<b\cdot\tau$. DCTCP flows oscillate around the marking threshold $K>\frac{b\cdot\tau}{7}$ which depends on BDP~\cite{alizadeh2010data}. This does not satisfy our stringent requirement in Eq.~\ref{eq:Desire}. While ECN-based schemes reduce the amount of standing queue required, we still consider the standing queue to be unacceptable given the stringent latency requirements.

It can be shown that there exists a unique equilibrium point for queue length and delay approaches (voltage-based CC) defined by Eq.~\ref{eq:simplifiedModel}.
However, current-based CC and, in particular, RTT-gradient approaches do not have a unique equilibrium point suggesting a lack of control over queue lengths. Intuitively, RTT-gradient approaches quickly adapt the sending rate to stabilize the RTT-gradient ($\dot{\theta}=\frac{\dot{q}}{b}$) which in turn only stabilizes the queue length gradient $\dot{q}(t)$ but fails to control the absolute value of the queue length. It has indeed been shown that TIMELY, a current-based CC does not have a unique equilibrium~\cite{zhu2016ecn}.

Figure~\ref{fig:phaseplots} visualizes the system behavior according to the window dynamics in Eq.~\ref{eq:simplifiedModelDynamics} and the queue dynamics in Eq.~\ref{eq:queueRate}.
In Figure~\ref{fig:voltagePhaseplot} we can see that voltage-based CC eventually reaches a unique equilibrium point.
In contrast, in Figure~\ref{fig:currentPhaseplot} we see that current-based CC reaches different final points for different initial points, indicating that there exists no unique equilibrium point thereby violating the desired equilibrium state properties (Eq.~\ref{eq:Desire}).
To give more context on this observation, in Figure~\ref{fig:drawbacks} we show the reactions of different schemes for observed queue lengths and queue buildup rate. In Figure~\ref{fig:trendDrawbacks}, we can see that current-based CC has the same reaction for different queue lengths but exhibits a proportional reaction to queue buildup rate (Figure~\ref{fig:currentDrawbacks}); consequently, current-based CC cannot stabilize at a unique equilibrium point.
Due to space constraints, we move the detailed proof of equilibrium points to Appendix ~\ref{sec:justification}.

\textit{\textbf{Takeaway.} While voltage-based CC can in principle meet the desired equilibrium state requirements in Eq.~\ref{eq:Desire}, current-based CC cannot.}

\myitem{Response to perturbation:}
We observe an orthogonal behavior in the responses of voltage-based CC and current-based CC.
In Figure~\ref{fig:trendDrawbacks} we show that voltage-based CC has a proportional reaction to increased queue lengths but a current-based CC approach has the same response for any queue length.
Further in Figure~\ref{fig:currentDrawbacks} we observe that current-based CC has a proportional reaction to the rate at which queue is building up but a voltage-based CC has the same reaction for any rate of queue build up.
This orthogonality in existing schemes often results in scenarios with either insufficient reaction or overreaction. 
To underline our observation, we use the system of differential equations (Eq.~\ref{eq:simplifiedModelDynamics} and Eq.~\ref{eq:queueRate}) to observe the trajectories taken by different control laws after a perturbation.
We show the trajectories in Figure~\ref{fig:phaseplots}. 
Specifically, Figure~\ref{fig:voltagePhaseplot} shows that voltage-based CC (queue length or delay based) eventually reaches a unique equilibrium point but overreacts in the response and losing throughput (window $<$ BDP and $q(t)=0$) almost for every initial point.
In Figure~\ref{fig:currentPhaseplot} we observe that current-based CC (RTT-gradient) reaches different end points for different initial states and consequently does not have a single equilibrium point.
However, we see that the initial response is faster with current-based CC due to their use of RTT-gradient which is arguably a superior signal to detect congestion onset even at low queue lengths.

\textit{\textbf{Takeaway.} Current-based CC is superior in terms of fast reaction but lacks equilibrium state properties while voltage-based CC eventually reaches a unique equilibrium but overreacts in its response for almost any initial state resulting in long trajectories from initial state to equilibrium state.}

\subsection{Lessons Learned and Design Goals}
\label{sec:lessonsGoals}

From our analysis we derive two key observations.
First, both voltage and current-based CC have individual benefits. 
Particularly, voltage-based CC is desirable for the stringent equilibrium properties we require and current-based CC is desirable for fast reaction.
Second, both voltage and current-based CC have drawbacks.
On one hand, voltage-based CC is oblivious to congestion onset at low queue lengths and on the other hand current-based CC is oblivious to the absolute value of queue lengths.
Moreover, voltage-based CC overreacts when the queue drains essentially losing throughput immediately after.

Based on these observations, our goal is to design a control law that systematically combines both voltage and current for every window update action. Specifically our aim is to design a congestion control algorithm with \first equilibrium properties from Eq.~\ref{eq:Desire} exhibited by voltage-based CC and \second fast response to perturbation exhibited by current-based CC. The challenges are to avoid inheriting the drawbacks of both types of CC, stability and fairness. 
However in order to design such a control law we face the following challenges:

\begin{itemize}
    \item Finding an accurate relationship between window, voltage and current. \hfill \raggedright\textcolor{blue}{\Comment{Property~\ref{prop:window}}}
    \item Ensuring stability, convergence and fairness. \\
    \hfill\raggedleft\textcolor{blue}{\Comment{Theorem~\ref{theorem:stability},~\ref{theorem:convergence},~\ref{theorem:fairness}}}
\end{itemize}

\section{Power-Based Congestion Control}
\label{sec:ourApproach}

Reflecting on our observations in \S\ref{sec:motivation}, 
we seek to design a congestion control algorithm that systematically reacts to both the absolute value of the bottleneck queue length and its rate of change.
Our aim is to address today's datacenter performance requirements in terms of high throughput, low latency, and fast reaction to bursts and bandwidth fluctuations. 

\subsection{The Notion of Power}
\label{sec:power}
To address the challenges faced by prior datacenter congestion control algorithms and to optimize along both dimensions, we introduce the notion of \emph{power} associated with the network pipe. Following the bottleneck link model from literature~\cite{hollot2001control,misra2000fluid}, from Eq.~\ref{eq:queueRate} we observe that the window size is indeed related to the product of network voltage and network current which we call \textit{power} (Table~\ref{table:analogy}).
This corresponds to the product of \first total sending rate $\lambda$ (current) and \second the sum of BDP plus the accumulated bytes $q$ at the bottleneck link (voltage), formally expressed in Eq.~\ref{eq:power}. 
\begin{equation}
\label{eq:power}
\underbrace{\Gamma(t)}_{\text{power}}  = \underbrace{(q(t) + b \cdot \tau)}_{\text{voltage}} \cdot \underbrace{\lambda(t-t^f)}_{\text{current}} 
\end{equation}
Notice that the unit of power is $\frac{bit^2}{second}$. 
We will show the useful properties of power specifically under congestion.
\noindent Using Eq.~\ref{eq:queueRate}, we can rewrite Eq.~\ref{eq:power} in terms of queue length gradient $\dot{q}$ and the transmission rate $\mu$ as,
\begin{equation}
\label{eq:powerExtended}
\Gamma(t) = (q(t) + b \cdot \tau) \cdot (\dot{q}(t)+\mu(t))
\end{equation}
\noindent We now derive a useful property of power using Eq.~\ref{eq:powerExtended} and Eq.~\ref{eq:queueRate} showing an accurate relationship of power and window.\begin{property}[Relationship of Power and Congestion Window]
\label{prop:window}
Power is the bandwidth-window product
\[
\Gamma(t) = b\cdot w(t-t^f)
\]
\end{property}

\noindent Note that the property is over the aggregate window size \ie the sum of window sizes of all flows sharing the common bottleneck. 
We emphasize that our notion of power is intended for the networking context and cannot be applied to other domains of science.
In the following, we outline the benefits of considering the notion of power and how Property~\ref{prop:window} can be  useful in the context of congestion control.

\subsection{Benefits of Power-Based CC}\label{sec:power-benefits}
A power-based control law can exploit Property~\ref{prop:window} to precisely update per flow window sizes.
Accurately controlling aggregate window size is a key challenge for an end-host congestion control algorithm.
A power-based CC overcomes this challenge by gaining precise knowledge about the aggregate window size from measured power. 
First, using power enables the window update action to account for the bottleneck queue lengths as well as the queue build-up rate.
As a result, a power-based CC can rapidly detect congestion onset even at very low queue lengths.
At the same time, a power-based CC also reacts to the absolute value of queue lengths, effectively dampening perturbations.
Second, calculating power at the end-host requires no extra measurement and feedback mechanisms compared to INT based schemes such as HPCC~\cite{li2019hpcc}.

\subsection{The \name Algorithm}
\label{sec:algorithm}
Driven by our observations, we carefully designed our control law based on power, capturing a systematic reaction to voltage (related to bottleneck queue length), as well as to current (related to variations in the bottleneck queue length).

\myitem{Control law: }\name is a window-based congestion control algorithm and updates its window size upon receipt of an acknowledgment.
For a flow $i$, every window update is based on \first current window size $w_i(t)$, \second additive increase $\beta$, \third window size at the time of transmission of the acknowledged segment $w_i(t-\theta(t))$, and \fourth power measured from the feedback information.
We refer the reader to Table~\ref{table:notations} for the general notations being used. 
Formally, \name's control law can be expressed as
\begin{equation}
\label{eq:controlLawMain}
w_i(t) \leftarrow \gamma\cdot \left( w_i(t-\theta(t))\cdot \frac{e}{f(t)} + \beta \right) + (1-\gamma)\cdot w_i(t)
\end{equation}
\[
e=b^2\cdot \tau; \ \ \ \ f(t) = \Gamma(t-\theta(t)+t^f)
\]
\noindent where $\gamma\in(0,1]$ and $\beta$ are parameters to the control law.
The base round trip time $\tau$ must be configured at compile time.
If baseRTT is not precisely known, an alternative is to keep track of minimum observed RTT.
We first describe how power $\Gamma$ is computed and then present the pseudocode of \name in Algorithm~\ref{alg:tcp}.

\myitem{Feedback: } \name's control law is based on power.
Note that power (Eq.~\ref{eq:power}) is only related to variables at the bottleneck link.
In order to measure power, we leverage in-band network telemetry.
Specifically, the workings of INT and the header fields required are the same as in HPCC (Figure. 4 in~\cite{li2019hpcc}).
When a TCP sender sends out a packet \scalerel*{\includegraphics{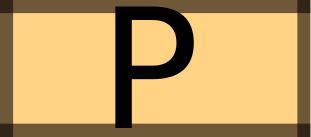}}{B} into the network, it additionally inserts an INT header \scalerel*{\includegraphics{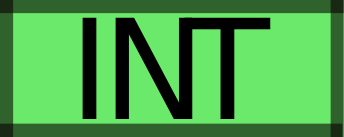}}{B} into the packet.
Each switch along the path then pushes metadata containing the egress queue length ($qlen$), timestamp ($ts$), so far transmitted bytes ($txBytes$), and bandwidth ($b$). All values correspond to the time when the packet is scheduled for transmission.
At the receiver, the received packet \scalerel*{\includegraphics{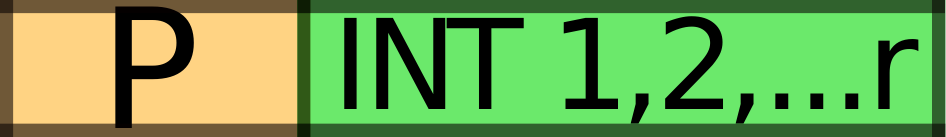}}{B} is read and the INT information is copied to the acknowledgment $ACK$ packet  \scalerel*{\includegraphics{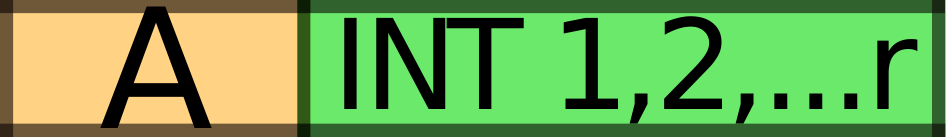}}{B}.
The sender then receives an $ACK$ with an INT header and metadata inserted by all the switches along the path from sender to receiver and back to sender \scalerel*{\includegraphics{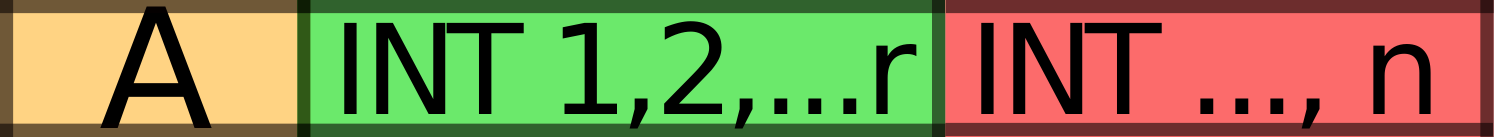}}{B}.
Here, the INT header and meta-data pushed by switches along the path serve as feedback and as an input to the CC algorithm.

\myitem{Accounting for the old window sizes: } \name's control law (Eq.~\ref{eq:controlLawMain}) uses the past window size in addition to the current window size to compute the new window size. \name accounts for old window size by remembering current window size once per RTT.

\myitem{Algorithm: }
Putting it all together, we now present the workflow of \name in Algorithm~\ref{alg:tcp}. Upon the receipt of a new acknowledgment (line~\ref{line:newack}), \name 
: \first retrieves the old $cwnd$ (line~\ref{line:getcwnd}), \second computes the normalized power (line~\ref{line:normpower}) \ie $\frac{f(t)}{e}$ in Eq.~\ref{eq:controlLawMain}, \third updates $cwnd$ (line~\ref{line:updatewindow}), \fourth sets the pacing rate (line~\ref{line:rate}), and \fifth remembers the INT header metadata and updates the old $cwnd$ once per RTT based on the $ack$ sequence number (line~\ref{line:prevint}). 

Specifically, power is calculated in the function call to \textsc{\textcolor{red}{normPower}}.
First, the gradient of queue lengths is obtained from the difference in queue lengths and difference in timestamps corresponding to an egress port (line~\ref{line:queueGradient}).
Then the transmission rate of the egress port is calculated from the difference in $txBytes$ and timestamps (line~\ref{line:txrate}).
Current is calculated by adding the queue gradient and transmission rate (line~\ref{line:current}).
Then, the sum of BDP and the queue length gives voltage (line~\ref{line:voltage}).
Finally, power is calculated by multiplying current and voltage (line~\ref{line:power}).
We calculate the base power (line~\ref{line:basePower}) and obtain the normalized power (line~\ref{line:normpower}).
The normalized power is calculated for each egress port along the path and the maximum value is smoothed and used as an input to the control law.

Finally, the congestion window is updated in the function call to \textsc{\textcolor{red}{updateWindow}} (line~\ref{line:fnUpdatewindow}) where $\gamma$ is the exponential moving average parameter and $\beta$ is the additive increase parameter, both being parameters to the control law (Eq.~\ref{eq:controlLawMain})

\begin{algorithm}[h]
    \SetKwFunction{newAck}{\textsc{\textcolor{red}{newAck}}}
    \SetKwFunction{updateWindow}{\textsc{\textcolor{red}{updateWindow}}}
    \SetKwFunction{normPower}{\textsc{\textcolor{red}{normPower}}}

    \SetKwProg{Fn}{function}{:}{}
    \SetKwProg{Proc}{procedure}{:}{}
    \SetKwInOut{KwIn}{Input}
    \SetKwInOut{KwOut}{Output}

    \tcc{ack contains an INT header with sequence of per-hop egress port meta-data accessed as ack.H[i]}
    \KwIn{\ $ack$ and $prevInt$}
    \KwOut{\ $cwnd$, $rate$}

    \Proc{\newAck{$ack$}}{\label{line:newack}
      
      $cwnd_{old}$ = \textsc{getCwnd}($ack.seq$) \label{line:getcwnd}
      
      $normPower =$ \textsc{normPower}($ack$) \label{line:normpower}

      \textsc{updateWindow}($normPower$, $cwnd_{old}$) \label{line:updatewindow}

      $rate = \frac{cwnd}{\tau}$ \label{line:rate}

      $prevInt = ack.H$; \textsc{updateOld}$(cwnd, ack.seq)$ \label{line:prevint}

    }

    \Fn{\normPower{ack}}{
      $\Gamma_{norm} = 0$
      
      \For{each egress port i on the path}{
      
        $ dt  = ack.H[i].ts - prevInt[i].ts$ 

        $\dot{q} = \frac{ack.H[i].qlen - prevInt[i].qlen}{dt}$ \Comment{$\frac{dq}{dt}$} \label{line:queueGradient}

        $\mu = \frac{ack.H[i].txBytes - prevInt[i].txBytes}{dt} $ \Comment{txRate}  \label{line:txrate}

        $\lambda = \dot{q} + \mu$ \Comment{$\lambda:$ Current} \label{line:current}

        $BDP = ack.H[i].b\times\tau $ \label{line:bdp}

        $\nu = ack.H[i].qlen + BDP$ \Comment{$\nu:$ Voltage} \label{line:voltage}

        $\Gamma^{'} = \lambda\times\nu $ \Comment{$\Gamma^{'}:$ Power} \label{line:power}
        
        $e = (ack.H[i].b)^2 \times \tau$ \label{line:basePower}

        $\Gamma^{'}_{norm} = \frac{\Gamma^{'}}{e} $ \Comment{$\Gamma^{'}_{norm}:$Normalized power} \label{line:normpower}

        \If{ $ \Gamma^{'} > \Gamma_{norm}$ }{

          $\Gamma_{norm} = \Gamma^{'}$; $\Delta t = dt$

        }
      }
      $\Gamma_{smooth} = \frac{(\Gamma_{smooth}\cdot(\tau - \Delta t)+(\Gamma_{norm}\cdot \Delta t)}{\tau}$ \Comment{Smoothing}
      
      \KwRet{$ \Gamma_{smooth}$}
    }

    \Fn{\updateWindow{$power$, $ack$}}{\label{line:fnUpdatewindow}

      $cwnd = \gamma\times( \frac{cwnd_{old}}{normPower} + \beta ) + (1-\gamma)\times cwnd $

      \Comment{$\gamma:$ EWMA parameter}

      \Comment{$\beta$: Additive Increase}

      \KwRet{$cwnd$}
    }
    
    \caption{\name}
    \label{alg:tcp}
\end{algorithm}

\myitem{Parameters: } \name has only two parameters, that is the EWMA parameter $\gamma$ and the additive increase parameter $\beta$.
$\gamma$ dictates the balance in reaction time and sensitivity to noise.
We recommend $\gamma=0.9$ based on our parameter sweep over wide range of scenarios including traffic patterns that induce rapid fluctuations in the bottleneck queue lengths.
Reflecting the intuition for additive increase in prior work~\cite{li2019hpcc}, we set $\beta=\frac{HostBw\times \tau}{N}$ where $N$ is the expected number of flows sharing host NIC, $HostBw$ is the NIC bandwidth at the host and $\tau$ is the base-RTT.
This is to avoid queuing at the local interface or, in other words, to avoid making the host NIC a bottleneck, assuming a maximum of $N$ flows share the host NIC bandwidth.
Finally, all flows transmit at line rate in the first RTT and use $cwnd_{init}=HostBw\times\tau$.
By transmitting at line rate, a new flow is able to discover the bottleneck link state and reduce its $cwnd$ accordingly without getting throttled due to the presence of existing flows.

\subsection{Properties of \name}
\label{sec:theoryProps}
\name comes with strong theoretical guarantees. We show that \name's control law achieves asymptotic stability with a unique equilibrium point that satisfies our desired equilibrium state properties (Eq.~\ref{eq:Desire}).
\name also guarantees rapid convergence to equilibrium and achieves fairness at the same time.
In the following we outline \name's properties and defer the proofs to Appendix~\ref{sec:AppendixAnalysis}.

\begin{restatable}[Stability]{theorem}{stabilitytheorem}
\label{theorem:stability}
\name's control law is Lyapunov-stable as well as asymptotically stable with a unique equilibrium point.
\vspace{-3mm}
\end{restatable}
\begin{restatable}[Convergence]{theorem}{convergencetheorem}
\label{theorem:convergence}
After a perturbation, \name's control law exponentially converges to equilibrium with a time constant $\frac{\delta t}{\gamma}$ where $\delta t$ is the window update interval. 
\vspace{-6mm}
\end{restatable}
\begin{restatable}[Fairness]{theorem}{fairnesstheorem}
\label{theorem:fairness}
\name is $\beta_i$ weighted proportionally fair, where $\beta_i$ is the additive increase used by a flow $i$.
\vspace{-3mm}
\end{restatable}

Theorem~\ref{theorem:stability} and Theorem~\ref{theorem:convergence} state the key properties of \name.
First, the convergence with time constant of $\frac{\delta t}{\gamma}$ shows the fast reaction to perturbations.
Second, the system being asymptotically stable at low queue lengths satisfies our stringent equilibrium property discussed in \S\ref{sec:motivation}.
Indeed, \textbf{power} and Property~\ref{prop:window} play a key role in the proof of Theorem~\ref{theorem:stability} and Theorem~\ref{theorem:convergence} (Appendix~\ref{sec:AppendixAnalysis}) revealing its importance in congestion control.
In Figure~\ref{fig:gradientTcpPhaseplot}, we see the trajectories of \name from different initial states to a unique equilibrium without violating throughput and latency requirements, showing the accurate control enabled by power-based congestion control.

\subsection{\nameapprox: Standalone Version}
\name's control law requires in-network queue length information which can be obtained by using techniques such as INT. In order to widen its applicability, \name can still be deployed in datacenters with legacy, non-programmable switches through accurate RTT measurement capabilities at the end-host.
In this case, we rearrange term  $\frac{e}{f}$ in Eq.~\ref{eq:controlLawMain} as follows,
\[
\frac{e}{f} = \frac{b^2\cdot\tau}{\Gamma} = \frac{b^2\cdot\tau}{(\dot{q}+b)\cdot(q+b\cdot\tau)}=\frac{\tau}{(\frac{\dot{q}}{b}+1)\cdot(\frac{q}{b}+\tau)}
\]
\noindent finally, using the fact that $\frac{q}{b}+\tau = \theta$ (RTT) and $\frac{\dot{q}}{b}=\dot{\theta}$ (RTT-gradient), we reduce $\frac{e}{f}$ to,

\begin{equation}
\label{eq:noIntControl}
\frac{e}{f} = \frac{\tau}{(\dot{\theta}+1)\cdot(\theta)}
\end{equation}

\noindent where $\dot{\theta}$ is the RTT-gradient and $\theta$ is RTT. Using Eq.~\ref{eq:noIntControl} in Eq.~\ref{eq:controlLawMain} allows for deployment even when INT is not supported by switches in the datacenter.
Due to space constraints we moved the algorithm to Appendix~\ref{appendix:approxPowerTCP}, presenting \nameapprox in Algorithm~\ref{alg:tcpnoInt}. This algorithm demonstrates how \name's control law can be mimicked by using a delay signal without the need for switch support.
However, as we will show later in our evaluation, there are drawbacks in using RTT instead of queue lengths.
First, notice how queue lengths are changed to RTT, where we assume bottleneck $txRate$ ($\mu$) as bandwidth ($b$).
The implication is that, when using $txRate$ which is essentially obtained from INT, the control law knows the exact transmission rate and rapidly fills the available bandwidth.
But, when using RTT, the control law assumes the bottleneck is at maximum transmission rate and does not react by multiplicative increase and rather relies on slow additive increase to fill the available bandwidth.
Secondly, in multi-bottleneck scenarios, the control law precisely reacts to the most bottlenecked link when using INT but reacts to the sum of queuing delays when using RTT. 
Nevertheless, under congestion, both \name and \nameapprox have the same properties in a single-bottleneck scenario.

\begin{figure*}
\centering
\begin{subfigure}{0.24\linewidth}
\includegraphics[width=0.9\linewidth]{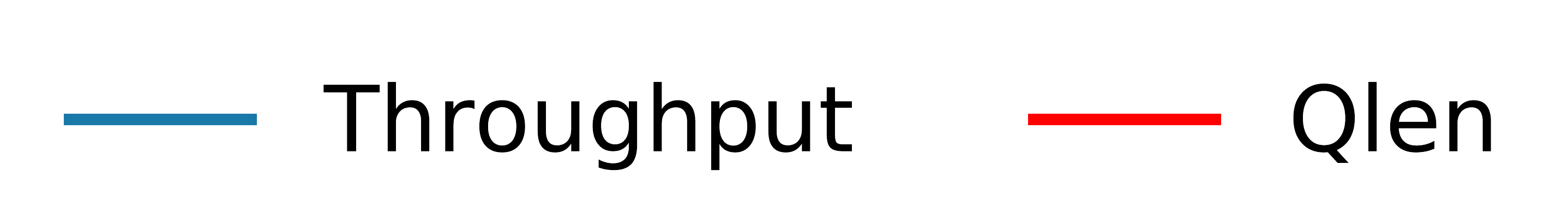}
\end{subfigure}
\begin{subfigure}{0.24\linewidth}
\includegraphics[width=0.9\linewidth]{plots/Powertcp-NSDI/burst/burst-legend.pdf}
\end{subfigure}
\begin{subfigure}{0.24\linewidth}
\includegraphics[width=0.9\linewidth]{plots/Powertcp-NSDI/burst/burst-legend.pdf}
\end{subfigure}
\begin{subfigure}{0.24\linewidth}
\includegraphics[width=0.9\linewidth]{plots/Powertcp-NSDI/burst/burst-legend.pdf}
\end{subfigure}

\begin{subfigure}{0.19\linewidth}
\includegraphics[width=1\linewidth]{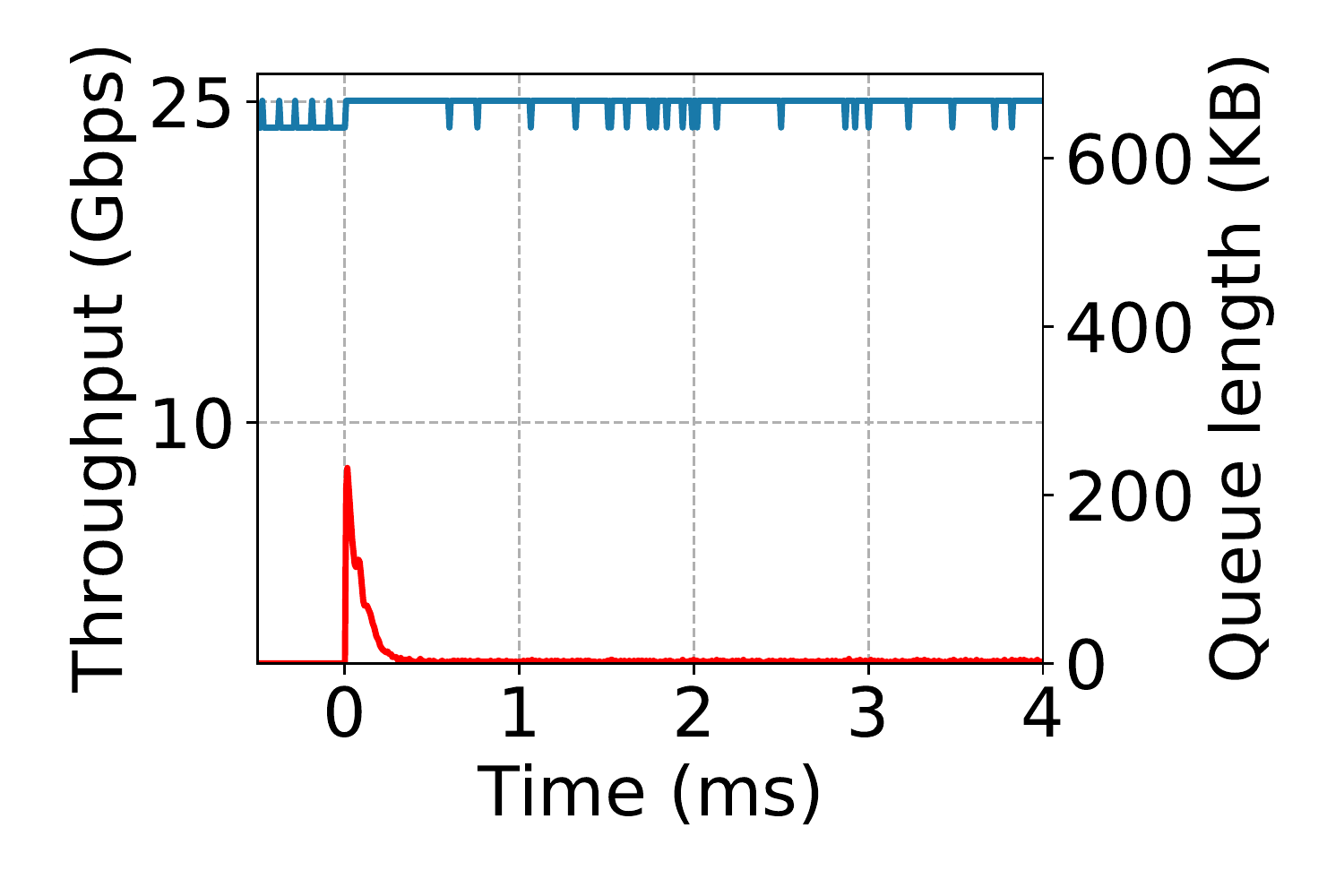}
\label{fig:bursts-powerint}
\end{subfigure}
\begin{subfigure}{0.19\linewidth}
\includegraphics[width=1\linewidth]{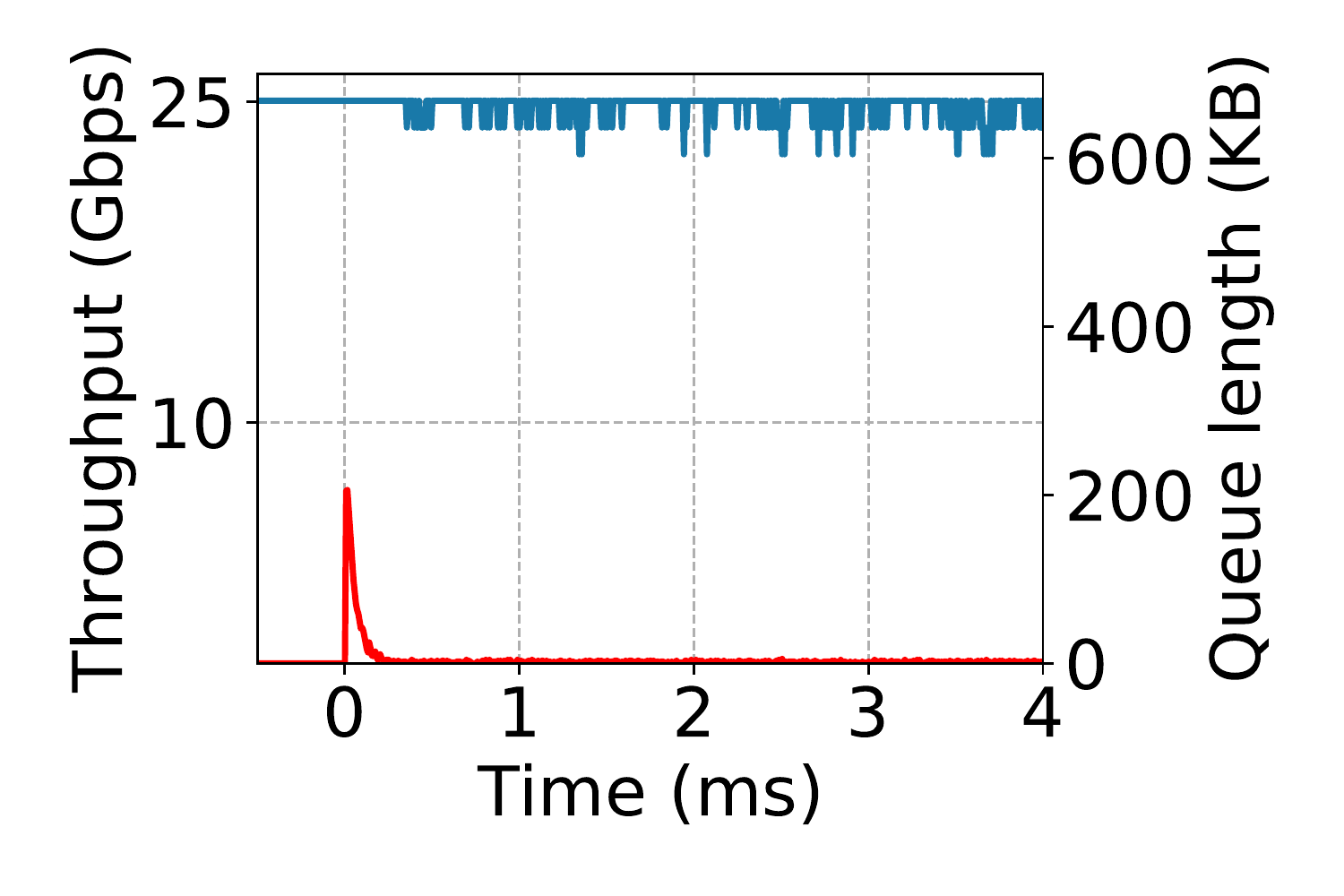}
\label{fig:bursts-powerdelay}
\end{subfigure}
\begin{subfigure}{0.19\linewidth}
\includegraphics[width=1\linewidth]{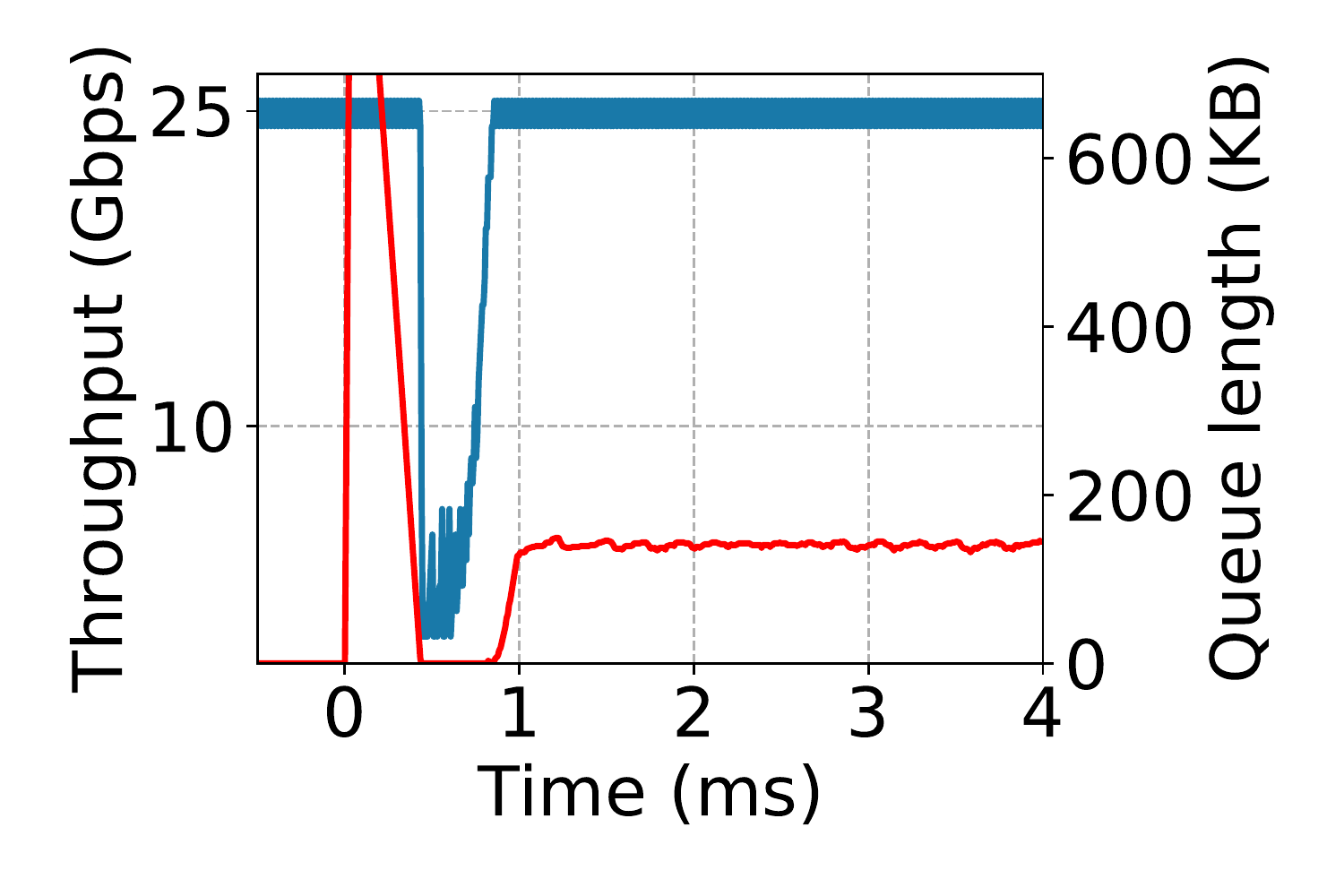}
\label{fig:bursts-timely}
\end{subfigure}
\begin{subfigure}{0.19\linewidth}
\includegraphics[width=1\linewidth]{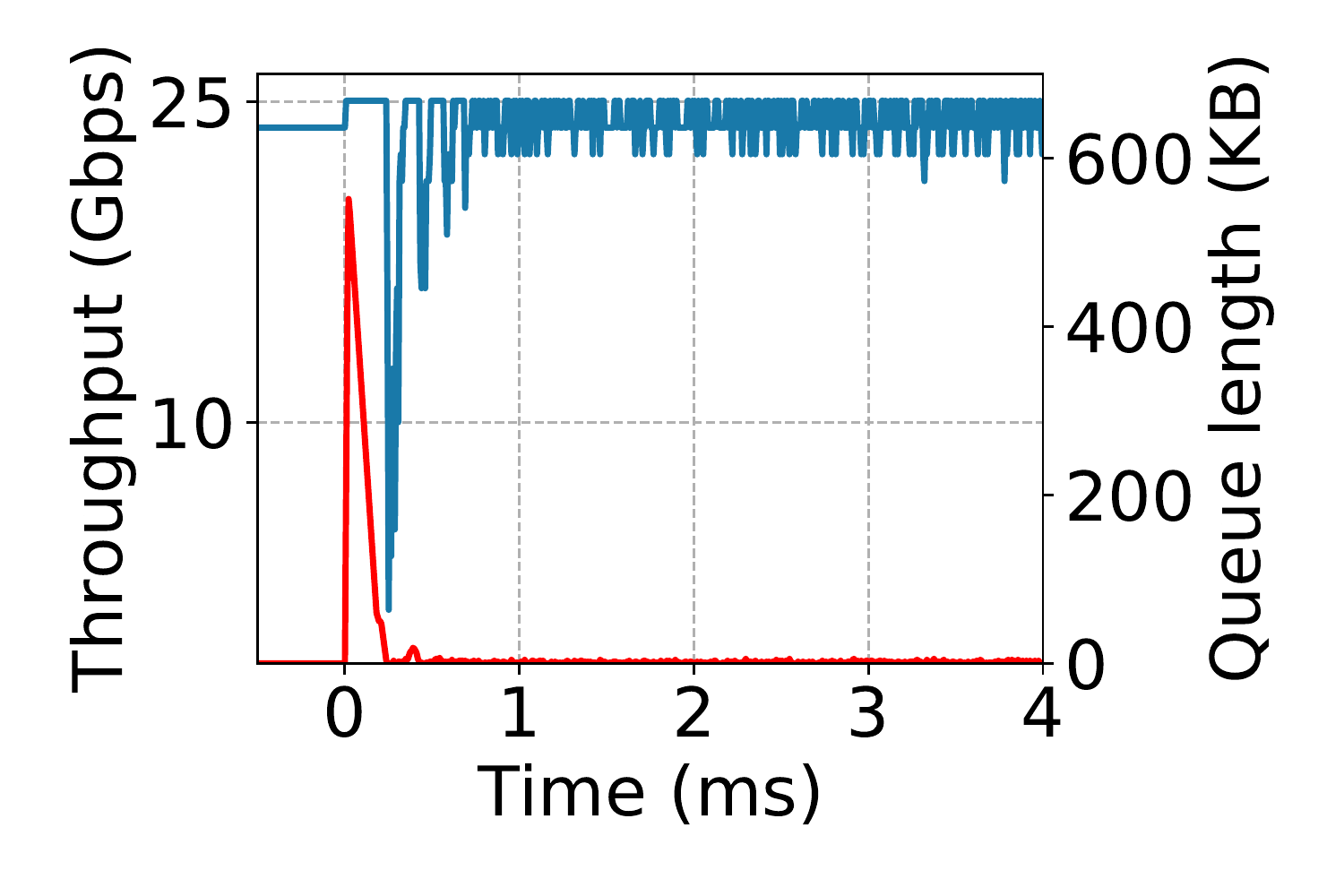}
\label{fig:bursts-hpcc}
\end{subfigure}
\begin{subfigure}{0.19\linewidth}
\includegraphics[width=1\linewidth]{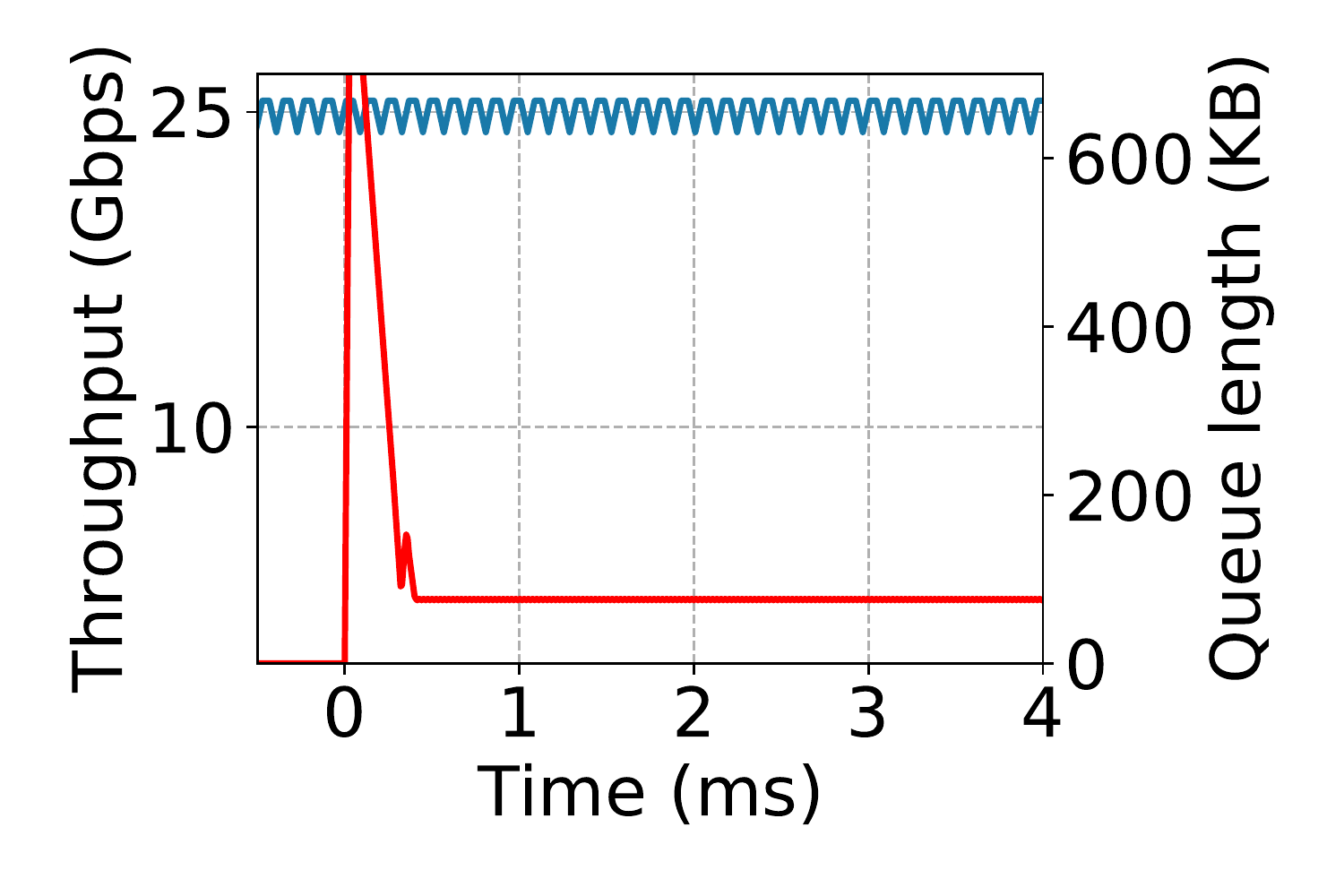}
\label{fig:bursts-homa}
\end{subfigure}\vspace{-6mm}

\begin{subfigure}{0.19\linewidth}
\includegraphics[width=1\linewidth]{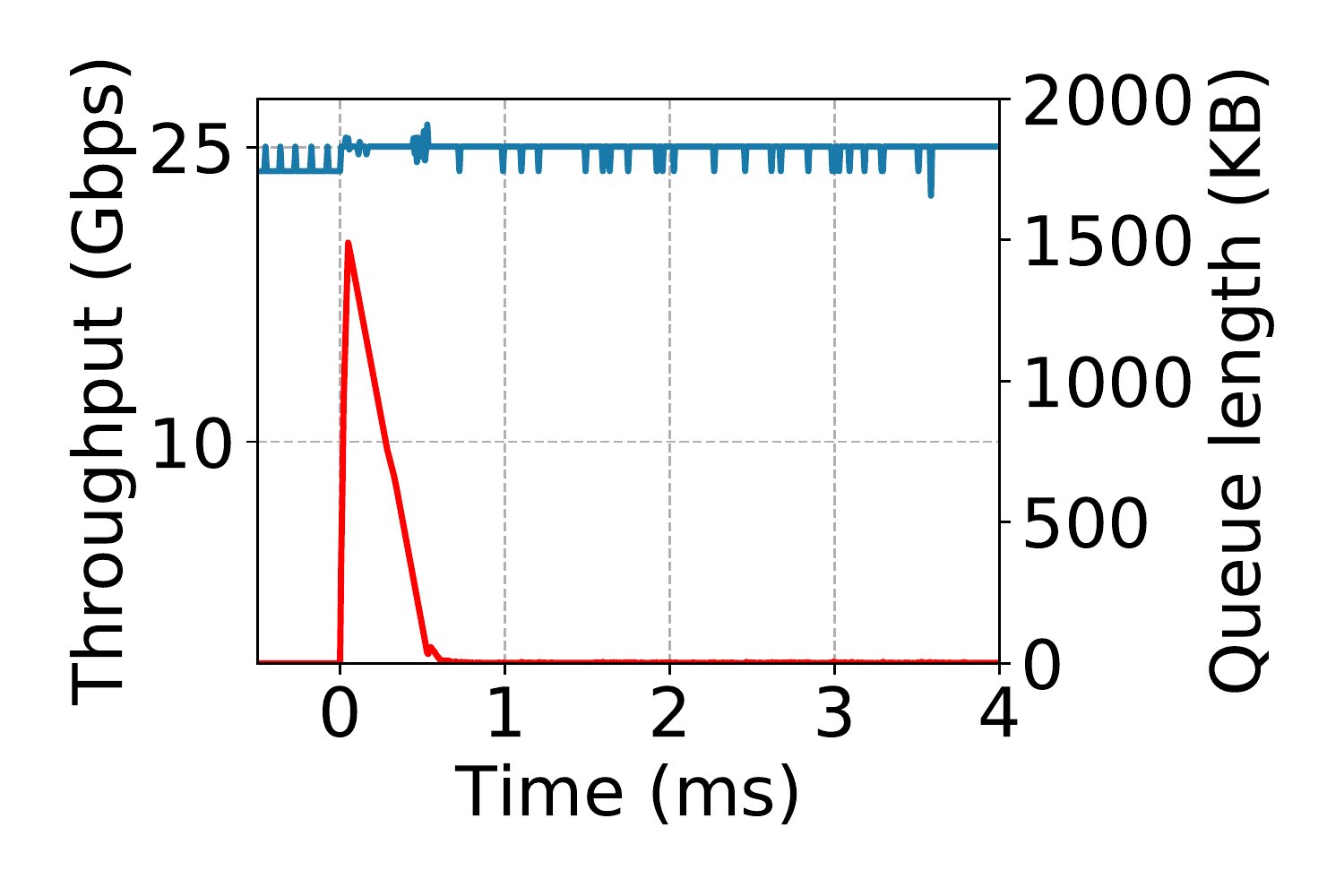}
\caption{\name}
\label{fig:bursts-powerint}
\end{subfigure}
\begin{subfigure}{0.19\linewidth}
\includegraphics[width=1\linewidth]{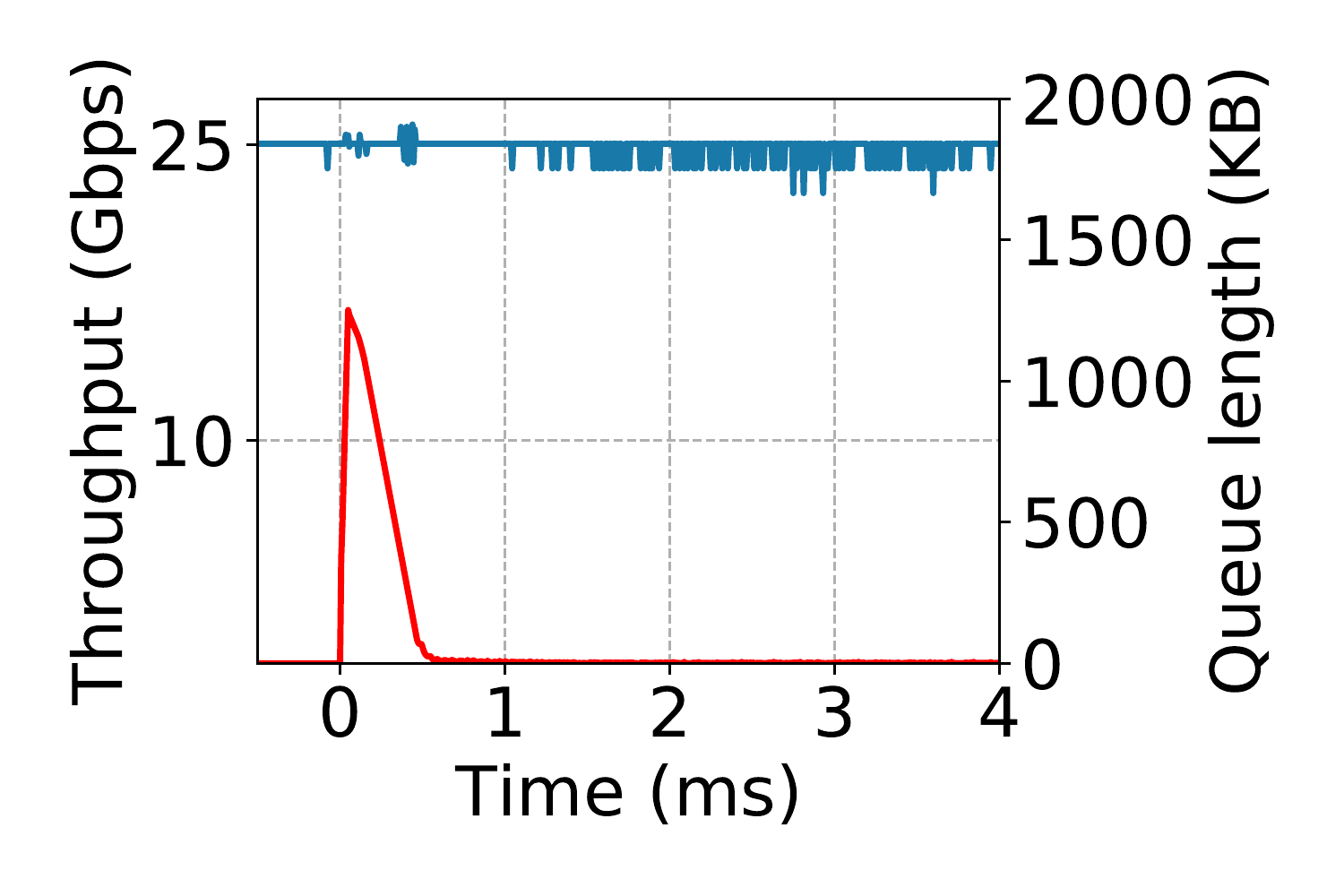}
\caption{\nameapprox}
\label{fig:bursts-powerdelay}
\end{subfigure}
\begin{subfigure}{0.19\linewidth}
\includegraphics[width=1\linewidth]{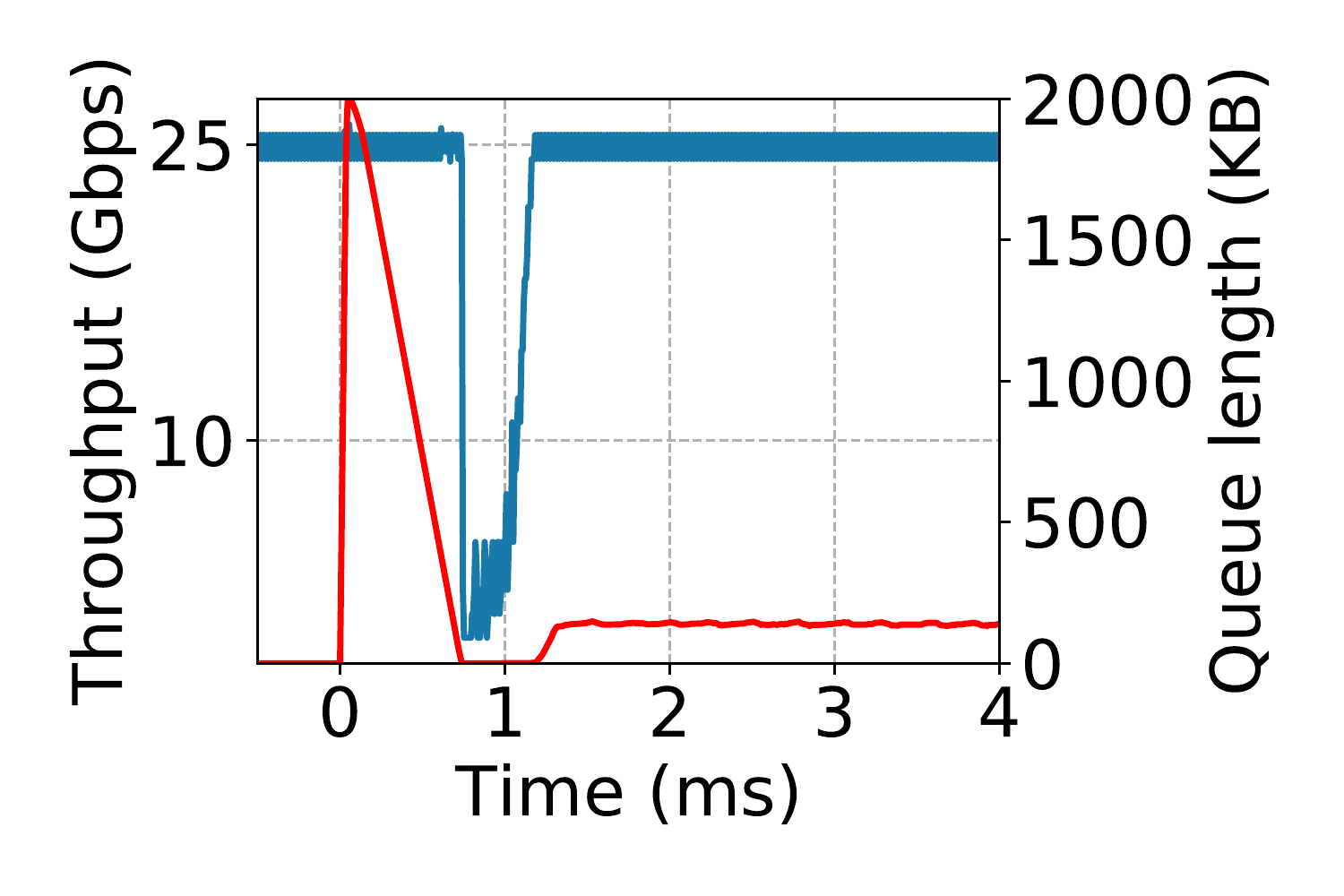}
\caption{TIMELY}
\label{fig:bursts-timely}
\end{subfigure}
\begin{subfigure}{0.19\linewidth}
\includegraphics[width=1\linewidth]{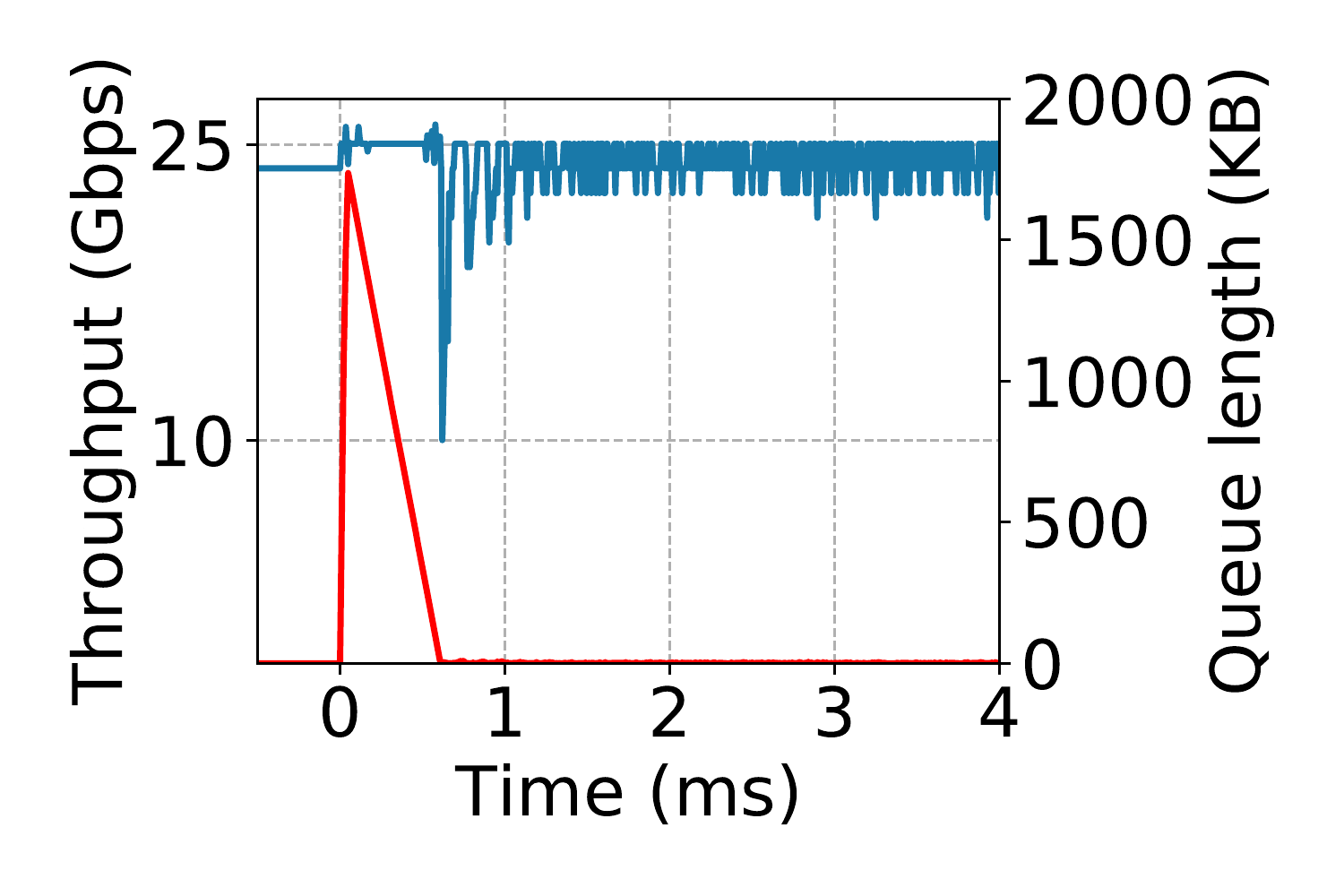}
\caption{HPCC}
\label{fig:bursts-hpcc}
\end{subfigure}
\begin{subfigure}{0.19\linewidth}
\includegraphics[width=1\linewidth]{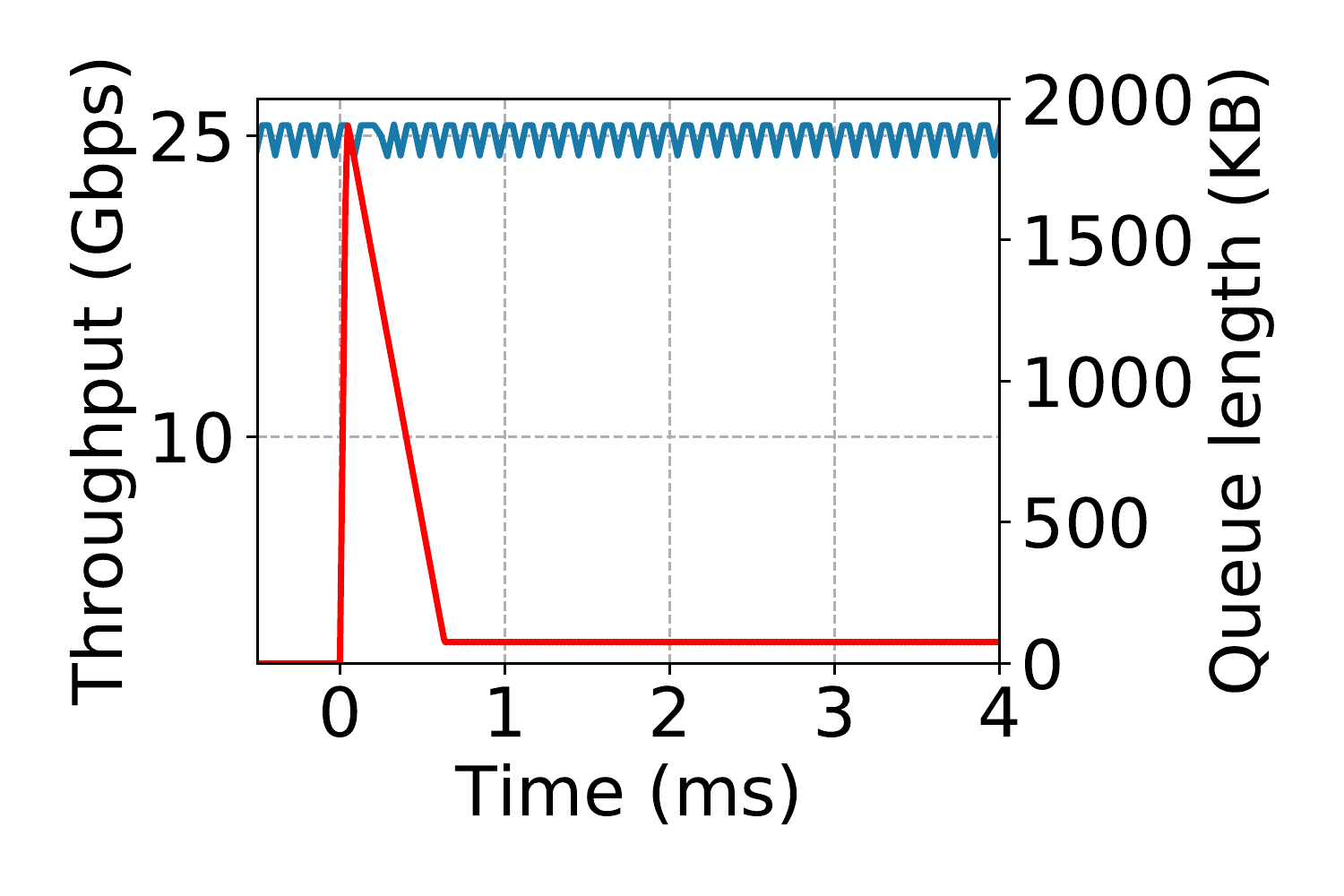}
\caption{HOMA}
\label{fig:bursts-homa}
\end{subfigure}
\caption{State-of-the-art congestion control algorithms vs \name in response to an incast. For each algorithm, we show the corresponding reaction to $10:1$ incast in the top row and to $255:1$ incast in the bottom row.}
\label{fig:bursts}
\vspace{-5mm}
\end{figure*}

\subsection{Deploying \name}

Modern programmable switches are able to export user-defined header fields and device metrics \cite{tofino, kim2015band}.
These metrics can be embedded into data packets, a mechanism commonly referred to as in-band network telemetry (INT).
\name leverages INT to obtain fine-grained, per-packet feedback about queue occupancies, traffic counters, and link configurations within the network.
For deployment with legacy networking equipment, we have proposed \nameapprox which only requires accurate timestamps to measure the RTT.

We imagine \name and \nameapprox to be deployed on low-latency kernel-bypass stacks such as SNAP~\cite{arashloo2016snap} or using NIC offload.
Yet, in this work, instead of implementing our algorithms for these platforms, we show how \name and \nameapprox can readily be deployed by merely changing the control logic of existing congestion control algorithms.
In particular, we compare our work to HPCC~\cite{li2019hpcc} which is based on INT feedback and SWIFT~\cite{kumar2020swift} which is based on delay feedback.

\name requires the same switch support and header format as HPCC, as well as packet pacing support from the NIC.
Additionally, it does not maintain additional state compared to HPCC but requires one extra parameter $\gamma$, the moving average parameter for window updates.
Similar to SWIFT and TIMELY, \nameapprox requires accurate packet timestamps from the NIC but it does not require any switch support.
The simpler logic of \nameapprox (compared to \name) only reacts once per RTT and reduces the number of congestion control function calls.

The core contribution of this paper is the design of a novel control law and we do not explore implementation challenges further at this point since \name does not add additional complexity compared to existing algorithms.
Still, to confirm the practical feasibility of our approach, we implemented \name as a Linux kernel congestion control module.
We also implemented the INT component as a proof of concept for the Intel Tofino switch ASIC \cite{tofino}.

The switch implementation is written in P4 and uses a direct counter associated with the egress port to maintain the so far transmitted bytes and appends this metric together with the current queue occupancy upon dequeue from the traffic manager to each segment.
We leverage a custom TCP option type to encode this data and append 64 bit per-hop headers to a 32 bit base header.
The implementation uses less than one out of 12 stages of the Tofino's ingress pipeline (where the headers are prepared and appended) and less than one out of 12 stages in the egress pipeline (where the measurements are taken and inserted).
The processing logic runs at line rate of 3.2 Tbit per second.

\section{Evaluation}
\label{sec:evaluation}
We evaluate the performance of \name and \nameapprox and compare against existing CC algorithms.

\subsection{Setup}
\label{sec:setup}
Our evaluation is based on network simulator NS3~\cite{ns3}.

\myitem{Topology:} We consider a datacenter network based on a FatTree topology~\cite{al2008scalable} with 2 core switches and 256 servers organized into four pods.
Each pod consists of two ToR switches and two aggregation switches.
The capacity of all the switch-to-switch links are 100Gbps and server-to-switch links are all 25Gbps leading to $4:1$ oversubscription similar to prior work~\cite{10.1145/3387514.3405899}.
The links connecting to core switches have a propagation delay of $5\mu s$ and all the remaining links have a propagation delay of $1\mu s$.
We set up a shared memory architecture on all the switches and enable the Dynamic Thresholds algorithm~\cite{choudhury1998dynamic} for buffer management across all the ports, commonly enabled in datacenter switches~\cite{broadcom,broadcomNew}.
Finally we set the buffer sizes in our topology proportional to the bandwidth-buffer ratio of Intel Tofino switches~\cite{tofino}.

\myitem{Traffic mix:} We generate traffic using the web search~\cite{alizadeh2010data} flow size distribution to evaluate our algorithm using realistic workloads.
We evaluate an average load (on the ToR uplinks) in the range of $20\%-95\%$.
We also use a synthetic workload similar to prior work~\cite{alizadeh2013data} to generate incast traffic.
Specifically, the synthetic workload represents a distributed file system where each server requests a file from a set of servers chosen uniformly at random from a different rack. 
All the servers which receive the request respond at the same time by transmitting the requested part of the file.
As a result, each file request creates an incast scenario.
We evaluate across different request rates and request sizes.

\myitem{Comparisons and metrics:} We evaluate \name with and without switch support and compare to HPCC~\cite{li2019hpcc}, DCQCN~\cite{zhu2015congestion}, and TIMELY~\cite{mittal2015timely} representing sender-based control law approaches similar to \name and HOMA~\cite{10.1145/3230543.3230564} representing receiver-driver transport. We report flow completion times and switch buffer occupancy metrics.

\myitem{Configuration:} We set $\gamma=0.9$ for \name and \nameapprox. Both HPCC and \name are configured with base-RTT ($\tau$) set to the maximum RTT in our topology and $HostBw$ is set to the server NIC bandwidth. The product of base-RTT and $HostBw$ is configured as RTTBytes for HOMA and the over-commitment level is set to $1$ where HOMA performed best across different overcommitment levels in our setup. We report our results for all overcommitment levels ($1$-$6$) in Appendix~\ref{appendix:homa}.
We set the parameters for DCQCN following the suggestion in~\cite{li2019hpcc} which is based on experience and TIMELY parameters are set according to~\cite{mittal2015timely}.

\subsection{Results}
\label{sec:results}
\begin{figure*}
\centering
\begin{minipage}{0.39\linewidth}
\centering
\begin{minipage}{1\linewidth}
\includegraphics[width=1\linewidth]{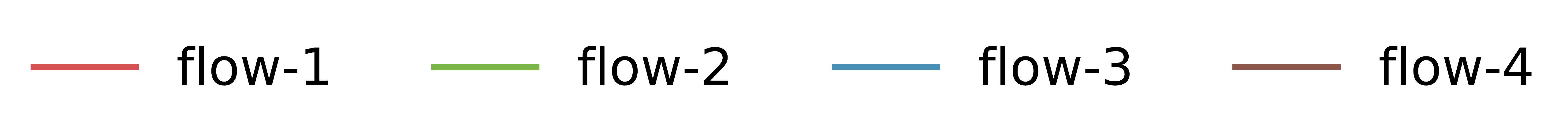}
\end{minipage}
\begin{minipage}{0.49\linewidth}
\includegraphics[width=1\linewidth]{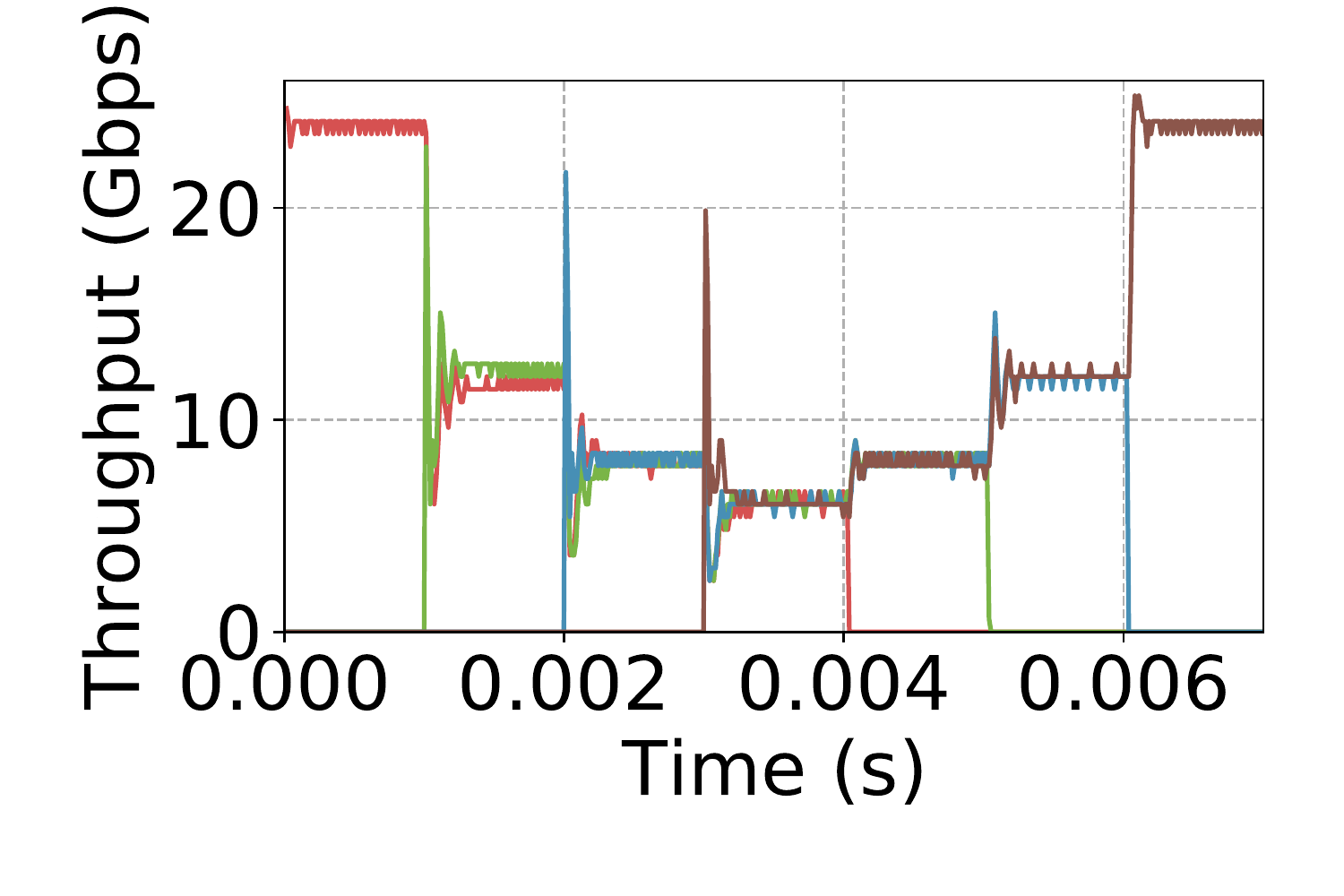}
\subcaption{\name ($ms$ scale)}
\label{fig:fairPowerTcpms}
\end{minipage}
\begin{minipage}{0.49\linewidth}
\includegraphics[width=1\linewidth]{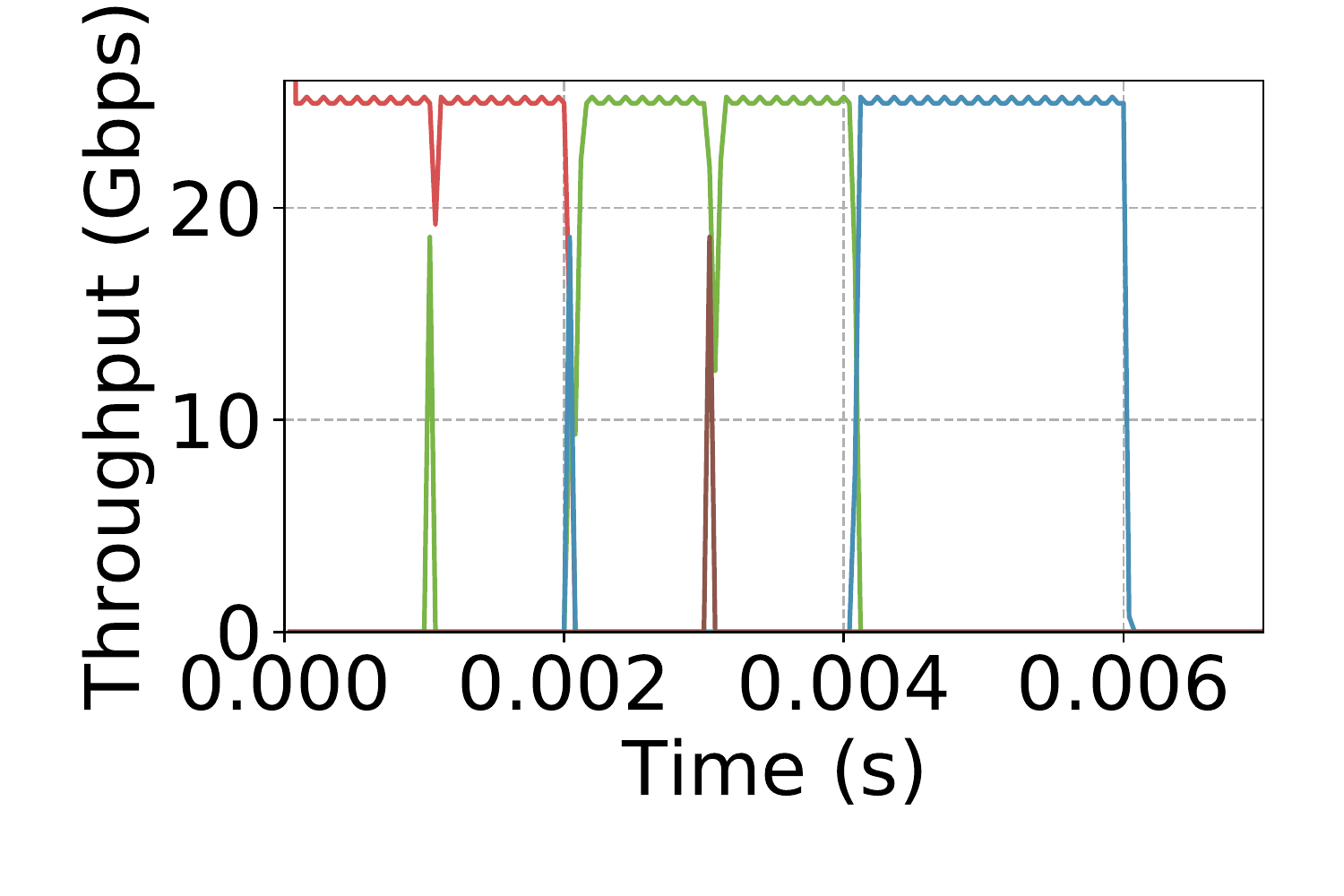}
\subcaption{HOMA ($ms$ scale)}
\label{fig:fairHomams}
\end{minipage}
\begin{minipage}{0.49\linewidth}
\includegraphics[width=1\linewidth]{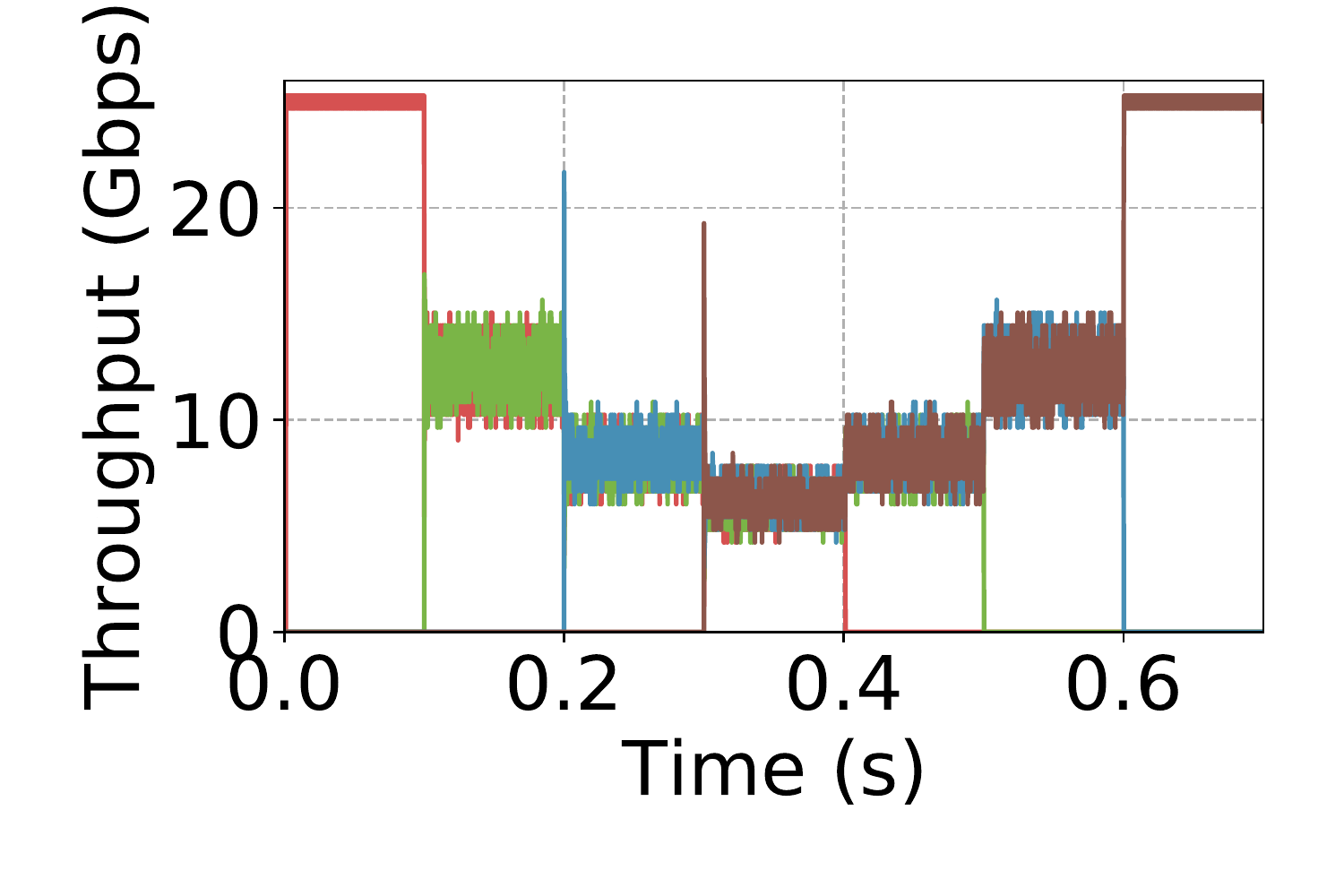}
\subcaption{\nameapprox}
\end{minipage}
\begin{minipage}{0.49\linewidth}
\includegraphics[width=1\linewidth]{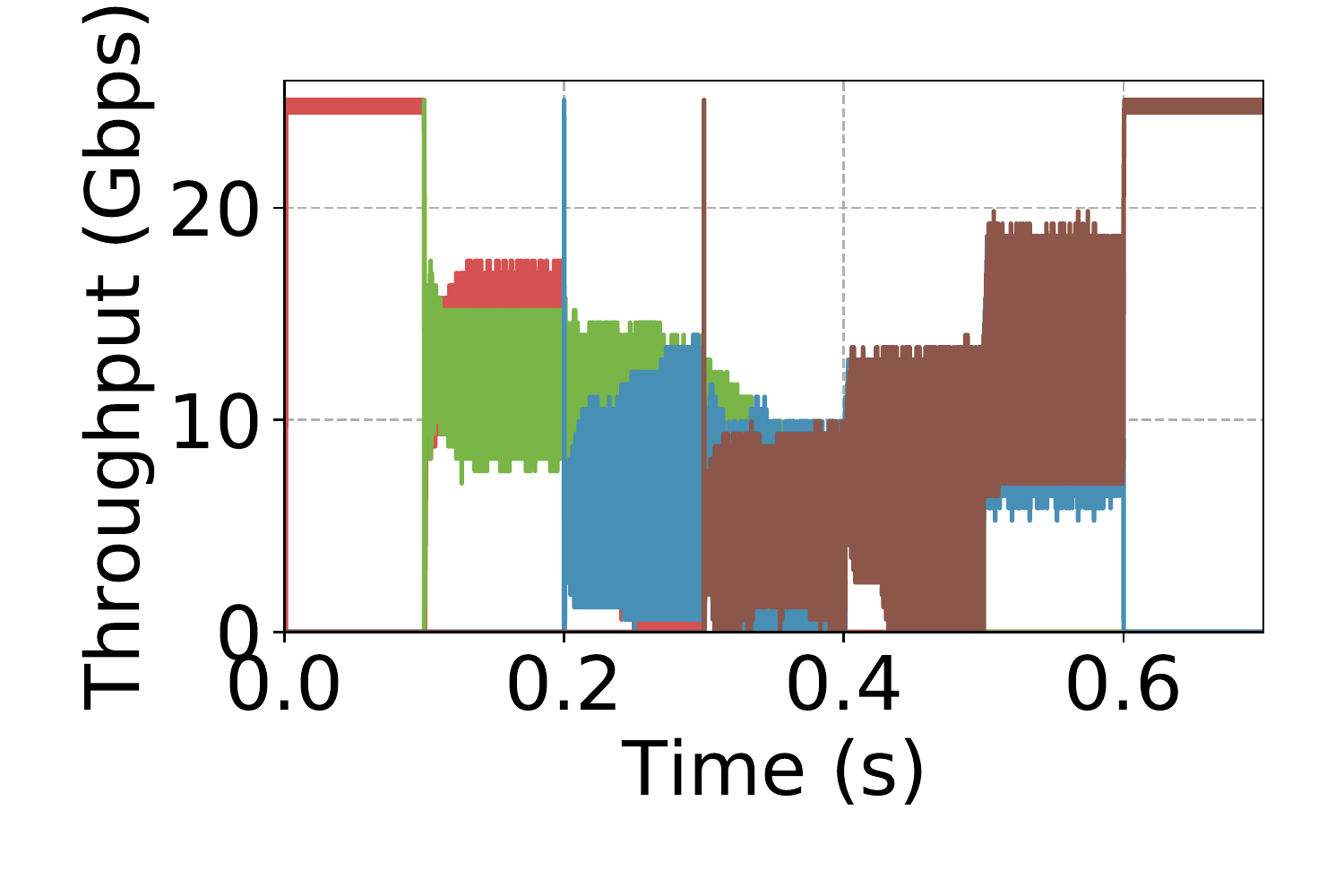}
\subcaption{TIMELY}
\end{minipage}
\caption{Fairness and stability}
\label{fig:fairness}
\end{minipage}\hfill
\begin{minipage}{0.58\linewidth}
\centering
\begin{minipage}{1\linewidth}
\centering
\includegraphics[width=0.9\linewidth]{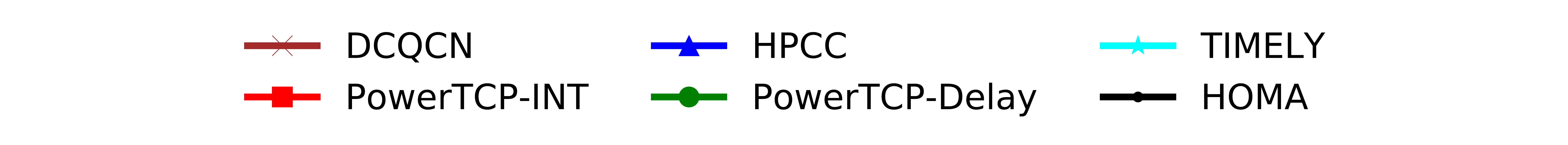}
\end{minipage}
\begin{minipage}{0.49\linewidth}
\centering
\includegraphics[width=1\linewidth]{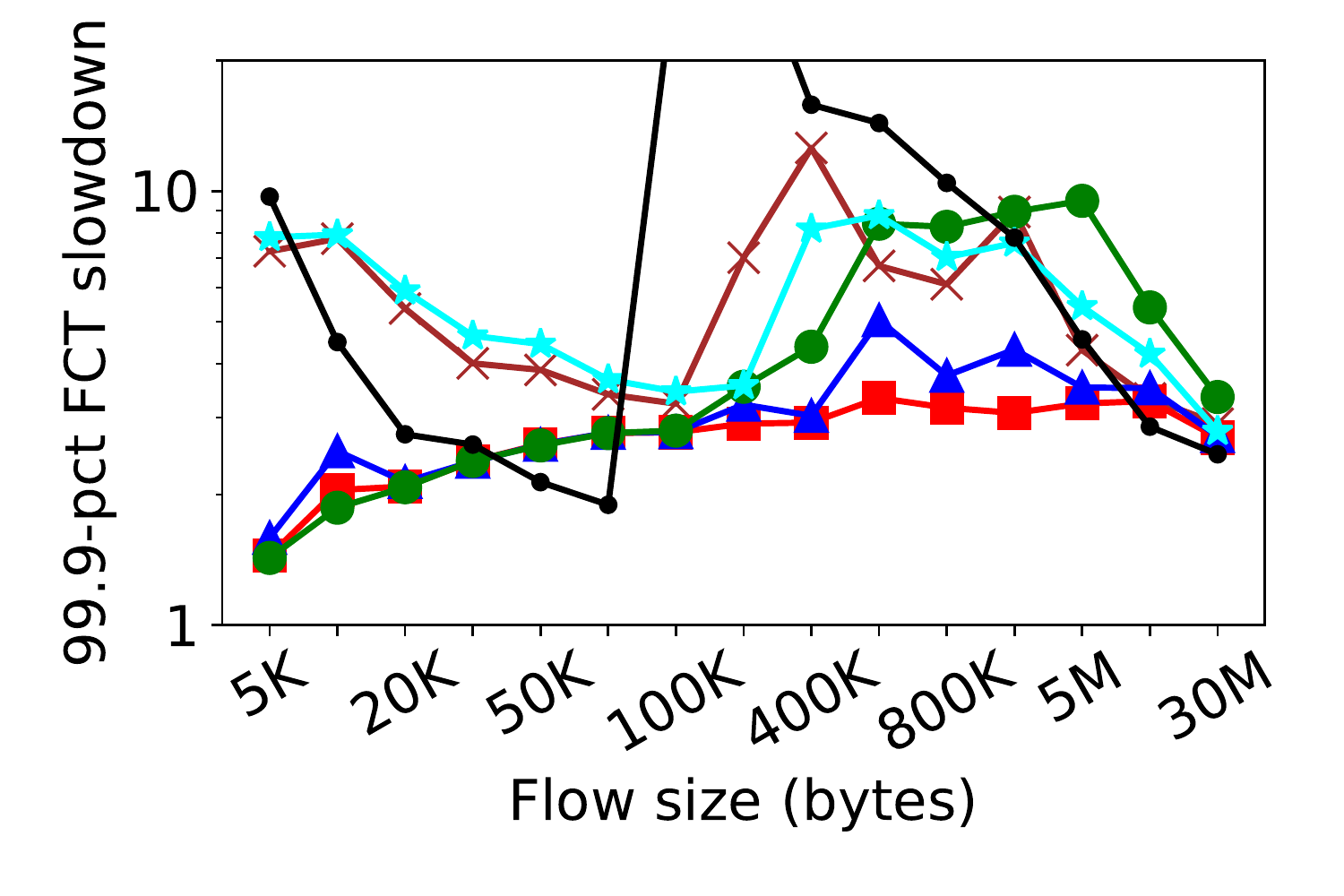}
\subcaption{$20\%$ load}
\label{fig:fctsload20}
\end{minipage}\hfill
\begin{minipage}{0.49\linewidth}
\centering
\includegraphics[width=1\linewidth]{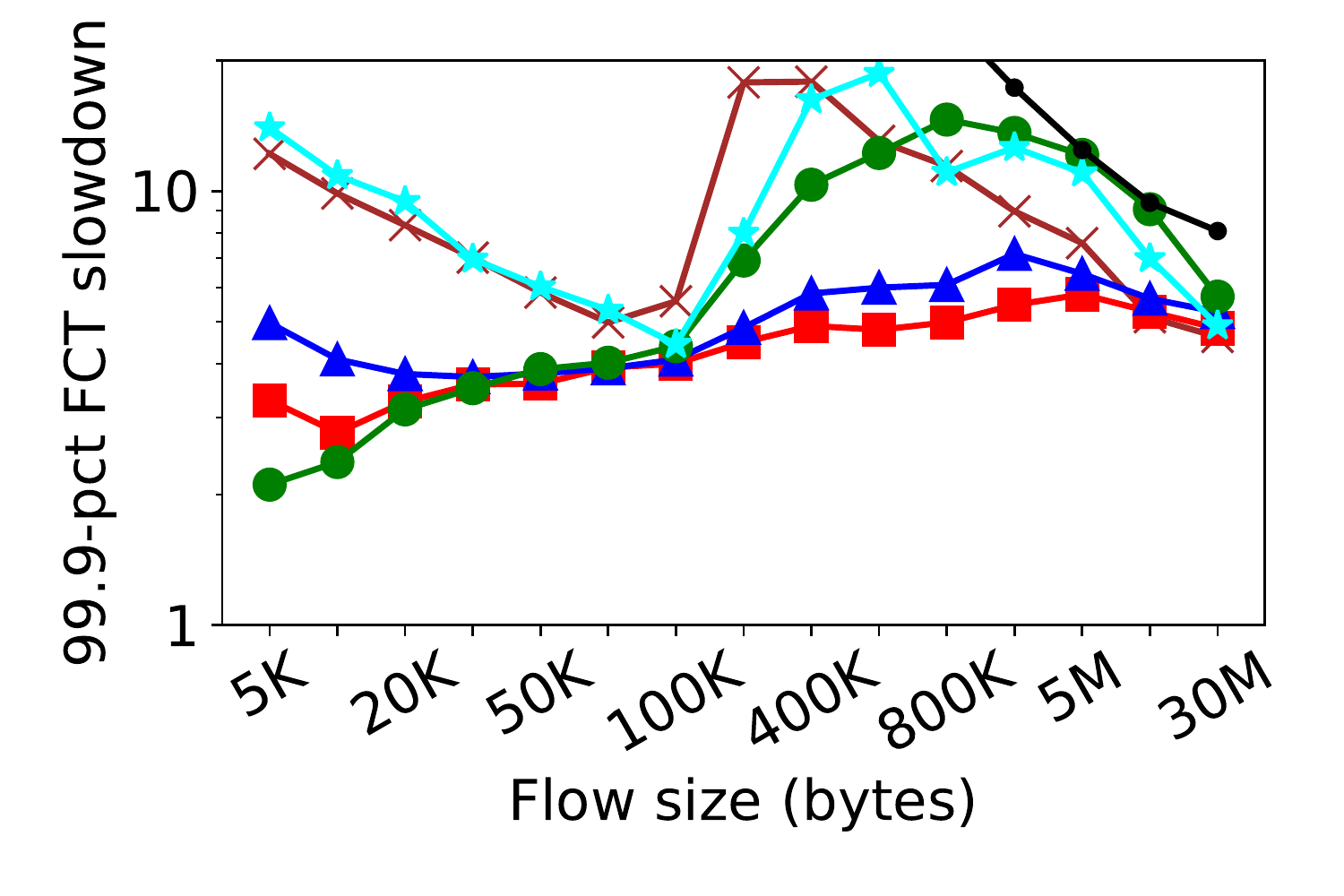}
\subcaption{$60\%$ load}
\label{fig:fctsload60}
\end{minipage}
\caption{$99.9$ percentile flow completion times with websearch workload \emph{(a)} even at low network load, \name outperforms existing algorithms and \emph{(b)} as the load increases the benefits of \name are enhanced. However, only short flows benefit from \nameapprox. }
\label{fig:fctsload}
\end{minipage}
\vspace{-5mm}
\end{figure*}
\myitem{\name reacts rapidly yet accurately to congestion:}
We evaluate \name's reaction to congestion in two scenarios: \first $10:1$ small-scale incast and \second $255:1$ large-scale incast. Figure~\ref{fig:bursts} shows the aggregate throughput and the buffer occupancy at the bottleneck link for \name, TIMELY, HPCC and HOMA. 
First, at time $t=0$, we launch ten flows simultaneously towards the receiver of a long flow leading to a \textbf{10:1} incast. 
We show in Figure~\ref{fig:bursts-powerint} and Figure~\ref{fig:bursts-powerdelay} that \name quickly mitigates the incast and reaches near zero queue lengths without losing throughput.
In Figure~\ref{fig:bursts-hpcc} we see that HPCC indeed reacts quickly to get back to near-zero queue lengths.
On one hand, however, HPCC does not react enough during the congestion onset and reaches higher buffer occupancy $\approx 2$x compared to \name and on the other hand loses throughput after mitigating the incast as opposed to \name's stable throughput.
TIMELY as shown in Figure~\ref{fig:bursts-timely} does not control the queue-lengths either and loses throughput after reacting to the incast. While HOMA sustains throughput, we observe from Figure~\ref{fig:bursts-homa} that HOMA does not accurately control bottleneck queue-lengths. 
Second, at time $t=0$, in addition to the $10:1$ incast, the $256^{th}$ server sends a query request (\S\ref{sec:setup}) to all the other $255$ servers which then respond at the same time, creating a \textbf{255:1} incast. From Figure~\ref{fig:bursts-powerint} and Figure~\ref{fig:bursts-powerdelay} (bottom row), we observe similar benefits from both \name and \nameapprox even at large-scale incast: both react quickly and converge to near-zero queue-lengths without losing throughput. In contrast, from Figure~\ref{fig:bursts-timely} and Figure~\ref{fig:bursts-hpcc} we see that TIMELY and HPCC lose throughput immediately after reacting to the increased queue length. From Figure~\ref{fig:bursts-homa} we observe that HOMA reaches approximately $500$KB higher queue-length compared to \name and cannot converge to near-zero queue-lengths quickly.

\myitem{\name is stable and achieves fairness: }
\name not only reacts rapidly to reduce queue lengths but also features excellent stability. Figure~\ref{fig:fairness} shows how bandwidth is shared by multiple flows as they arrive and leave.
We see that \name stabilizes to a fair share of bandwidth quickly, both when flows arrive and leave, confirming \name's fast reaction to congestion as well as the available bandwidth. 

Figure~\ref{fig:bursts-powerint} showing convergence and Figure~\ref{fig:fairPowerTcpms} showing fairness and stability confirm the theoretical guarantees of \name.
Hereafter, all our results are based on the setup described above, \S\ref{sec:setup}, using realistic workloads.

\myitem{\name significantly improves short flows FCTs:}
In Figure~\ref{fig:fctsload} we show the $99.9$-percentile flow completion times using \name and state-of-the-art datacenter congestion control algorithms.
At $20\%$ network load (Figure~\ref{fig:fctsload20}), \name and \nameapprox improve $99.9$-percentile flow completion times for short flows ($<10KB$) by $9\%$ compared to HPCC and by $80\%$ compared to TIMELY, DCQCN and HOMA.
Even at moderate load of $60\%$ (Figure~\ref{fig:fctsload60}), short flows significantly benefit from \name as well as \nameapprox.
Specifically, \name improves $99.9$ percentile flow completion times for short flows by $33\%$ compared to HPCC, by $99\%$ compared to HOMA and by $74\%$ compared to TIMELY and DCQCN.
\nameapprox provides even greater benefits to short flows showing an improvement of $36\%$ compared to HPCC and $82\%$ compared to TIMELY and DCQCN.
Indeed, web search workload being buffer-intensive, our results confirm the observations made in \S\ref{sec:motivation}.
TIMELY being a current-based CC, does not explicitly control queuing latency, while HPCC, a voltage-based CC, does not react as fast as \name to mitigate congestion resulting in higher flow completion times. Surprisingly, HOMA performs the worst, showing an order-of-magnitude higher FCTs for short flows at high loads as shown in Figure~\ref{fig:fctsload60}.

\begin{figure*}
\centering
\begin{subfigure}{0.49\linewidth}
\centering
\includegraphics[width=0.8\linewidth]{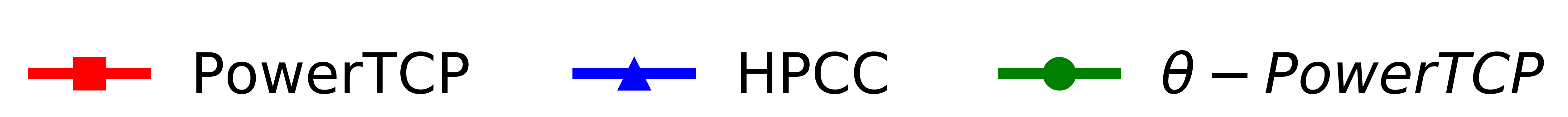}
\end{subfigure}\hfill
\begin{subfigure}{0.49\linewidth}
\centering
\includegraphics[width=0.8\linewidth]{plots/Powertcp-NSDI/workload/all-legend.pdf}
\end{subfigure}
\begin{subfigure}{0.22\linewidth}
\includegraphics[width=1\linewidth]{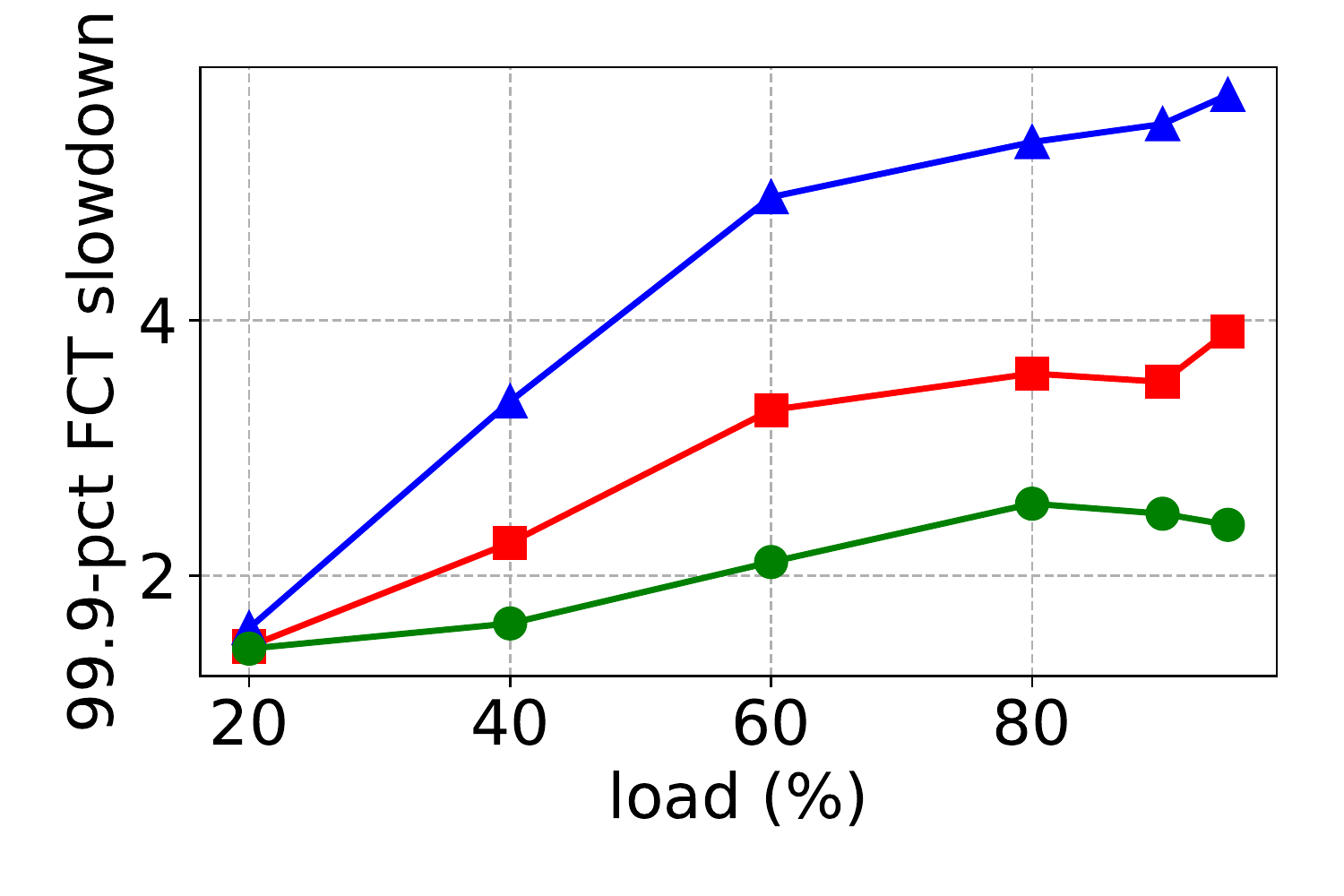}
\caption{Short flows FCT with websearch workload}
\label{fig:loadsShort}
\end{subfigure}\hfill
\begin{subfigure}{0.22\linewidth}
\includegraphics[width=1\linewidth]{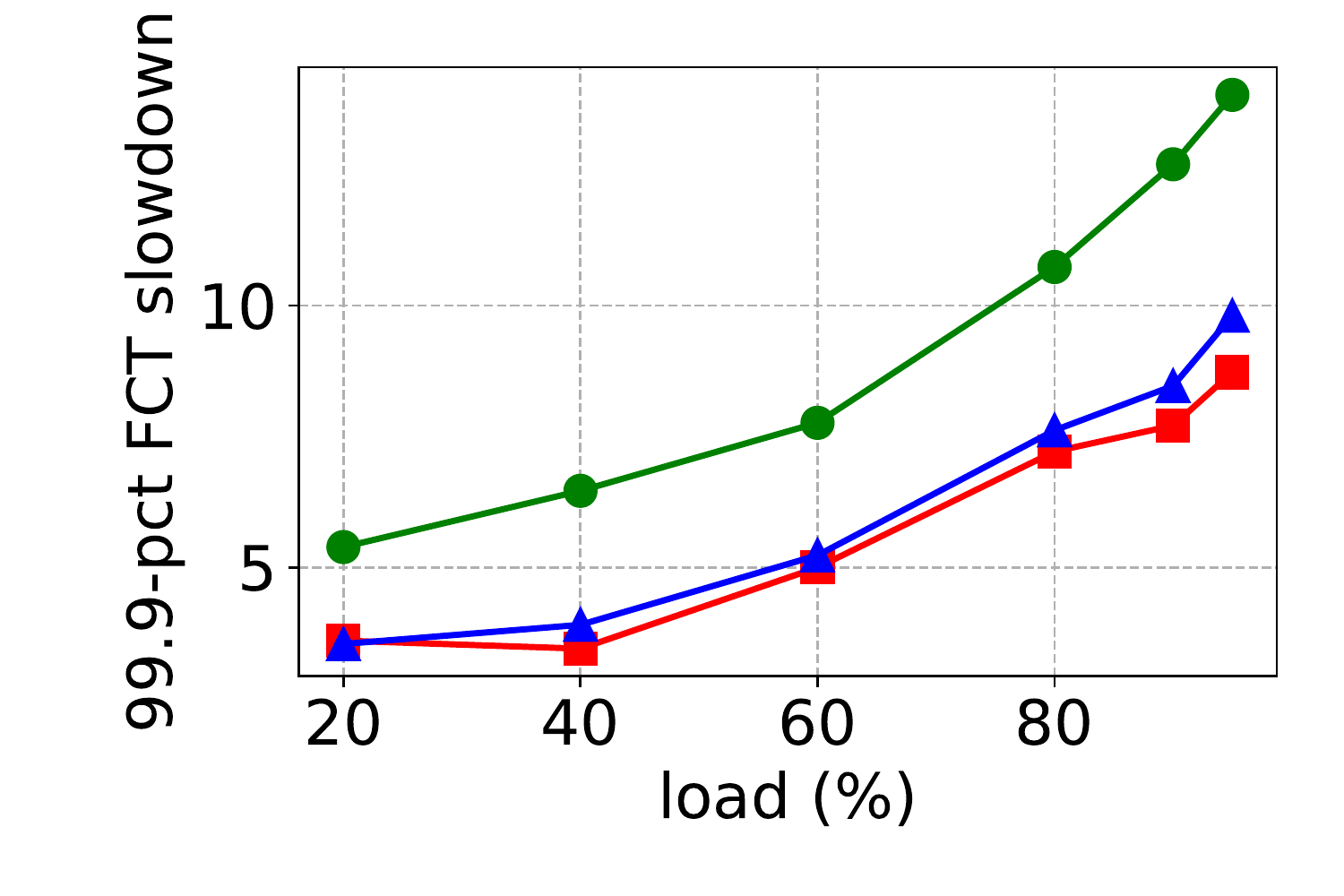}
\caption{Long flows FCT with websearch workload}
\label{fig:loadsLong}
\end{subfigure}\hfill
\begin{subfigure}{0.22\linewidth}
\includegraphics[width=1\linewidth]{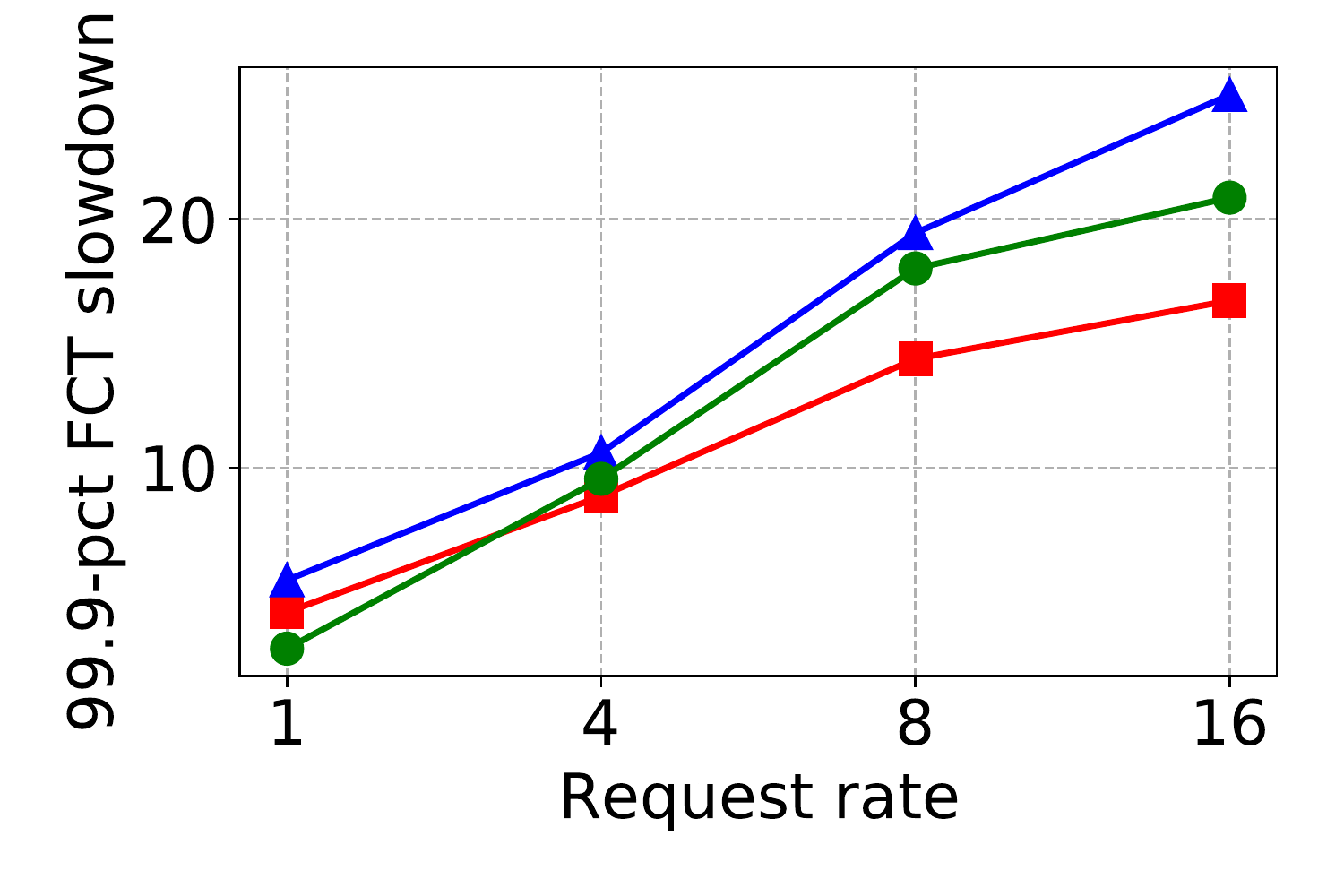}
\caption{Short flows FCTs with websearch + incasts}
\label{fig:burstRateShort}
\end{subfigure}\hfill
\begin{subfigure}{0.22\linewidth}
\includegraphics[width=1\linewidth]{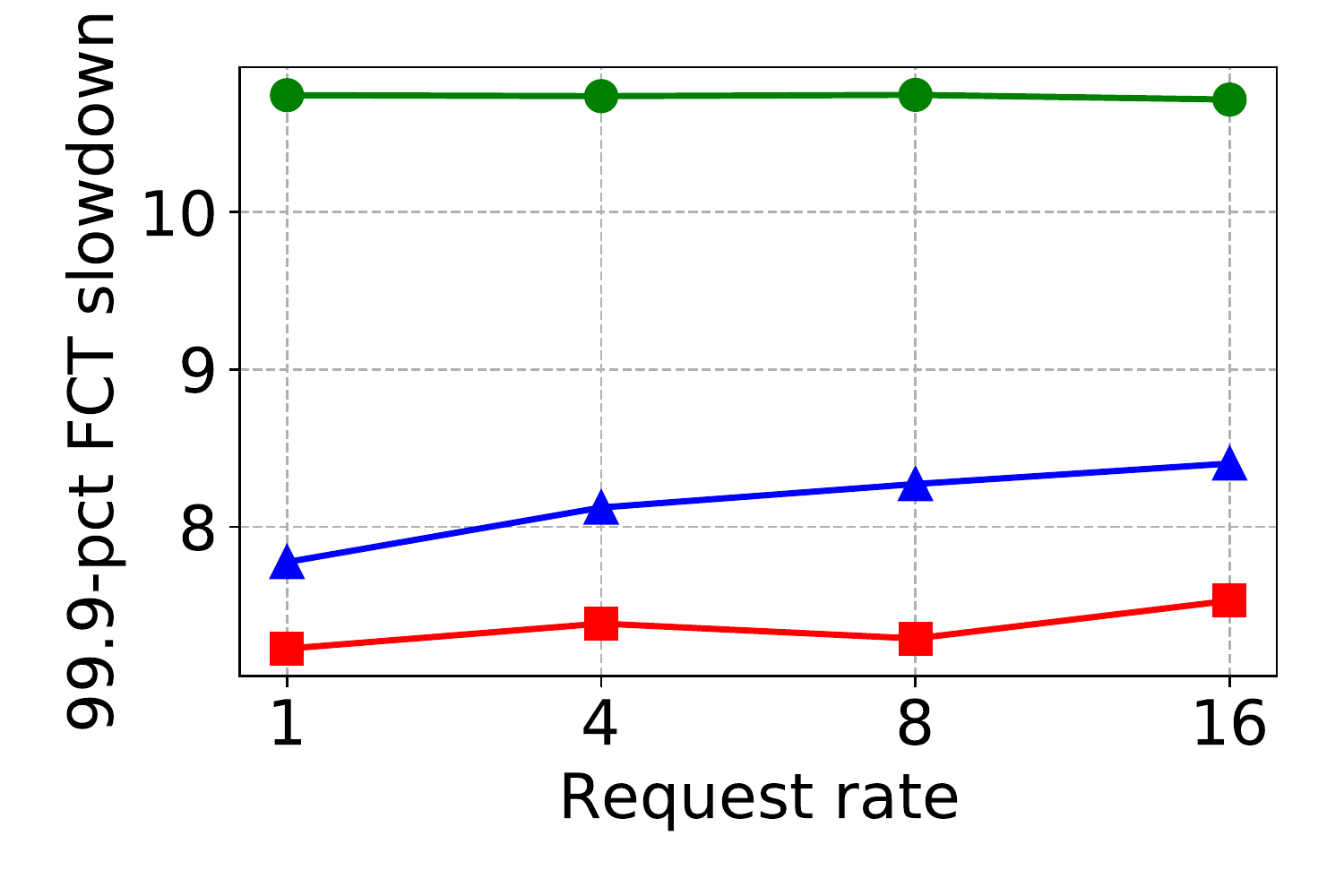}
\caption{Long flows FCT with websearch + incasts}
\label{fig:burstRateLong}
\end{subfigure}

\begin{subfigure}{0.22\linewidth}
\includegraphics[width=1\linewidth]{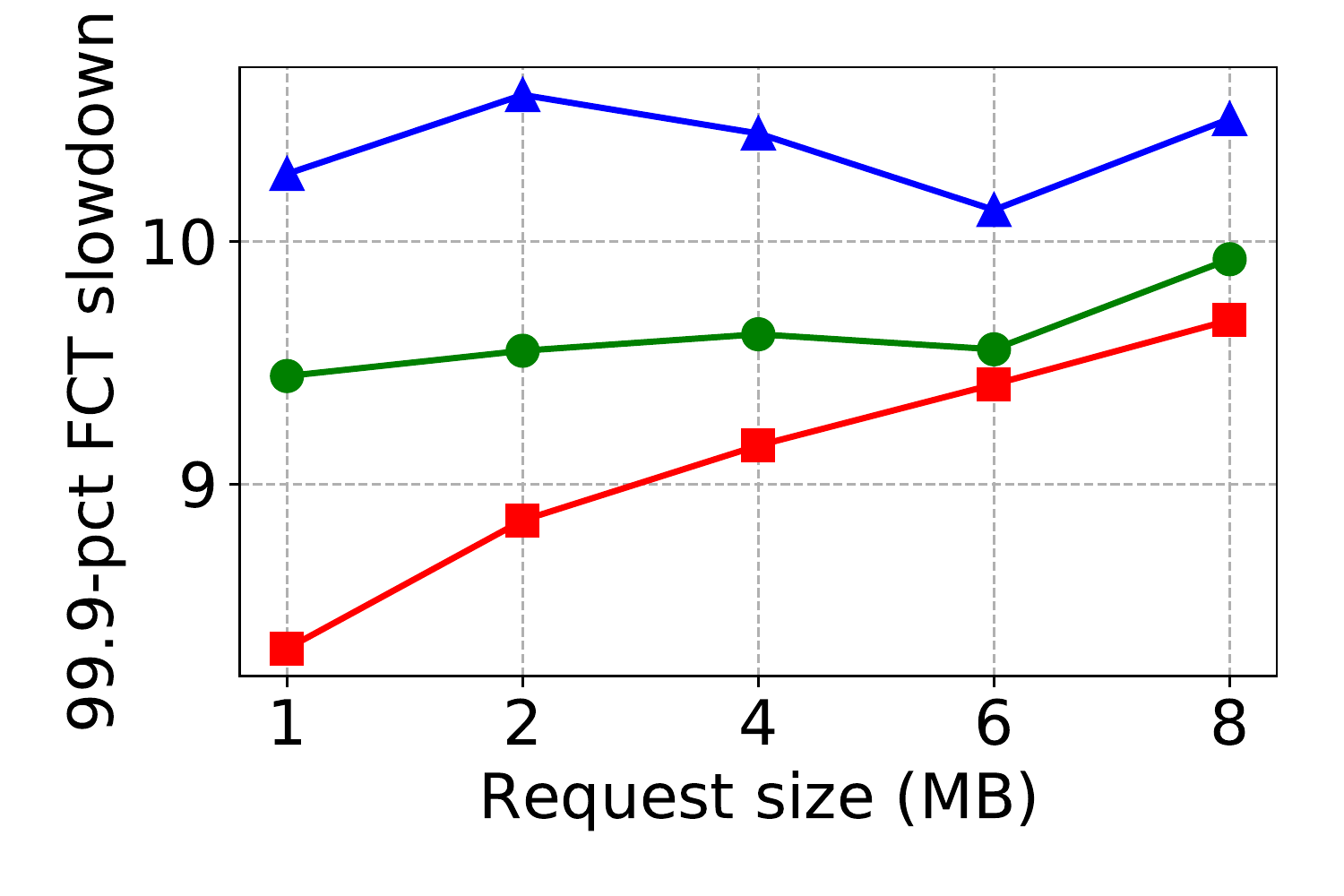}
\caption{Short flows FCTs with websearch + incasts}
\label{fig:burstSizeShort}
\end{subfigure}\hfill
\begin{subfigure}{0.22\linewidth}
\includegraphics[width=1\linewidth]{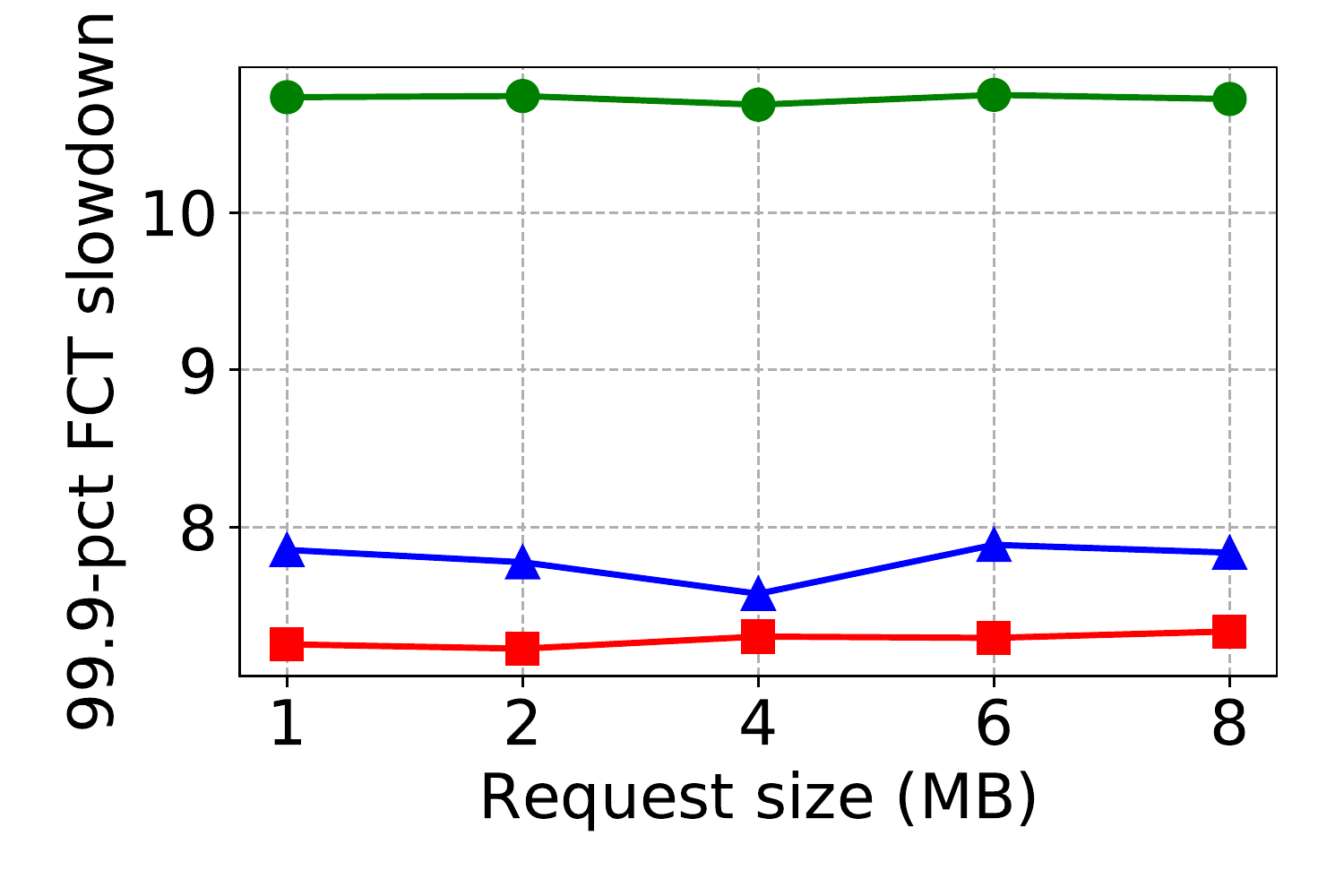}
\caption{Long flows FCT with websearch + incasts}
\label{fig:burstSizeLong}
\end{subfigure}\hfill
\begin{subfigure}{0.22\linewidth}
\includegraphics[width=1\linewidth]{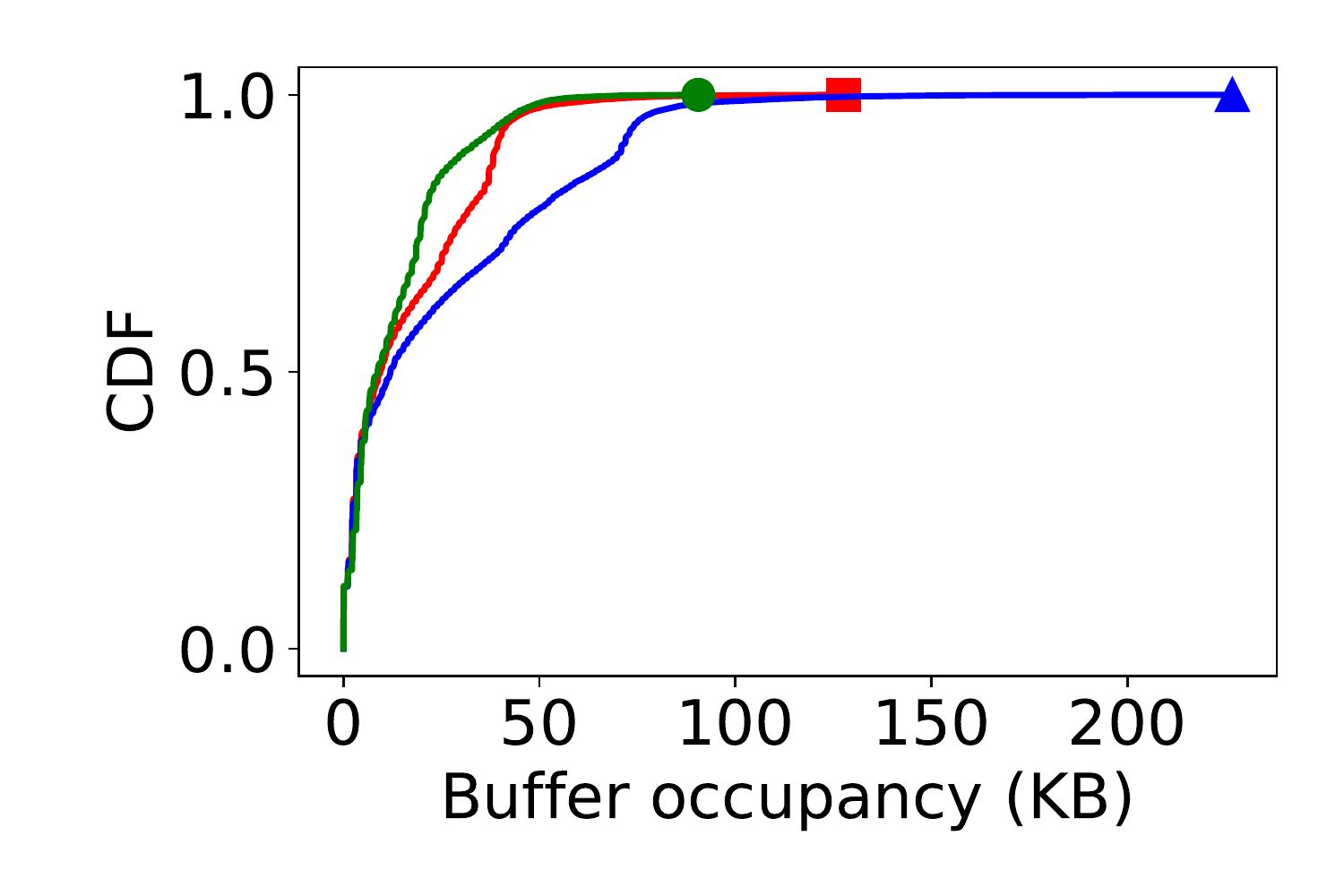}
\caption{Buffer occupancy with websearch workload at $80\%$ load}
\label{fig:bufferLoad}
\end{subfigure}\hfill
\begin{subfigure}{0.22\linewidth}
\includegraphics[width=1\linewidth]{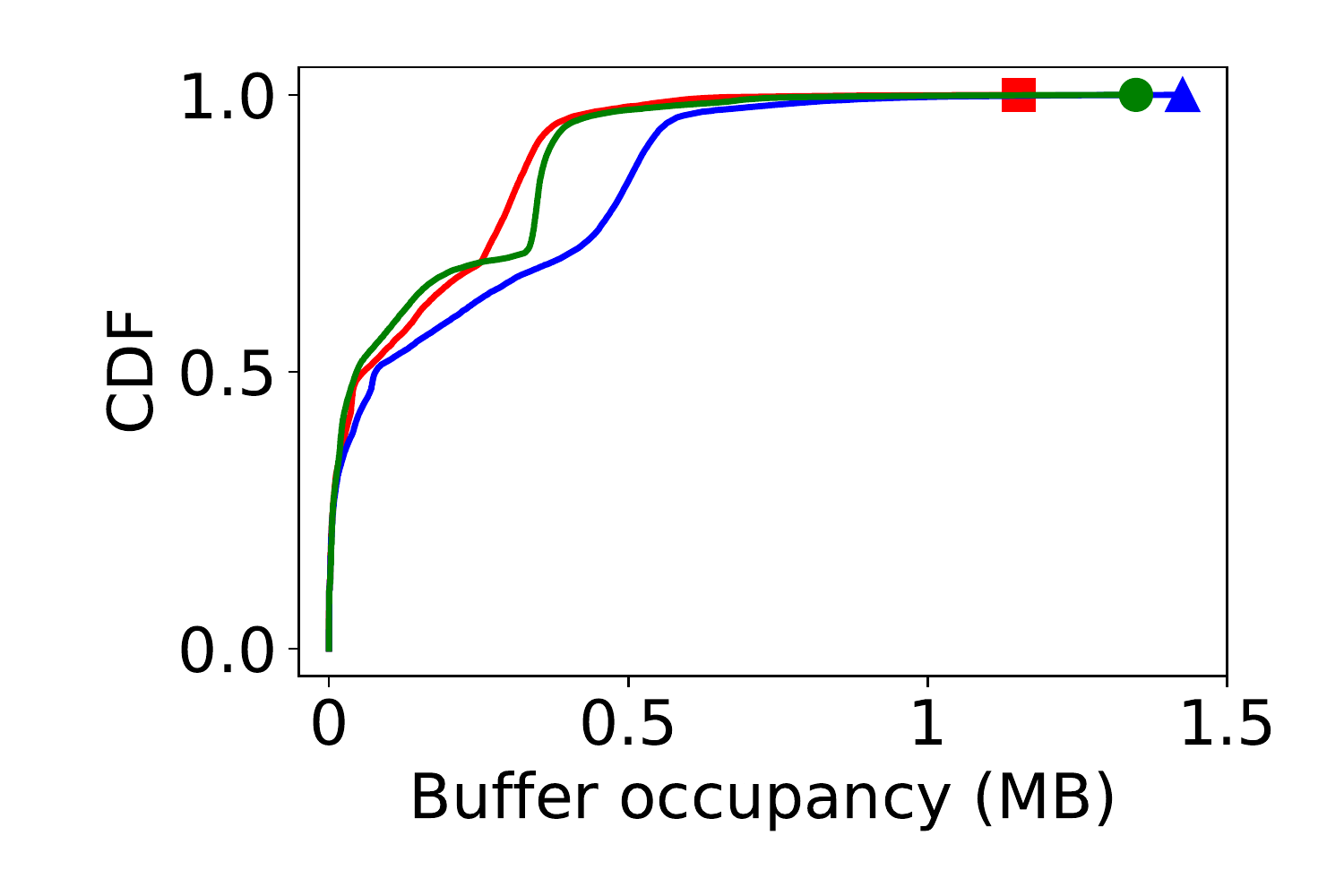}
\caption{Buffer occupancy with websearch + incasts}
\label{fig:bufferBurst}
\end{subfigure}
\caption{A detailed comparison of \name, \nameapprox and the state-of-the-art showing the benefits of \name and the trade-offs of \nameapprox. Particularly \name outperforms the state-of-the-art across a range of network loads even under bursty traffic. However, \nameapprox performs well for short flows but long flows cannot benefit from \nameapprox.}
\vspace{-3mm}
\end{figure*}

We also evaluate across various loads in the range $20\%-95\%$ and show the $99.9$-percentile flow completion times for short flows in Figure~\ref{fig:loadsShort}.
In particular, we see that the benefits of \name and \nameapprox are further enhanced as the network load increases.
\name (and \nameapprox) improve the flow completion times of short flows by $36\%$ (and $55\%$) compared to HPCC. Short flows particularly benefit from \name due its accurate control of buffer occupancies close to zero. In Figure~\ref{fig:bufferLoad} we show the CDF of buffer occupancies at $80\%$ load. \name consistently maintains lower buffer occupancy and cuts the tail buffer occupancy by $50\%$ compared to HPCC.

\myitem{Medium sized flows also benefit from \name:}
We find that \name not only improves short flow performance but also improves the $99.9$-percentile flow completion times for medium sized flows ($100KB-1M$). 
In Figure~\ref{fig:fctsload} we see that \name consistently achieves better flow completion times for medium sized flows. 
Specifically, at $20\%$ network load (Figure~\ref{fig:fctsload20}), \name improves $99.9$-percentile flow completion times for medium flows by $33\%$ compared to HPCC, by $76\%$ compared to HOMA and by $62\%$ (and $50\%$) compared to TIMELY (and DCQCN). 
In Figure~\ref{fig:fctsload60}, we observe similar benefits even at $60\%$ load.

We notice from Figure~\ref{fig:fctsload20} and Figure~\ref{fig:fctsload60} that the performance of \nameapprox deteriorates sharply for medium sized flows. 
\nameapprox uses RTT for window update calculations. 
While RTT can be a good congestion signal, it does not signal under-utilization as opposed to INT that explicitly notifies the exact utilization. 
As a result, medium flows with \nameapprox experience $60\%$ worse performance on average compared to \name and HPCC.
We also observe similar performance for TIMELY that uses RTT as a congestion signal.
Although delay is simple and effective for short flows performance even at the tail, our results show that delay as a congestion signal is not ideal if not worse for medium sized flows.

\myitem{\name does not penalize long flows:}
Fast reaction to available bandwidth makes \name ideal for best performance across all flow sizes.
We observe from Figure~\ref{fig:fctsload} that \name achieves flow completion times comparable to existing algorithms, indicating that \name does not trade throughput for low latency. 
Further, in Figure~\ref{fig:loadsLong} we show the $99.9$-percentile flow completion times for long flows across various loads. 
At low load, \name performs similar to HPCC and performs $9\%$ better compared to HPCC at $90\%$ network loads. 
However, we see that \nameapprox is consistently $35\%$ worse on average across various loads compared to \name and HPCC.

\myitem{\name outperforms under bursty traffic:}
We generate incast-like traffic described in \S\ref{sec:setup} in addition to the web search workload at $80\%$ load. In Figure~\ref{fig:burstRateShort} and Figure~\ref{fig:burstRateLong} we show the $99.9$-percentile flow completion times for short and long flows across different request rates for a request size of $2MB$. 
Note that by varying request rates, we are essentially varying the frequency of incasts.
We observe that even under bursty traffic, \name improves $99.9$-percentile flow completion times on average for short flows by $24\%$ and for long flows by $10\%$ compared to HPCC.
Further \name outperforms at high request rates showing $33\%$ improvement over HPCC for short flows.
On the other hand, \nameapprox improves flows completion times for short flows but performs worse across all request rates compared to HPCC.

We further vary the request size at a request rate of four per second. Note that by varying the request size, we also vary the duration of congestion. In Figure~\ref{fig:burstSizeShort} and Figure~\ref{fig:burstSizeLong}, we show the $99.9$-percentile flow completion times for short and long flows.
Specifically, in Figure~\ref{fig:burstSizeShort} we observe that flow completion times with \name gradually increase with request size.
\name, compared to HPCC, improves flow completion times of short flows by $20\%$ at $1MB$ request size and improves by $7\%$ at $8MB$ request size. 
At the same time, \name does not sacrifice long flows performance under bursty traffic. 
\name improves flow completion times for long flows by $5\%$ on average compared to HPCC. 
\nameapprox's performance similar to previous experiments is on average $30\%$ worse for long flows but $9\%$ better for short flows compared to HPCC. 
We show the CDF of buffer occupancies under bursty traffic with $2MB$ request size and 16 per second request rate. Both \name and \nameapprox reduce the $99$ percentile buffer by $31\%$ compared to HPCC.

We note that HOMA's performance in our evaluation is not in line with the evaluation results presented in~\cite{10.1145/3230543.3230564}. Recent work~\cite{10.1145/3387514.3405878} reports similar performance issues with HOMA. 
We suspect two possible reasons: \first HOMA's accuracy in controlling congestion is specifically limited in our network setup with an oversubscribed Fat-Tree topology where congestion at the ToR uplinks is a possibility which cannot be controlled by a receiver-driven approach such as HOMA. \second As pointed out by~\cite{10.1145/3387514.3405878}, HOMA's original evaluation considered practically infinite buffers at the switches whereas switches in our setup are limited in buffer space and additionally use Dynamic Thresholds to share buffer.

\section{Case Study: Reconfigurable DCNs}
\label{sec:casestudy}

\begin{figure*}
\centering
\begin{subfigure}{0.7\linewidth}
\centering
\includegraphics[width=1\linewidth]{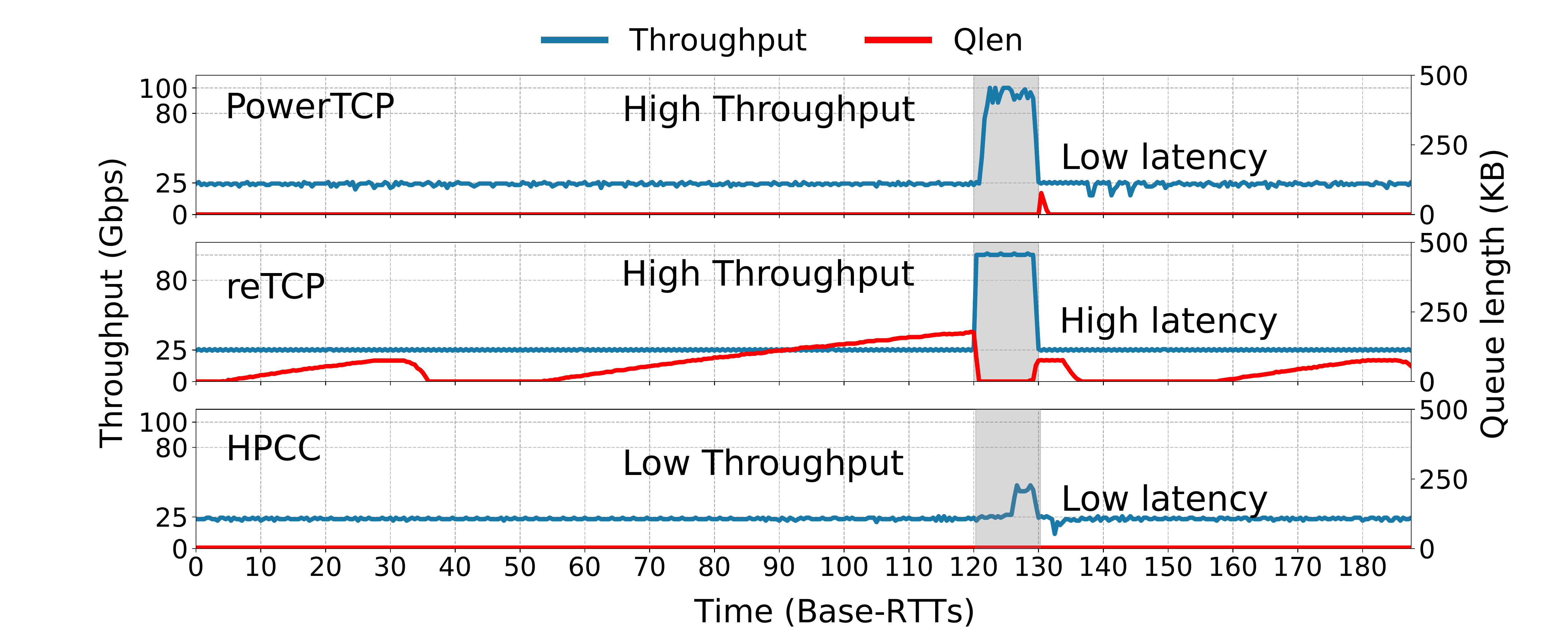}
\caption{\name reacts rapidly to the available bandwidth achieving good circuit utilization.}
\label{fig:timeseriesRdcn}
\end{subfigure}\hfill
\begin{subfigure}{0.28\linewidth}
\centering
\includegraphics[width=0.75\linewidth]{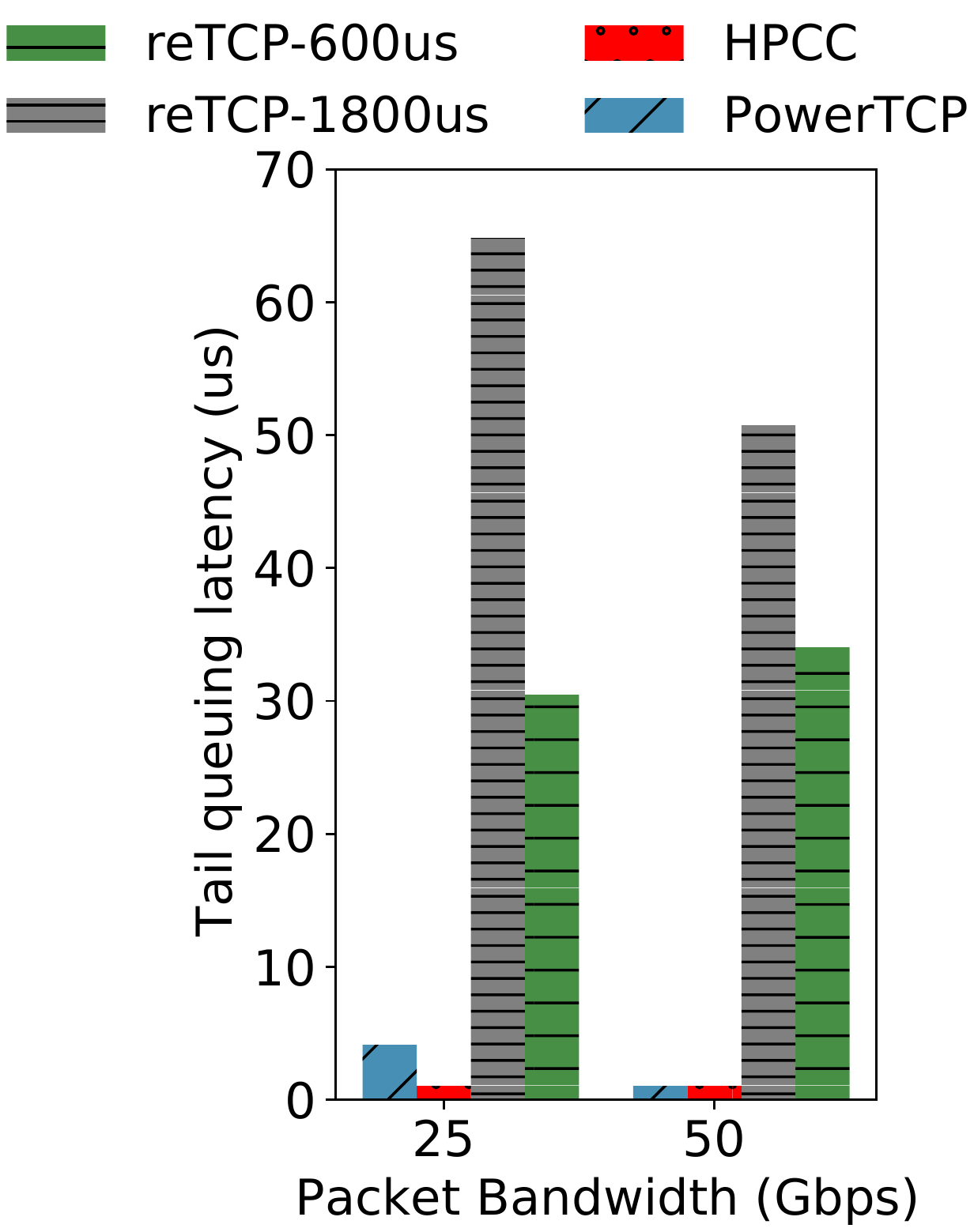}
\caption{\name significantly reduces the tail latency}
\label{fig:tailLatencyRdcn}
\end{subfigure}

\caption{The benefits of \name in reconfigurable datacenter networks showing its ability to achieve good circuit utilization while significantly reducing the tail latency compared to reTCP.}
\vspace{-5mm}
\end{figure*}

Given \name's rapid reaction to congestion and available bandwidth, we believe that \name is well suited for emerging reconfigurable datacenter networks (RDCN)~\cite{NANCEHALL2021100621}. 
We now examine \name's applicability in this context through a case study. 
Congestion control in RDCNs is especially challenging as the available bandwidth rapidly fluctuates due to changing circuits.
In this section, we evaluate the performance of \name and compare against the state-of-the-art reTCP~\cite{mukerjee2020adapting} and HPCC using packet-level simulations in NS3.
We implement both \name and HPCC in the transport layer and limit their window updates to once per RTT for a fair comparison with reTCP. 
\name and HPCC flows initialize the TCP header with the unused option number 36.
Switches are configured to append INT metadata to TCP options.
It should be noted that TCP options are limited to 40 bytes.
As a result, our implementation can only support at most four hops round-trip path length. 

We evaluate in a topology with 25 ToR switches with 10 servers each and a single optical circuit switch connected to all the ToR switches.
ToR switches are also connected to a separate packet switched network with 25Gbps links.
The optical switch internally connects each input port to an output port and cycles across 24 matchings in a permutation schedule where the switch stays in a specific matching for $225\mu s$ (one day) and takes $20\mu s$ to reconfigure to the next matching (one night).
In this setting, each pair of ToR switches has direct connectivity through the circuit switch once over a length of 24 matchings (one week). 
We use single-hop routing in the circuit network and a maximum base RTT is $24\mu s$. 
Note that circuit-on time (\ie one day) is approximately $10$ RTTs.
The links between servers and ToR switches are 25Gbps and circuit links are 100Gbps.
We configure the ToR switches to forward packets exclusively on the circuit network when available.
Switches are further equipped with per-destination virtual output queues (VOQs). Our setup is in line with prior work~\cite{mukerjee2020adapting}.
We set reTCP's prebuffering to $1800\mu s$ based on the suggestions in~\cite{mukerjee2020adapting} and set to $600\mu s$ based on our parameter sweep for the minimum required prebuffering in our topology.
We compare against both versions.

In Figure~\ref{fig:timeseriesRdcn}, we show the time series of throughput and VOQ length for a pair of ToR switches.
Specifically, the gray-shaded area in Figure~\ref{fig:timeseriesRdcn} highlights the availability of high bandwidth through the circuit-switched network.
On one hand, reTCP instantly fills the available bandwidth but incurs high latency due to prebuffering before the circuit is available.
On the other hand, HPCC maintains low queue lengths but does not fill the available bandwidth.
In contrast, \name fills the available bandwidth within one RTT and maintains near-zero queue lengths and thereby achieves both high throughput and low latency.
We show the tail queuing latency incurred by reTCP, HPCC and \name in Figure~\ref{fig:tailLatencyRdcn}.
We observe that \name improves the tail queuing latency at least by $5\times$ compared to reTCP.
Our case study reveals that fine-grained congestion control algorithms such as \name can alleviate the circuit utilization problem in RDCNs without trading latency for throughput.

\section{Related Work}
\label{sec:relatedwork}

Dealing with congestion has been an active research topic for decades with a wide spectrum of
approaches, including buffer management~\cite{choudhury1998dynamic,apostolaki2019fab,cisco9000} and scheduling~\cite{alizadeh2013pfabric,hong2012finishing,perry2014fastpass,perry2017flowtune}.
In the following, we will focus on the most closely related works on 
end-host congestion control.

Approaches such as \cite{alizadeh2010data,zhu2015congestion,vamanan2012deadline} (\eg DCTCP, D$^2$TCP) rely on ECN as the congestion signal and react proportionally.
Such algorithms require the bottleneck queue to grow up to a certain threshold, which results in queuing delays. 
ECN-based schemes remain oblivious to congestion onset and intensity. 
Protocols such as TIMELY~\cite{mittal2015timely}, SWIFT~\cite{kumar2020swift}, CDG~\cite{hayes2011revisiting}, DX~\cite{lee2015accurate} rely on RTT measurements for window update calculations.
TIMELY and CDG partly react to congestion based on delay gradients, remaining oblivious to absolute queue lengths.
TIMELY, for instance, uses a threshold to fall back to proportional reaction to delay instead of delay gradient.
SWIFT, a successor of TIMELY, only reacts proportionally to delay.
As a result, SWIFT cannot detect congestion onset and intensity unless the distance from target delay significantly increases. In contrast, \nameapprox also being a delay-based congestion control algorithm updates the window sizes using the notion of power. As a result, \nameapprox accurately detects congestion onset and intensity even at near-zero queue lengths.

XCP~\cite{katabi2002congestion}, D$^3$~\cite{wilson2011better}, $RCP$~\cite{dukkipati2006flow} rely on explicit network feedback based on rate calculations within the network. However, the rate calculations are based on heuristics and require parameter tuning to adjust for different goals such as fairness and utilization. HPCC~\cite{li2019hpcc} introduces a novel use of in-band network telemetry and significantly improves the fidelity of feedback. Our work builds on the same INT capabilities to accurately measure the bottleneck link state. However, as we show analytically and empirically, HPCC's control law then adjusts rate and window size solely based on observed queue lengths and lacks control accuracy compared to \name.
Our proposal \name uses the same feedback signal but uses the notion of power to update window sizes leading to significantly more fine-grained and accurate reactions.

Receiver-driven transport protocols such as NDP~\cite{10.1145/3098822.3098825}, HOMA~\cite{10.1145/3230543.3230564}, and Aeolus~\cite{10.1145/3387514.3405878} have received much attention lately. Such approaches are conceptually different from classic transmission control at the sender. Importantly, receiver-driven transport approaches make assumptions on the uniformity in datacenter topologies and oversubscription~\cite{10.1145/3098822.3098825}. \name is a sender-based classic CC approach that uses our novel notion of power and achieves fine-grained control over queuing delays without sacrificing throughput.

\section{Conclusion}
\label{sec:conclusion}

We presented \name, a novel fine-grained congestion control algorithm.
By reacting to both the current state of the network as well as its trend (i.e., power), \name improves throughput, reduces latency, and keeps queues within the network short.
We proved that \name has a set of desirable properties, such as fast convergence and stability allowing it to significantly improve flow completion times compared to the state-of-the-art.
Its fast reaction makes \name attractive for many dynamic network environments
including emerging reconfigurable datacenters which served us as a case study in this paper. In our future work, we plan to explore more such use cases.

\label{bodyLastPage}

\bibliographystyle{plain}
\bibliography{references}

\begin{thebibliography}{10}

\bibitem{broadcom}
Broadcom. 12.8 tb/s strataxgs tomahawk 3 ethernet switch series.
\newblock
  \url{https://www.broadcom.com/products/ethernet-connectivity/switching/strataxgs/bcm56980-series}.

\bibitem{broadcomNew}
Broadcom. 2020. 25.6 tb/s strataxgs tomahawk 4 ethernet switch series.
\newblock
  \url{https://www.broadcom.com/products/ethernet-connectivity/switching/strataxgs/bcm56990-series}.

\bibitem{cisco9000}
Cisco nexus 9000 series switches.
\newblock
  \url{https://www.cisco.com/c/en/us/products/collateral/switches/nexus-9000-series-switches/white-paper-c11-738488.html}.

\bibitem{ns3}
Ns3 network simulator.
\newblock \url{https://www.nsnam.org/}.

\bibitem{al2008scalable}
Mohammad Al-Fares, Alexander Loukissas, and Amin Vahdat.
\newblock A scalable, commodity data center network architecture.
\newblock {\em ACM SIGCOMM computer communication review}, 38(4):63--74, 2008.

\bibitem{alizadeh2013data}
Mohammad Alizadeh and Tom Edsall.
\newblock On the data path performance of leaf-spine datacenter fabrics.
\newblock In {\em 2013 IEEE 21st annual symposium on high-performance
  interconnects}, pages 71--74. IEEE, 2013.

\bibitem{alizadeh2010data}
Mohammad Alizadeh, Albert Greenberg, David~A Maltz, Jitendra Padhye, Parveen
  Patel, Balaji Prabhakar, Sudipta Sengupta, and Murari Sridharan.
\newblock Data center tcp (dctcp).
\newblock In {\em Proceedings of the ACM SIGCOMM 2010 conference}, pages
  63--74, 2010.

\bibitem{10.1145/2007116.2007125}
Mohammad Alizadeh, Adel Javanmard, and Balaji Prabhakar.
\newblock Analysis of dctcp: Stability, convergence, and fairness.
\newblock {\em SIGMETRICS Perform. Eval. Rev.}, 39(1):73–84, June 2011.

\bibitem{alizadeh2013pfabric}
Mohammad Alizadeh, Shuang Yang, Milad Sharif, Sachin Katti, Nick McKeown,
  Balaji Prabhakar, and Scott Shenker.
\newblock pfabric: Minimal near-optimal datacenter transport.
\newblock {\em ACM SIGCOMM Computer Communication Review}, 43(4):435--446,
  2013.

\bibitem{apostolaki2019fab}
Maria Apostolaki, Laurent Vanbever, and Manya Ghobadi.
\newblock Fab: Toward flow-aware buffer sharing on programmable switches.
\newblock In {\em Proceedings of the 2019 Workshop on Buffer Sizing}, pages
  1--6, 2019.

\bibitem{arashloo2016snap}
Mina~Tahmasbi Arashloo, Yaron Koral, Michael Greenberg, Jennifer Rexford, and
  David Walker.
\newblock {SNAP}: Stateful network-wide abstractions for packet processing.
\newblock In {\em SIGCOMM '16}. ACM, 2016.

\bibitem{sigmetrics20complexity}
Chen Avin, Manya Ghobadi, Chen Griner, and Stefan Schmid.
\newblock On the complexity of traffic traces and implications.
\newblock In {\em Proc. ACM SIGMETRICS}, 2020.

\bibitem{apocs21renets}
Chen Avin and Stefan Schmid.
\newblock Renets: Statically-optimal demand-aware networks.
\newblock In {\em Proc. SIAM Symposium on Algorithmic Principles of Computer
  Systems (APOCS)}, 2021.

\bibitem{sirius}
Hitesh Ballani, Paolo Costa, Raphael Behrendt, Daniel Cletheroe, Istvan Haller,
  Krzysztof Jozwik, Fotini Karinou, Sophie Lange, Kai Shi, Benn Thomsen, et~al.
\newblock Sirius: A flat datacenter network with nanosecond optical switching.
\newblock In {\em Proceedings of the Annual conference of the ACM Special
  Interest Group on Data Communication on the applications, technologies,
  architectures, and protocols for computer communication}, pages 782--797,
  2020.

\bibitem{brakmo1994tcp}
Lawrence~S Brakmo, Sean~W O'Malley, and Larry~L Peterson.
\newblock Tcp vegas: New techniques for congestion detection and avoidance.
\newblock In {\em Proceedings of the conference on Communications
  architectures, protocols and applications}, pages 24--35, 1994.

\bibitem{chen2009understanding}
Yanpei Chen, Rean Griffith, Junda Liu, Randy~H Katz, and Anthony~D Joseph.
\newblock Understanding tcp incast throughput collapse in datacenter networks.
\newblock In {\em Proceedings of the 1st ACM workshop on Research on enterprise
  networking}, pages 73--82, 2009.

\bibitem{choudhury1998dynamic}
Abhijit~K Choudhury and Ellen~L Hahne.
\newblock Dynamic queue length thresholds for shared-memory packet switches.
\newblock {\em IEEE/ACM Transactions On Networking}, 6(2):130--140, 1998.

\bibitem{tofino}
Intel Corporation.
\newblock {Intel Tofino}, 2020.
\newblock Retrieved Dec. 29, 2020 from
  \url{https://www.intel.com/content/www/us/en/products/network-io/programmable-ethernet-switch/tofino-series/tofino.html}.

\bibitem{dukkipati2006flow}
Nandita Dukkipati and Nick McKeown.
\newblock Why flow-completion time is the right metric for congestion control.
\newblock {\em ACM SIGCOMM Computer Communication Review}, 36(1):59--62, 2006.

\bibitem{projector}
Monia Ghobadi, Ratul Mahajan, Amar Phanishayee, Nikhil Devanur, Janardhan
  Kulkarni, Gireeja Ranade, Pierre-Alexandre Blanche, Houman Rastegarfar,
  Madeleine Glick, and Daniel Kilper.
\newblock Projector: Agile reconfigurable data center interconnect.
\newblock In {\em Proceedings of the 2016 ACM SIGCOMM Conference}, pages
  216--229, 2016.

\bibitem{ha2008cubic}
Sangtae Ha, Injong Rhee, and Lisong Xu.
\newblock Cubic: a new tcp-friendly high-speed tcp variant.
\newblock {\em ACM SIGOPS operating systems review}, 42(5):64--74, 2008.

\bibitem{10.1145/3098822.3098825}
Mark Handley, Costin Raiciu, Alexandru Agache, Andrei Voinescu, Andrew~W.
  Moore, Gianni Antichi, and Marcin W\'{o}jcik.
\newblock Re-architecting datacenter networks and stacks for low latency and
  high performance.
\newblock SIGCOMM '17, page 29–42, New York, NY, USA, 2017. Association for
  Computing Machinery.

\bibitem{hayes2011revisiting}
David~A Hayes and Grenville Armitage.
\newblock Revisiting tcp congestion control using delay gradients.
\newblock In {\em International Conference on Research in Networking}, pages
  328--341. Springer, 2011.

\bibitem{hollot2001control}
Christopher~V Hollot, Vishal Misra, Don Towsley, and Wei-Bo Gong.
\newblock A control theoretic analysis of red.
\newblock In {\em Proceedings IEEE INFOCOM 2001. Conference on Computer
  Communications. Twentieth Annual Joint Conference of the IEEE Computer and
  Communications Society (Cat. No. 01CH37213)}, volume~3, pages 1510--1519.
  IEEE, 2001.

\bibitem{hong2012finishing}
Chi-Yao Hong, Matthew Caesar, and P~Brighten Godfrey.
\newblock Finishing flows quickly with preemptive scheduling.
\newblock {\em ACM SIGCOMM Computer Communication Review}, 42(4):127--138,
  2012.

\bibitem{10.1145/3387514.3405878}
Shuihai Hu, Wei Bai, Gaoxiong Zeng, Zilong Wang, Baochen Qiao, Kai Chen, Kun
  Tan, and Yi~Wang.
\newblock Aeolus: A building block for proactive transport in datacenters.
\newblock In {\em Proceedings of the Annual Conference of the ACM Special
  Interest Group on Data Communication on the Applications, Technologies,
  Architectures, and Protocols for Computer Communication}, SIGCOMM '20, page
  422–434, New York, NY, USA, 2020. Association for Computing Machinery.

\bibitem{10.1145/52325.52356}
V.~Jacobson.
\newblock Congestion avoidance and control.
\newblock {\em SIGCOMM Comput. Commun. Rev.}, 18(4):314–329, August 1988.

\bibitem{jin2004fast}
Cheng Jin, David~X Wei, and Steven~H Low.
\newblock Fast tcp: motivation, architecture, algorithms, performance.
\newblock In {\em IEEE INFOCOM 2004}, volume~4, pages 2490--2501. IEEE, 2004.

\bibitem{kandula2009nature}
Srikanth Kandula, Sudipta Sengupta, Albert Greenberg, Parveen Patel, and Ronnie
  Chaiken.
\newblock The nature of data center traffic: measurements \& analysis.
\newblock In {\em Proceedings of the 9th ACM SIGCOMM conference on Internet
  measurement}, pages 202--208, 2009.

\bibitem{katabi2002congestion}
Dina Katabi, Mark Handley, and Charlie Rohrs.
\newblock Congestion control for high bandwidth-delay product networks.
\newblock In {\em Proceedings of the 2002 conference on Applications,
  technologies, architectures, and protocols for computer communications},
  pages 89--102, 2002.

\bibitem{keshav2012mathematical}
Srinivasan Keshav.
\newblock {\em Mathematical foundations of computer networking}.
\newblock Addison-Wesley, 2012.

\bibitem{kim2015band}
Changhoon Kim, Anirudh Sivaraman, Naga Katta, Antonin Bas, Advait Dixit, and
  Lawrence~J Wobker.
\newblock In-band network telemetry via programmable dataplanes.
\newblock In {\em SIGCOMM '15 Demos}. ACM, 2015.

\bibitem{spaa21rdcn}
Janardhan Kulkarni, Stefan Schmid, and Pawel Schmidt.
\newblock Scheduling opportunistic links in two-tiered reconfigurable
  datacenters.
\newblock In {\em 33rd ACM Symposium on Parallelism in Algorithms and
  Architectures (SPAA)}, 2021.

\bibitem{kumar2020swift}
Gautam Kumar, Nandita Dukkipati, Keon Jang, Hassan~MG Wassel, Xian Wu, Behnam
  Montazeri, Yaogong Wang, Kevin Springborn, Christopher Alfeld, Michael Ryan,
  et~al.
\newblock Swift: Delay is simple and effective for congestion control in the
  datacenter.
\newblock In {\em Proceedings of the Annual conference of the ACM Special
  Interest Group on Data Communication on the applications, technologies,
  architectures, and protocols for computer communication}, pages 514--528,
  2020.

\bibitem{lee2015accurate}
Changhyun Lee, Chunjong Park, Keon Jang, Sue Moon, and Dongsu Han.
\newblock Accurate latency-based congestion feedback for datacenters.
\newblock In {\em 2015 $\{$USENIX$\}$ Annual Technical Conference
  ($\{$USENIX$\}$$\{$ATC$\}$ 15)}, pages 403--415, 2015.

\bibitem{li2019hpcc}
Yuliang Li, Rui Miao, Hongqiang~Harry Liu, Yan Zhuang, Fei Feng, Lingbo Tang,
  Zheng Cao, Ming Zhang, Frank Kelly, Mohammad Alizadeh, et~al.
\newblock Hpcc: High precision congestion control.
\newblock In {\em Proceedings of the ACM Special Interest Group on Data
  Communication}, pages 44--58. 2019.

\bibitem{980245}
S.H. Low, F.~Paganini, and J.C. Doyle.
\newblock Internet congestion control.
\newblock {\em IEEE Control Systems Magazine}, 22(1):28--43, 2002.

\bibitem{opera}
William~M Mellette, Rajdeep Das, Yibo Guo, Rob McGuinness, Alex~C Snoeren, and
  George Porter.
\newblock Expanding across time to deliver bandwidth efficiency and low
  latency.
\newblock In {\em 17th $\{$USENIX$\}$ Symposium on Networked Systems Design and
  Implementation ($\{$NSDI$\}$ 20)}, pages 1--18, 2020.

\bibitem{rotornet}
William~M Mellette, Rob McGuinness, Arjun Roy, Alex Forencich, George Papen,
  Alex~C Snoeren, and George Porter.
\newblock Rotornet: A scalable, low-complexity, optical datacenter network.
\newblock In {\em Proceedings of the Conference of the ACM Special Interest
  Group on Data Communication}, pages 267--280, 2017.

\bibitem{misra2000fluid}
Vishal Misra, Wei-Bo Gong, and Don Towsley.
\newblock Fluid-based analysis of a network of aqm routers supporting tcp flows
  with an application to red.
\newblock In {\em Proceedings of the conference on Applications, Technologies,
  Architectures, and Protocols for Computer Communication}, pages 151--160,
  2000.

\bibitem{mittal2015timely}
Radhika Mittal, Vinh~The Lam, Nandita Dukkipati, Emily Blem, Hassan Wassel,
  Monia Ghobadi, Amin Vahdat, Yaogong Wang, David Wetherall, and David Zats.
\newblock Timely: Rtt-based congestion control for the datacenter.
\newblock {\em ACM SIGCOMM Computer Communication Review}, 45(4):537--550,
  2015.

\bibitem{10.1145/3230543.3230564}
Behnam Montazeri, Yilong Li, Mohammad Alizadeh, and John Ousterhout.
\newblock Homa: A receiver-driven low-latency transport protocol using network
  priorities.
\newblock In {\em Proceedings of the 2018 Conference of the ACM Special
  Interest Group on Data Communication}, SIGCOMM '18, page 221–235, New York,
  NY, USA, 2018. Association for Computing Machinery.

\bibitem{mukerjee2020adapting}
Matthew~K Mukerjee, Christopher Canel, Weiyang Wang, Daehyeok Kim, Srinivasan
  Seshan, and Alex~C Snoeren.
\newblock Adapting $\{$TCP$\}$ for reconfigurable datacenter networks.
\newblock In {\em 17th $\{$USENIX$\}$ Symposium on Networked Systems Design and
  Implementation ($\{$NSDI$\}$ 20)}, pages 651--666, 2020.

\bibitem{NANCEHALL2021100621}
Matthew {Nance Hall}, Klaus-Tycho Foerster, Stefan Schmid, and Ramakrishnan
  Durairajan.
\newblock A survey of reconfigurable optical networks.
\newblock {\em Optical Switching and Networking}, 41:100621, 2021.

\bibitem{perry2017flowtune}
Jonathan Perry, Hari Balakrishnan, and Devavrat Shah.
\newblock Flowtune: Flowlet control for datacenter networks.
\newblock In {\em 14th $\{$USENIX$\}$ Symposium on Networked Systems Design and
  Implementation ($\{$NSDI$\}$ 17)}, pages 421--435, 2017.

\bibitem{perry2014fastpass}
Jonathan Perry, Amy Ousterhout, Hari Balakrishnan, Devavrat Shah, and Hans
  Fugal.
\newblock Fastpass: A centralized" zero-queue" datacenter network.
\newblock In {\em Proceedings of the 2014 ACM conference on SIGCOMM}, pages
  307--318, 2014.

\bibitem{phanishayee2008measurement}
Amar Phanishayee, Elie Krevat, Vijay Vasudevan, David~G Andersen, Gregory~R
  Ganger, Garth~A Gibson, and Srinivasan Seshan.
\newblock Measurement and analysis of tcp throughput collapse in cluster-based
  storage systems.
\newblock In {\em FAST}, volume~8, pages 1--14, 2008.

\bibitem{roy2015inside}
Arjun Roy, Hongyi Zeng, Jasmeet Bagga, George Porter, and Alex~C Snoeren.
\newblock Inside the social network's (datacenter) network.
\newblock In {\em Proceedings of the 2015 ACM Conference on Special Interest
  Group on Data Communication}, pages 123--137, 2015.

\bibitem{10.1145/3387514.3405899}
Ahmed Saeed, Varun Gupta, Prateesh Goyal, Milad Sharif, Rong Pan, Mostafa
  Ammar, Ellen Zegura, Keon Jang, Mohammad Alizadeh, Abdul Kabbani, and Amin
  Vahdat.
\newblock Annulus: A dual congestion control loop for datacenter and wan
  traffic aggregates.
\newblock In {\em Proceedings of the Annual Conference of the ACM Special
  Interest Group on Data Communication on the Applications, Technologies,
  Architectures, and Protocols for Computer Communication}, SIGCOMM '20, page
  735–749, New York, NY, USA, 2020. Association for Computing Machinery.

\bibitem{ton15splay}
Stefan Schmid, Chen Avin, Christian Scheideler, Michael Borokhovich, Bernhard
  Haeupler, and Zvi Lotker.
\newblock Splaynet: Towards locally self-adjusting networks.
\newblock {\em IEEE/ACM Transactions on Networking (ToN)}, 2016.

\bibitem{vamanan2012deadline}
Balajee Vamanan, Jahangir Hasan, and TN~Vijaykumar.
\newblock Deadline-aware datacenter tcp (d2tcp).
\newblock {\em ACM SIGCOMM Computer Communication Review}, 42(4):115--126,
  2012.

\bibitem{wilson2011better}
Christo Wilson, Hitesh Ballani, Thomas Karagiannis, and Ant Rowtron.
\newblock Better never than late: Meeting deadlines in datacenter networks.
\newblock {\em ACM SIGCOMM Computer Communication Review}, 41(4):50--61, 2011.

\bibitem{woodruff2019measuring}
Jackson Woodruff, Andrew~W Moore, and Noa Zilberman.
\newblock Measuring burstiness in data center applications.
\newblock 2019.

\bibitem{zarchy2019axiomatizing}
Doron Zarchy, Radhika Mittal, Michael Schapira, and Scott Shenker.
\newblock Axiomatizing congestion control.
\newblock {\em Proceedings of the ACM on Measurement and Analysis of Computing
  Systems}, 3(2):1--33, 2019.

\bibitem{zhang2017high}
Qiao Zhang, Vincent Liu, Hongyi Zeng, and Arvind Krishnamurthy.
\newblock High-resolution measurement of data center microbursts.
\newblock In {\em Proceedings of the 2017 Internet Measurement Conference},
  pages 78--85, 2017.

\bibitem{zhu2015congestion}
Yibo Zhu, Haggai Eran, Daniel Firestone, Chuanxiong Guo, Marina Lipshteyn,
  Yehonatan Liron, Jitendra Padhye, Shachar Raindel, Mohamad~Haj Yahia, and
  Ming Zhang.
\newblock Congestion control for large-scale rdma deployments.
\newblock {\em ACM SIGCOMM Computer Communication Review}, 45(4):523--536,
  2015.

\bibitem{zhu2016ecn}
Yibo Zhu, Monia Ghobadi, Vishal Misra, and Jitendra Padhye.
\newblock Ecn or delay: Lessons learnt from analysis of dcqcn and timely.
\newblock In {\em Proceedings of the 12th International on Conference on
  emerging Networking EXperiments and Technologies}, pages 313--327, 2016.

\end{thebibliography}

\appendix

\section{Analysis}
\label{sec:AppendixAnalysis}

Our analysis is based on a a single bottleneck link model widely used in the literature~\cite{zarchy2019axiomatizing,zhu2016ecn,hollot2001control,misra2000fluid}. Specifically, we assume that all senders use the same protocol, transmit long flows sharing a common bottleneck link with bandwidth $b$, and have a base round trip time $\tau$ (excluding queuing delays).
We denote at time $t$ queue length as $q(t)$, aggregate window size as $w(t)$, window size of a sender $i$ as $w_i(t)$, forward propagation delay between sender and bottleneck queue as $t^f$, the round-trip time as $\theta(t)$ and a base round-trip time as $\tau$. Here $w(t) = \sum_i w_i(t)$.

\begin{table}[t]
\begin{center}
\begin{tabular}{|c|c|}
\hline
\textbf{Notation} & \textbf{Description}\\
\hline
$b$ & bottleneck bandwidth \\
$q$ & bottleneck queue length \\
$\tau$ & base RTT \\
$t^f$ & sender to bottleneck delay \\
$\theta$ & round trip time RTT \\
$w_i$ & window size of a flow $i$\\
$w$ & aggregate window size (of all flows)\\
$\gamma$ & EWMA parameter \\
$\beta$ & additive increase \\
$e$ & desired equilibirum point \\
$f$ & feedback \\ 
$\lambda_i$ & sending rate of a flow $i$\\
$\lambda$ & Current: aggregate sending rate\\
$\nu$ & Voltage\\
$\Gamma$ & Power\\
\hline
\end{tabular}
\end{center}
\caption{Key notations used in this paper. Additionally for any variable say $x$, $\dot{x}$ denotes its derivate with respect to time \ie $\frac{dx}{dt}$.}
\label{table:notations}
\end{table}

\noindent We additionally use the traditional model of queue length dynamics which is independent of the control law~\cite{hollot2001control,misra2000fluid}
\begin{equation}
\label{eq:AqueueDyanamics}
\dot{q}(t) = \frac{w(t-t^f)}{\theta(t)} - b
\end{equation}

\noindent where $\theta(t)$ is given by,

\begin{equation}
\label{eq:Artt}
\theta(t) = \frac{q(t)}{b} + \tau
\end{equation}

\noindent Power at time $t$ denoted by $\Gamma(t)$ as defined in \S\ref{sec:power} is expressed as,

\begin{equation}
\label{eq:AGamma}
\Gamma(t) = \underbrace{(q(t) + b \cdot \tau)}_{{voltage}} \cdot \underbrace{(\dot{q}(t)+\mu(t))}_{current}
\end{equation}

\noindent \name's control law at a source $i$ is given by,

\begin{equation}
\label{eq:AcontrolLaw}
w_i(t+\delta t) = \gamma \cdot \left( \frac{w_i(t-\theta(t))\cdot e }{ f(t)  } + \beta \right) + (1-\gamma)\cdot w_i(t)
\end{equation}

\noindent where $e$ and $f(t)$ are given by,
\[
e=b^2\cdot \tau
\]
\[
f(t) = \Gamma(t-\theta(t)+t^f)
\]

\noindent and $\beta$ is the additive increase term and $\gamma\in (0,1]$ serves as the weight given for new updates using EWMA. Both $\beta$ and $\gamma$ are parameters to the control law.

Using the properties of power (Property~\ref{prop:window}), the aggregate window size at time $t-\theta(t)$ can be expressed in terms of power as,

\begin{equation}
\label{eq:ApropertyWindow}
w(t-\theta(t)) = \frac{\Gamma(t-\theta(t)+t^f)}{b} = \frac{f(t)}{b}
\end{equation}

Suppose an $ack$ arrives at time $t$ acknowledging a segement, time $t-\theta(t)$ corresponds to the time when the acknowledged segment was trasmitted.

\stabilitytheorem*
\proof

\noindent First, we rewrite Eq.~\ref{eq:AcontrolLaw} as follows to obtain the aggregate window $w$,

\[
\sum_i w_i(t+\delta t) = \sum_i \gamma \cdot \left( \frac{w_i(t-\theta(t))\cdot e }{ f(t)  } + \beta \right) + \sum_i(1-\gamma)\cdot w_i(t)
\]
let $\hat{\beta}=\sum_i \beta$

\[
w(t+\delta t) = \gamma \cdot \left( \frac{w(t-\theta(t))\cdot e }{ f(t)  } + \hat{\beta}\right) + (1-\gamma)\cdot w(t)
\]
by rearranging the terms in the above equation we obtain,

\[
w(t+\delta t) - w(t) = \gamma \cdot \left( -w(t) + \frac{w(t-\theta(t))\cdot e }{ f(t)  } + \hat{\beta} \right)
\]
dividing by $\delta t$ on both sides in the above equation and using Euler's first-order approximation, we derive the window dynamics for \name as follows,

\begin{equation}
\label{eq:Awdot}
\dot{w}(t) = \gamma_r \cdot \left( -w(t) + \frac{w(t-\theta(t))\cdot e }{ f(t)  } + \hat{\beta} \right)
\end{equation}
where $\gamma_r = \frac{\gamma}{\delta t}$.
Using Eq.~\ref{eq:ApropertyWindow} and substituting $e=b^2\cdot\tau$, Eq.~\ref{eq:Awdot} reduces to, 

\begin{equation}
\label{eq:AwdotReduced}
\dot{w}(t) = \gamma_r \cdot \left( -w(t) + b \cdot \tau+ \hat{\beta} \right)
\end{equation}

In the system defined by Eq.~\ref{eq:AqueueDyanamics} and Eq.~\ref{eq:Awdot}, when the window and the queue length stabilize \ie $\dot{w}(t)=0$ and $\dot{q}(t)=0$, it is easy to observe that there exists a unique equilibrium point $(w_e,q_e) = (b\cdot \tau + \hat{\beta}, \hat{\beta})$. 
We now apply a change of variable from $t$ to $t-t^f$ in Eq.~\ref{eq:AwdotReduced} and linearize Eq.~\ref{eq:AwdotReduced} and Eq.~\ref{eq:AqueueDyanamics} around $(w_e, q_e)$,

\begin{equation}
\delta \dot{w}(t-t^f) = -\gamma_r \cdot \delta w(t-t^f)
\end{equation}

\begin{equation}
\delta \dot{q}(t) = -\frac{\delta q(t)}{\tau} + \frac{\delta w(t-t^f)}{\tau}
\end{equation}

We now convert the above differential equations to matrix form,

\[
\begin{bmatrix}
\delta \dot{q}(t)\\
\delta \dot{w}(t)
\end{bmatrix}
=
\begin{bmatrix}
-\frac{1}{\tau} & \frac{1}{\tau}\\
0 & -\gamma_r
\end{bmatrix}
\times
\begin{bmatrix}
\delta q(t)\\
\delta w(t)
\end{bmatrix}
\]

It is then easy to observe that the eigenvalues of the system are $-\frac{1}{\tau}$ and $-\gamma_r$. Since $\tau$ (base RTT) and $\gamma_r=\frac{\gamma}{\delta t}$ are both positive, we see that both the eigenvalues are negative. This proves that the system is both lyapunav stable and asymptotically stable.

\endproof

\convergencetheorem*

\proof
A perturbation at time $t=0$ causes the window to shift from $w_e=c\cdot\tau + \hat{\beta}$ to say $w_{init}$.
We solve the differential equation in Eq.~\ref{eq:AwdotReduced} and obtain the following equation,

\begin{equation}
\label{eq:convergenceEq}
w(t) = w_e + \underbrace{(w_{init}-w_e)\cdot e^{-\gamma_r\cdot t}}_{exponential\ decay}
\end{equation}

\noindent From Eq.~\ref{eq:convergenceEq} we can see that, for any error $e=w_e-w_{init}$ caused by a perturbation, $e$ exponentially decays with a time constant $\frac{1}{\gamma_r}=\frac{\delta t}{\gamma}$. Hence for $e$ to decay $99.3\%$, it takes $\frac{5\cdot \delta}{\gamma}$ time.

\endproof

\fairnesstheorem*

\proof
Recall that \name's control law for each flow $i$ is defined as,
\[
w_i(t+\delta t) = \gamma \cdot \left( \frac{w_i(t-\theta(t))\cdot e }{ f(t)  } + \beta_i \right) + (1-\gamma)\cdot w_i(t)
\]

\noindent From the proof of Theorem~\ref{theorem:stability}, we know that the equilibrium point for aggregate window size and queue length is $(w_e,q_e)= (b\cdot \tau + \hat{\beta}, \hat{\beta})$. Using this equilibrium we can also obtain the equilibrium value for $f(t)$ as,

\[
f_e = (\hat{\beta}+b\cdot \tau)\cdot b
\]

\noindent We can then show that $w_i$ has an  equilibrium point.

\[
(w_i)_e = \frac{\hat{\beta}+b\cdot\tau}{\hat{\beta}} \cdot \beta_i
\]

We use the argument that window sizes and rates are synonymous especially that \name uses pacing with rate $r_i=\frac{w_i}{\tau}$. We can then easily observe that the rate allocation is approximately max-min fair if $\beta_i$ are small enough but $\beta_i$ proportionally fair in general.
\endproof

\section{\nameapprox}
We present \nameapprox: standalone version of \name which does not require switch support and only requires accurate packet timestamp support at the end-host. 
\label{appendix:approxPowerTCP}
\begin{algorithm}[!h]
    \SetKwFunction{newAck}{\textsc{\textcolor{red}{newAck}}}
    \SetKwFunction{updateWindow}{\textsc{\textcolor{red}{updateWindow}}}
    \SetKwFunction{normPower}{\textsc{\textcolor{red}{normPower}}}

    \SetKwProg{Fn}{function}{:}{}
    \SetKwProg{Proc}{procedure}{:}{}
    \SetKwInOut{KwIn}{Input}
    \SetKwInOut{KwOut}{Output}
    
    \tcc{$t_c$ is the timestamp when ACK is received}
    
    \KwIn{\ $ack$ }
    \KwOut{\ $cwnd$, $rate$}

    \Proc{\newAck{$ack$}}{
      
      $cwnd_{old}$ = \textsc{getCwnd}($ack.seq$)
      
      $normPower =$ \textsc{normPower}($ack$)

      \textsc{updateWindow}($normPower$, $cwnd_{old}$) 

      $rate = \frac{cwnd}{\tau}$ \label{line:rate}

      $prevRTT = RTT$ \label{line:prevRTT}
      
      $t_c^{prev} = t_c$; \textsc{updateOld}$(cwnd, ack.seq)$

    }

    \Fn{\normPower{ack}}{
        
        $ dt  = t_{c} - t^{prev}_{c}$ 
        
        $\dot{\theta} = \frac{RTT - prevRTT}{dt}$ \Comment{$\frac{dRTT}{dt}$}

        $\Gamma_{norm} = \frac{(\dot{\theta}+1)\times(RTT)}{\tau} $ \Comment{$\Gamma_{norm}:$Normalized power}

        $\Gamma_{smooth} = \frac{(\Gamma_{smooth}\cdot(\tau - \Delta t)+(\Gamma_{norm}\cdot \Delta t)}{\tau}$
    
      \KwRet{$\Gamma_{smooth}$}
    }

    \Fn{\updateWindow{$power$, $ack$}}{

      \If(\Comment{per RTT}){$ack.seq<lastUpdated$}{
      
        \KwRet{$cwnd$} 
      }
      $cwnd = \gamma\times( \frac{cwnd_{old}}{normPower} + \beta ) + (1-\gamma)\times cwnd $

      \Comment{$\gamma:$ EWMA parameter}

      \Comment{$\beta$: Additive Increase}
      
      $lastUpdated=snd\_nxt$

      \KwRet{$cwnd$}
    }
    
    \caption{\nameapprox (w/o switch support)}
    \label{alg:tcpnoInt}
\end{algorithm}

\section{Justifying the Simplified Model}
\label{sec:justification}

We considered a simplified control law model to study existing control laws in \S\ref{sec:motivation}. Here we justify how the simplified model approximately captures the 
existing control laws.
Our simplified model for congestion window update at time $t+\delta t$ is defined in Eq.~\ref{eq:AsimplifiedModel} as a function of current congestion window size, a target $e$, the feedback $f(t)$, an additive increase $\beta$ and an exponential moving average parameter $\gamma$.
\begin{equation}
\label{eq:AsimplifiedModel}
w_i(t+\delta t) = \underbrace{\gamma\cdot \underbrace{\left( w_i(t)\cdot \frac{e}{f(t)} + \beta \right)}_{update} + (1-\gamma)\cdot w_i(t)}_{EWMA}
\end{equation}

\noindent where $e$ and $f(t)$ are given by,

\begin{equation}
\label{eq:Aequilibrium}
e=
\begin{cases}
b\cdot\tau                    & \textit{queue-length based CC}\\
\tau                      & \textit{delay-based CC} \\
1                         & \textit{RTT-gradient based CC}\\
\end{cases}
\end{equation}

\begin{equation}
\label{eq:Aequilibrium}
f(t)=
\begin{cases}
q(t-\theta(t)+t^f)+b\cdot\tau               & \textit{queue-length based CC}\\
\frac{q(t-\theta(t)+t^f)}{b}+\tau           & \textit{delay-based CC} \\
\frac{\dot{q}(t-\theta(t)+t^f)}{b}+1          & \textit{RTT-gradient based CC}\\
\end{cases}
\end{equation}

\noindent We first use Euler's first order approximation and obtain the aggregate window $(\sum w)$ dynamics for the simplified model,
\begin{equation}
\label{eq:AsimplifiedModelDynamics}
\dot{w}(t) = \frac{\gamma}{\delta t}\cdot \left( w(t)\cdot \frac{e}{f(t)} - w(t) + \beta \right)
\end{equation}

\noindent In order for the system to stabilize, we require $\dot{q}(t)=0$ and $\dot{w}(t)=0$. Using Eq.~\ref{eq:AqueueDyanamics} and Eq.~\ref{eq:AsimplifiedModelDynamics} and applying equilibrium conditions and assuming that $f(t)$ stabilizes,

\begin{equation}
\label{eq:AqueueEquilibrium}
q_e = w_e - b\cdot \tau
\end{equation}

\begin{equation}
\label{eq:AwindowEquilibrium}
w_e = \frac{\hat{\beta}}{1-\frac{e}{f}}
\end{equation}

\noindent Recall that $\hat{\beta}=\sum_{\beta_i}$, the sum of additive increase terms of all flows sharing a bottleneck. To show whether there exists a unique equilibrium point, it remains to show whether Eq.~\ref{eq:AqueueEquilibrium} and Eq.~\ref{eq:AwindowEquilibrium} have a unique solution for $w_e$ and $q_e$. We now show how the simplified model captures existing control laws and show the equilibrium properties.

\myitem{Queue length or inflight-based control law:}

\noindent Substituting $e=b\cdot\tau$ and $f(t)=q(t-\theta(t)+t^f)+b\cdot\tau$, we express the simplified queue length based control law as,

\begin{equation}
\label{eq:AsimplifiedModelQueue}
w_i(t+\delta t) = \gamma\cdot \left( \frac{w_i(t)\cdot b\cdot\tau}{q(t-\theta(t)+t^f)+b\cdot\tau} + \beta \right) + (1-\gamma)\cdot w_i(t)
\end{equation}
notice that the update is an MIMD based on inflight bytes. Eq.~\ref{eq:AsimplifiedModelQueue} captures control laws based on inflight bytes; for example HPCC~\cite{li2019hpcc}.

A system defined by queue length based control law (Eq.~\ref{eq:AsimplifiedModelQueue} and the queue length dynamics (Eq.~\ref{eq:AqueueDyanamics}, there exists a unique equilibrium point. It can be observed that Eq.~\ref{eq:AwindowEquilibrium} for queue length based control law gives $w_e=b\cdot\tau+\hat{\beta}$ and $q_e=\hat{\beta}$.

\myitem{Delay-based control law:}
\noindent Substituting $e=\tau$ and $f(t)=\frac{q(t-\theta(t)+t^f)}{b}+\tau$, we express the simplified delay based control law as,
\begin{equation}
\label{eq:AsimplifiedModelDelay}
w_i(t+\delta t) = \gamma\cdot \left( \frac{w_i(t)\cdot \tau}{\frac{q(t-\theta(t)+t^f)}{b}+\tau} + \beta \right) + (1-\gamma)\cdot w_i(t)
\end{equation}
where the window update is an MIMD based on RTT. Eq.~\ref{eq:AsimplifiedModelDelay} captures control laws based on RTT; for example FAST~\cite{jin2004fast}.

Similar to queue-length based CC, a system defined by delay-based control law (Eq.~\ref{eq:AsimplifiedModelDelay} and the queue length dynamics (Eq.~\ref{eq:AqueueDyanamics}, there exists a unique equilibrium point. It can be observed that Eq.~\ref{eq:AwindowEquilibrium} for delay-based control law gives $w_e=b\cdot\tau+\hat{\beta}$ and $q_e=\hat{\beta}$.

\myitem{RTT-gradient based control law:}
\noindent Substituting $e=1$ and $f(t)=\frac{\dot{q}(t-\theta(t)+t^f)}{b}+1$, we express the simplified RTT-gradient based control law as,
\begin{equation}
\label{eq:AsimplifiedModelGradient}
w_i(t+\delta t) = \gamma\cdot \left( \frac{w_i(t)\cdot 1 }{\frac{\dot{q}(t-\theta(t)+t^f)}{b}+1} + \beta \right) + (1-\gamma)\cdot w_i(t)
\end{equation}
where the window update is an MIMD based on RTT-gradient. Eq.~\ref{eq:AsimplifiedModelGradient} by rearranging the terms, captures control laws based on RTT-gradient such as TIMELY~\cite{mittal2015timely}.

In contrast to queue-length and delay-based CC, RTT-gradient based CC has no unique equilibrium point since $f(t)=\frac{\dot{q}(t-\theta(t)+t^f)}{b}+1$ stabilizes when $\dot{q}=0$. However only $\dot{q}=0$ leads to window dynamics Eq.~\ref{eq:AsimplifiedModelGradient} also to stabilize $(\dot{w}=0)$ at any queue lengths. As a result under RTT-gradient control law, Eq.~\ref{eq:AqueueEquilibrium} and Eq.~\ref{eq:AwindowEquilibrium} do not have a unique solution and consequently we can state that RTT-gradient based CC has no unique equilibirum point.

\section{HOMA's Overcommitment}
\label{appendix:homa}

\begin{figure}[!h]
\centering
\begin{minipage}{1\linewidth}
\centering
\includegraphics[width=1\linewidth]{plots/Powertcp-NSDI/fairness/fair-legend.pdf}
\end{minipage}
\begin{subfigure}{0.48\linewidth}
\centering
\includegraphics[width=1\linewidth]{plots/Powertcp-NSDI/fairness/homa1-small.pdf}
\subcaption{Over-Commitment: $1$}
\end{subfigure}
\begin{subfigure}{0.48\linewidth}
\centering
\includegraphics[width=1\linewidth]{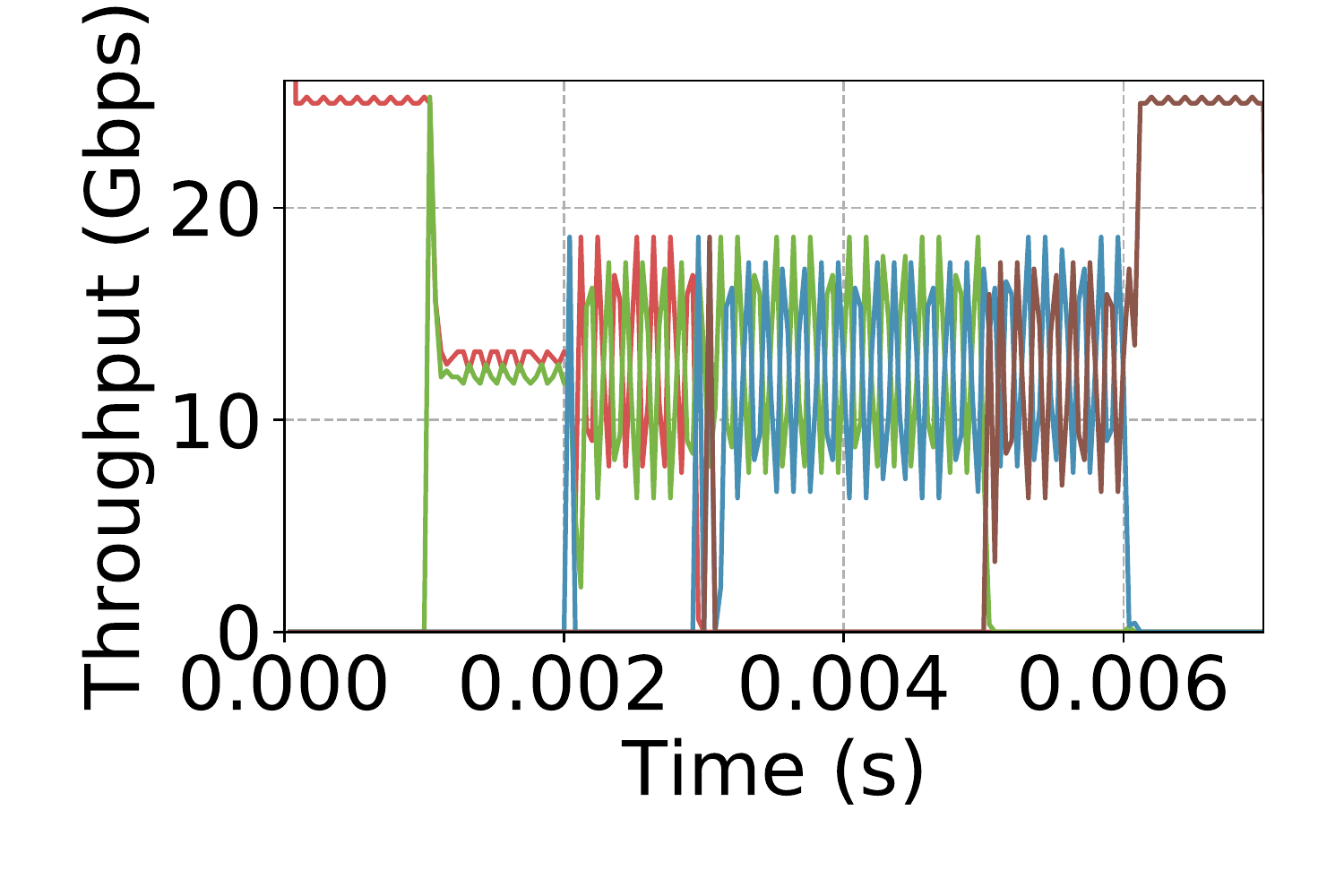}
\subcaption{Over-Commitment: $2$}
\end{subfigure}
\begin{subfigure}{0.48\linewidth}
\centering
\includegraphics[width=1\linewidth]{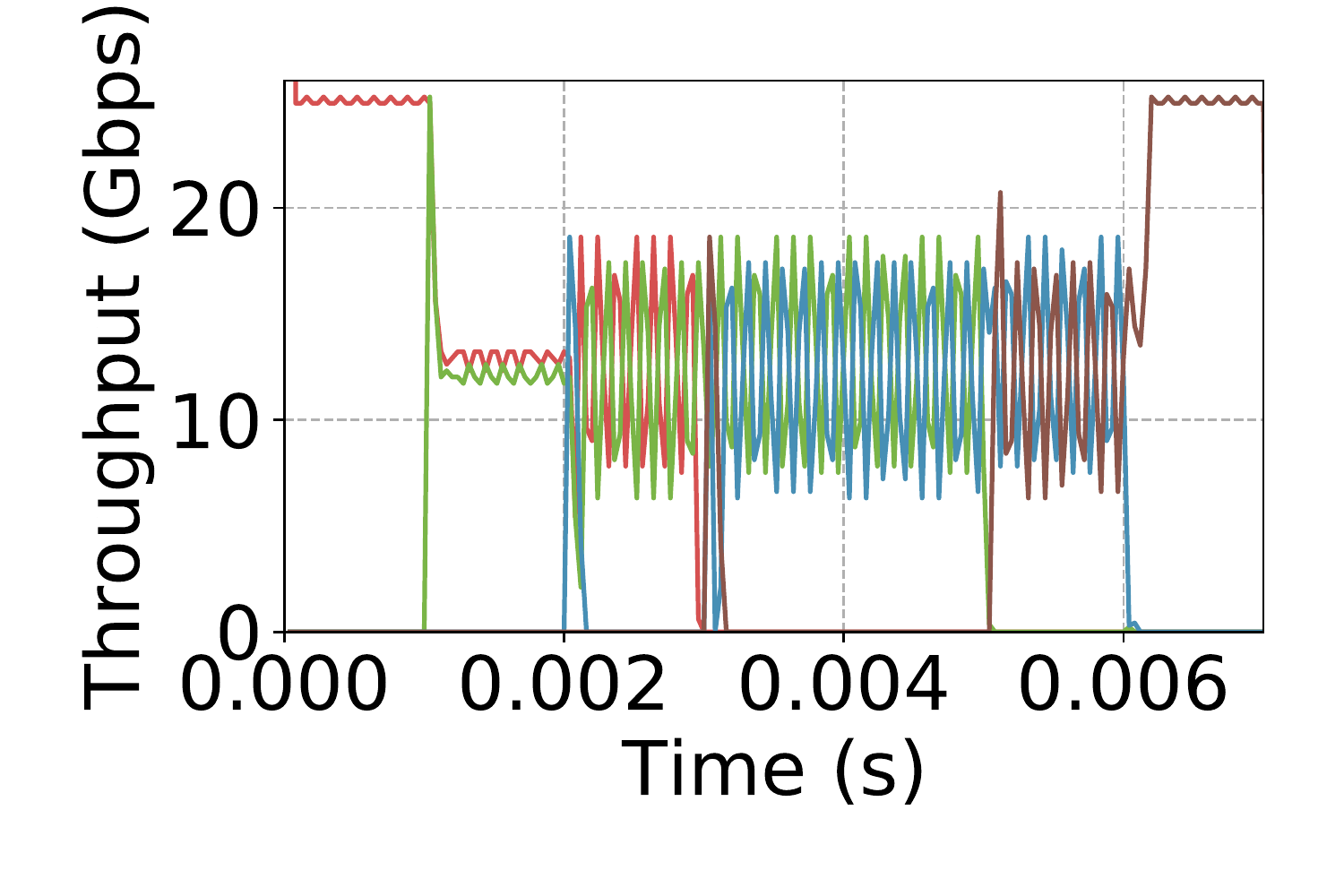}
\subcaption{Over-Commitment: $3$}
\end{subfigure}
\begin{subfigure}{0.48\linewidth}
\centering
\includegraphics[width=1\linewidth]{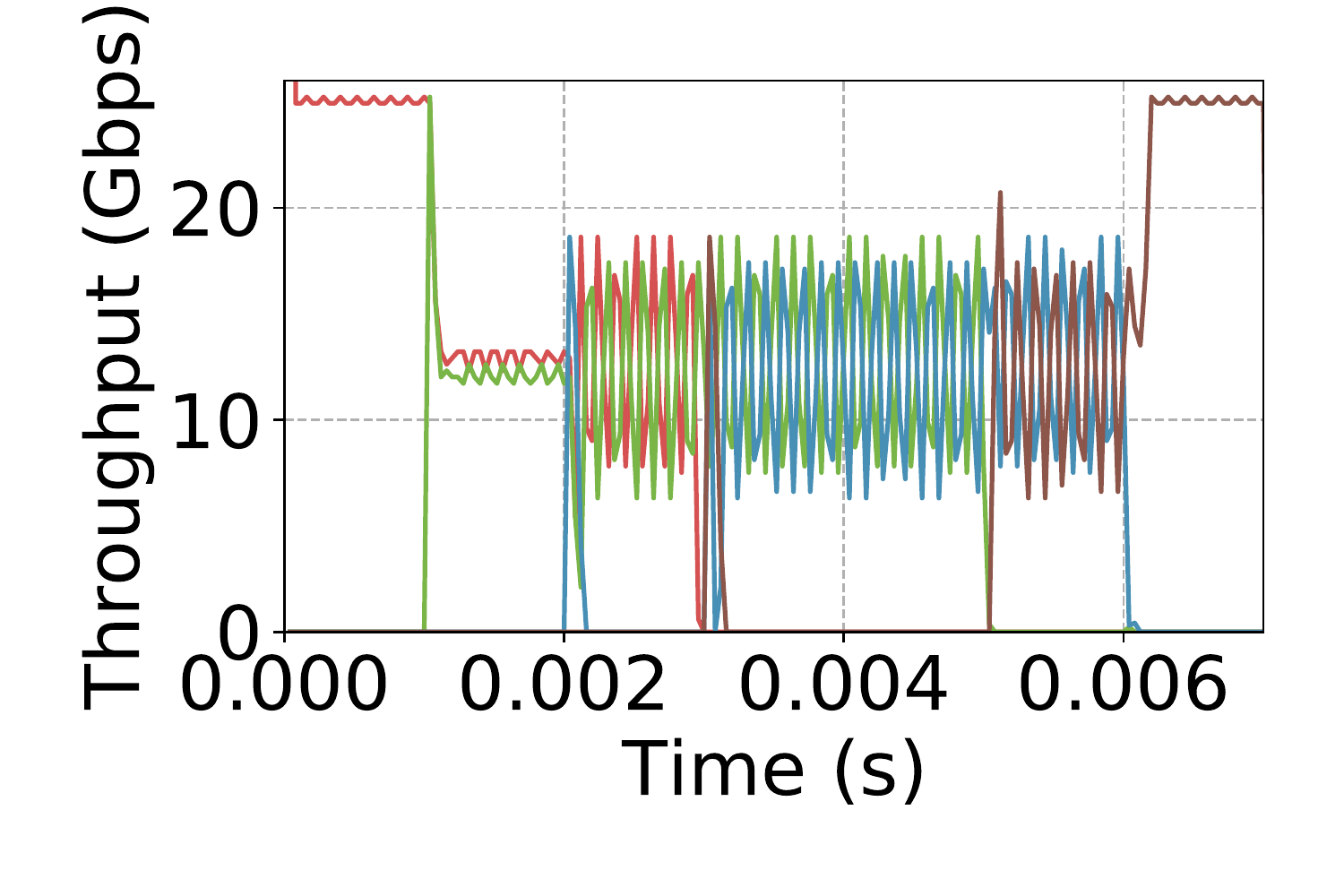}
\subcaption{Over-Commitment: $4$}
\end{subfigure}
\begin{subfigure}{0.48\linewidth}
\centering
\includegraphics[width=1\linewidth]{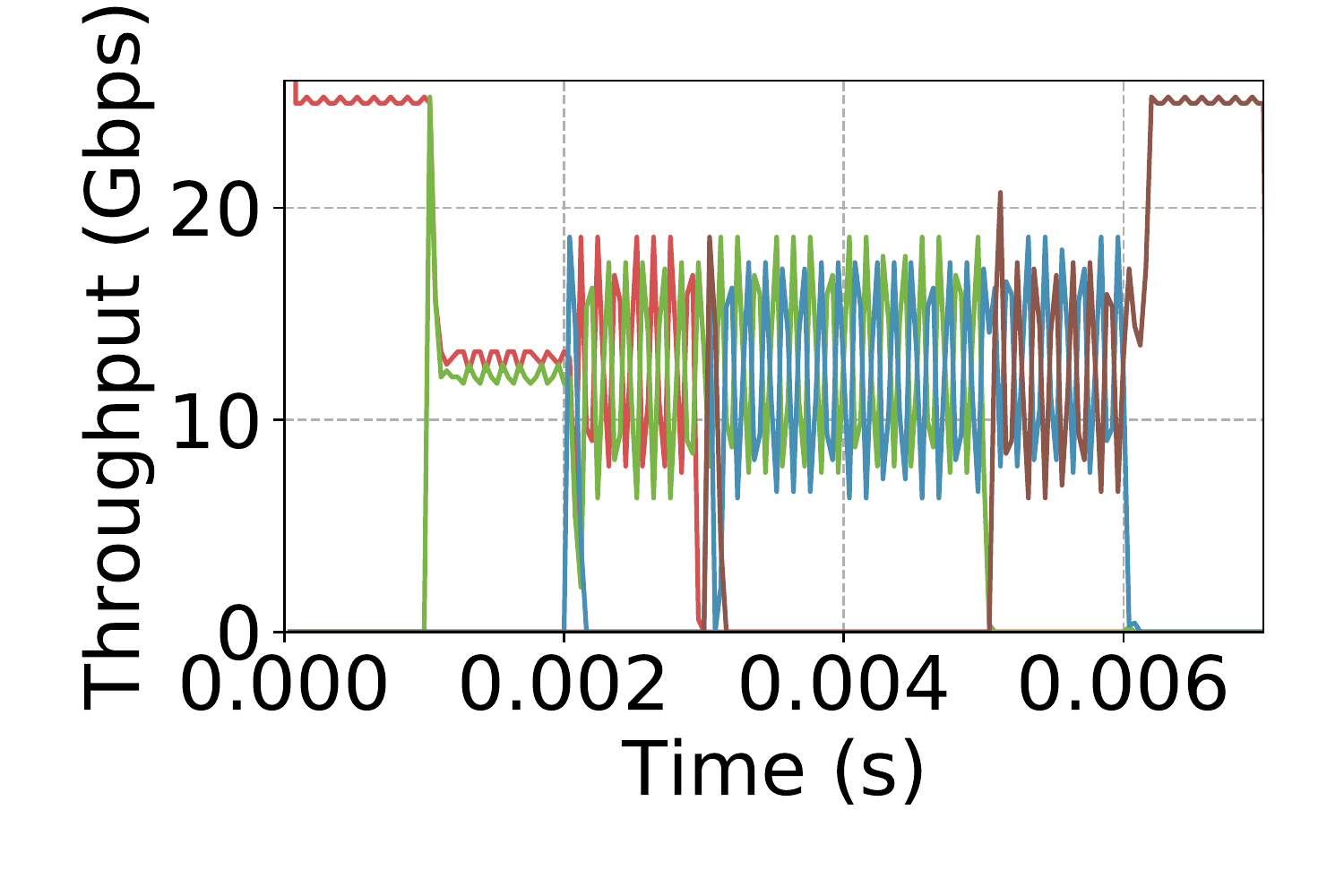}
\subcaption{Over-Commitment: $5$}
\end{subfigure}
\begin{subfigure}{0.48\linewidth}
\centering
\includegraphics[width=1\linewidth]{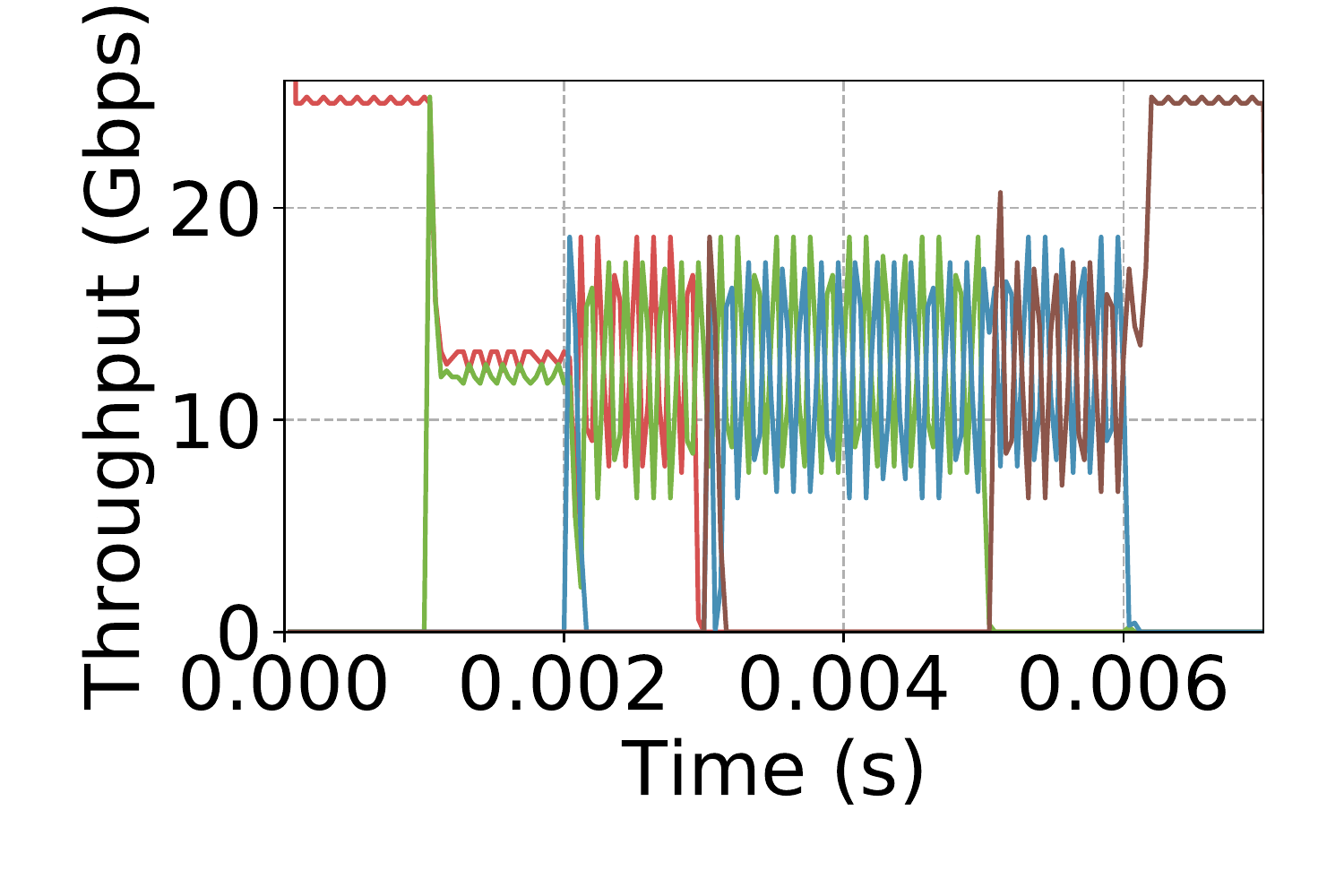}
\subcaption{Over-Commitment: $6$}
\end{subfigure}
\caption{HOMA's fairness and stability at different over-commitment levels. }
\label{fig:homa-fairness}
\end{figure}

\begin{figure*}
\centering
\begin{subfigure}{0.32\linewidth}
\includegraphics[width=0.9\linewidth]{plots/Powertcp-NSDI/burst/burst-legend.pdf}
\end{subfigure}
\begin{subfigure}{0.32\linewidth}
\includegraphics[width=0.9\linewidth]{plots/Powertcp-NSDI/burst/burst-legend.pdf}
\end{subfigure}
\begin{subfigure}{0.32\linewidth}
\includegraphics[width=0.9\linewidth]{plots/Powertcp-NSDI/burst/burst-legend.pdf}
\end{subfigure}
\begin{subfigure}{0.32\linewidth}
\centering
\includegraphics[width=1\linewidth]{plots/Powertcp-NSDI/burst/homa1-1.pdf}
\subcaption{Over-Commitment: $1$}
\end{subfigure}
\begin{subfigure}{0.32\linewidth}
\centering
\includegraphics[width=1\linewidth]{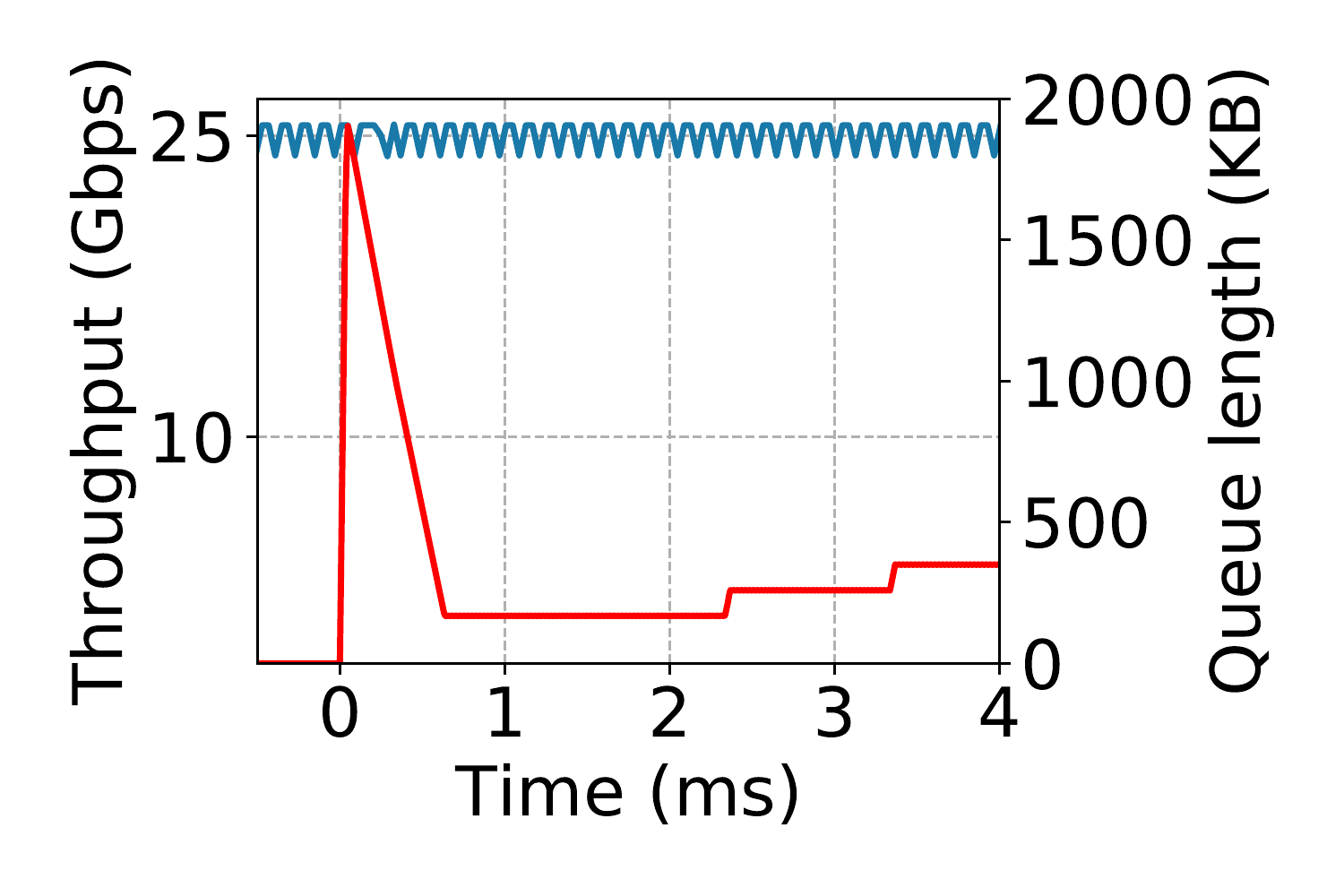}
\subcaption{Over-Commitment: $2$}
\end{subfigure}
\begin{subfigure}{0.32\linewidth}
\centering
\includegraphics[width=1\linewidth]{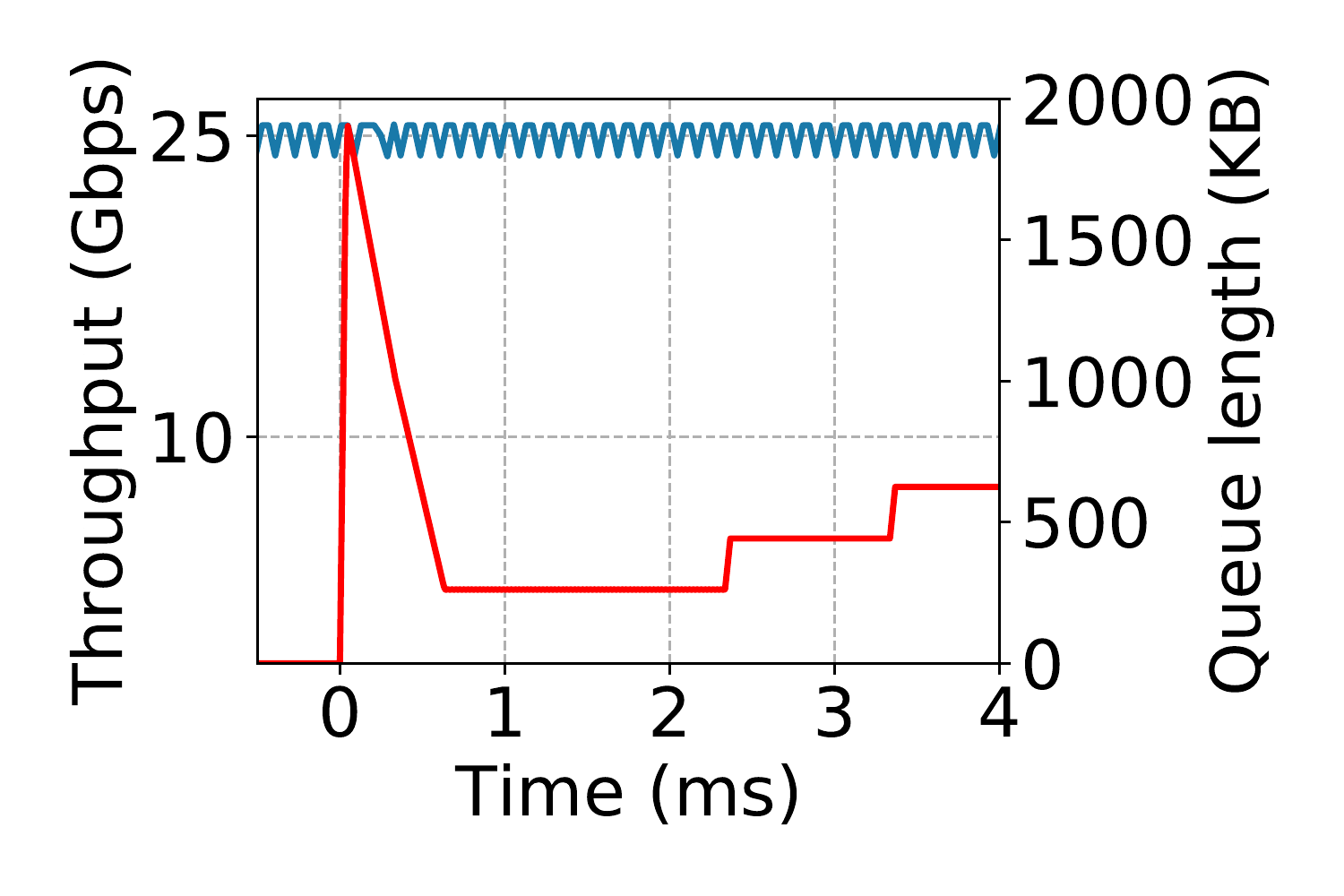}
\subcaption{Over-Commitment: $3$}
\end{subfigure}
\begin{subfigure}{0.32\linewidth}
\centering
\includegraphics[width=1\linewidth]{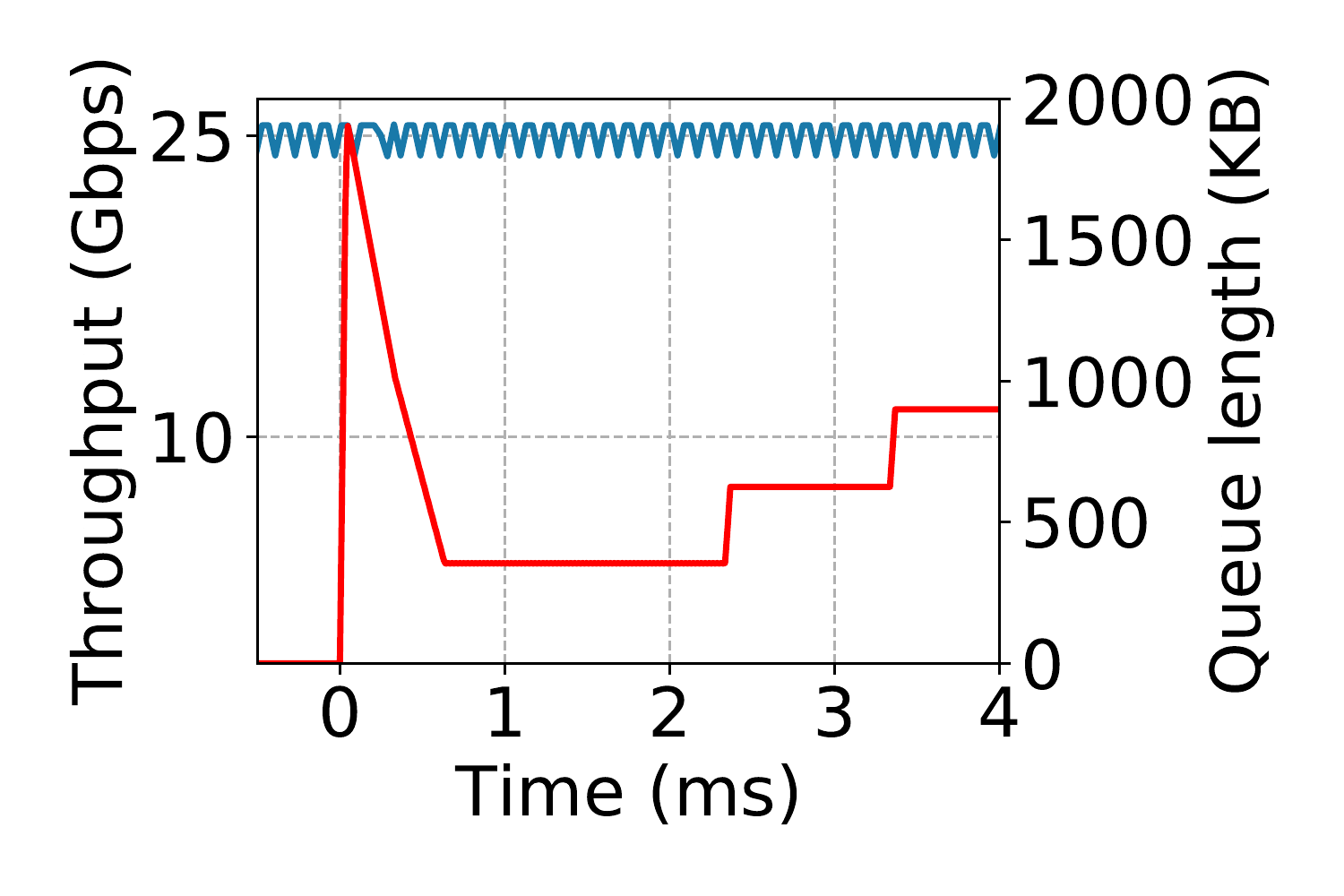}
\subcaption{Over-Commitment: $4$}
\end{subfigure}
\begin{subfigure}{0.32\linewidth}
\centering
\includegraphics[width=1\linewidth]{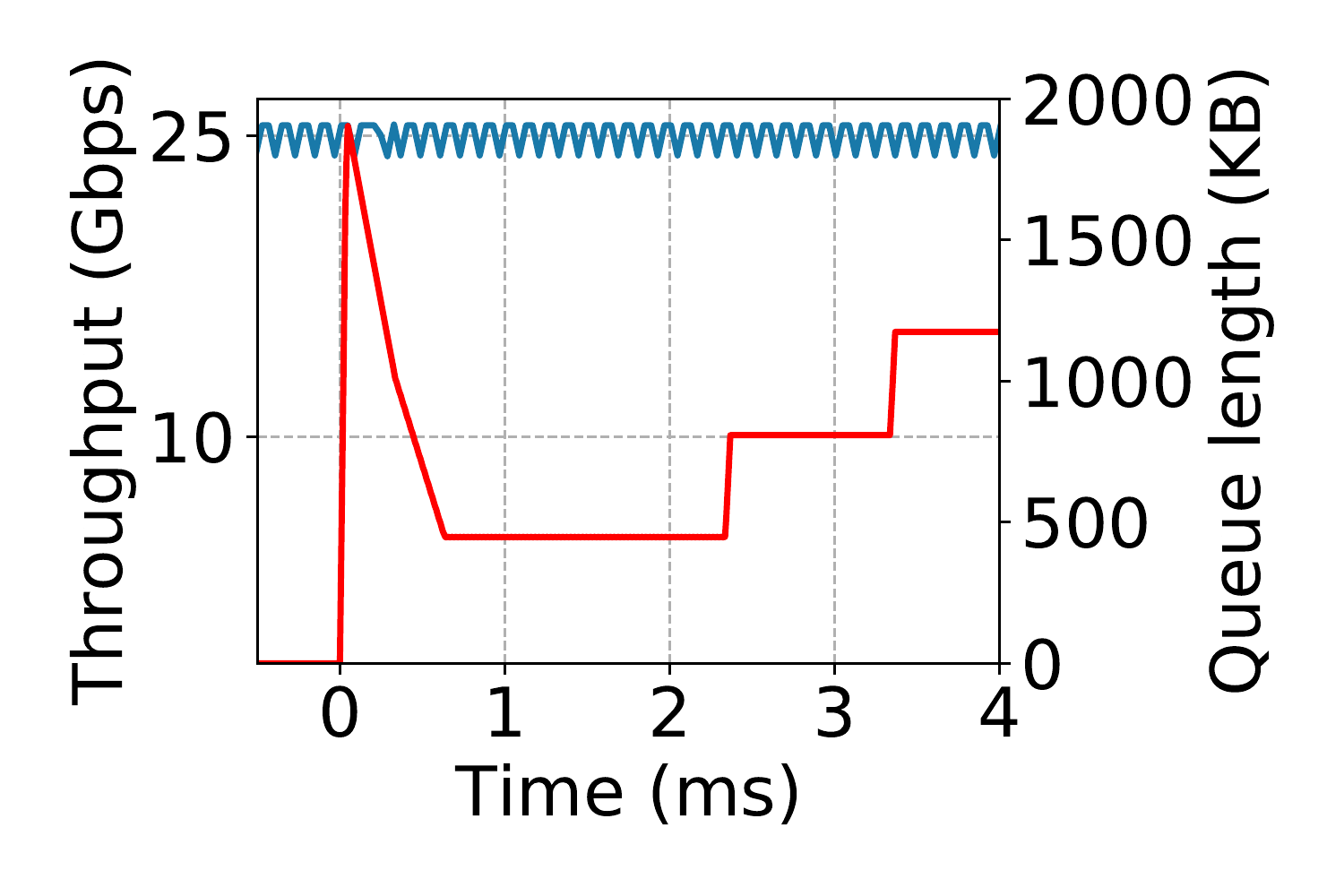}
\subcaption{Over-Commitment: $5$}
\end{subfigure}
\begin{subfigure}{0.32\linewidth}
\centering
\includegraphics[width=1\linewidth]{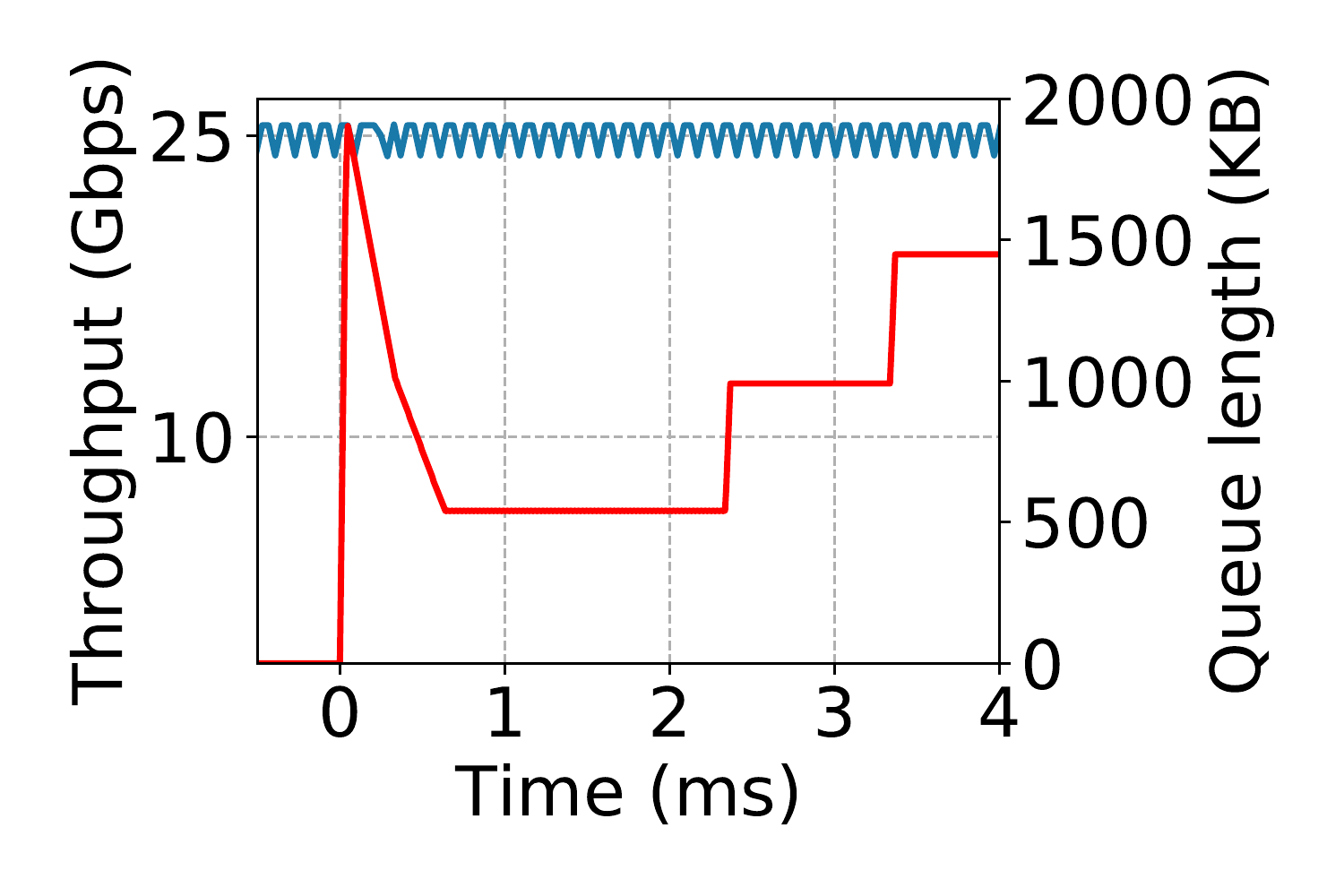}
\subcaption{Over-Commitment: $6$}
\end{subfigure}
\caption{HOMA's reaction to $255:1$ incast at different over-commitment levels. }
\label{fig:homa-255-incast}
\end{figure*}

\begin{figure*}
\centering
\begin{subfigure}{0.32\linewidth}
\includegraphics[width=0.9\linewidth]{plots/Powertcp-NSDI/burst/burst-legend.pdf}
\end{subfigure}
\begin{subfigure}{0.32\linewidth}
\includegraphics[width=0.9\linewidth]{plots/Powertcp-NSDI/burst/burst-legend.pdf}
\end{subfigure}
\begin{subfigure}{0.32\linewidth}
\includegraphics[width=0.9\linewidth]{plots/Powertcp-NSDI/burst/burst-legend.pdf}
\end{subfigure}
\begin{subfigure}{0.32\linewidth}
\centering
\includegraphics[width=1\linewidth]{plots/Powertcp-NSDI/burst/homa1-0.pdf}
\subcaption{Over-Commitment: $1$}
\end{subfigure}
\begin{subfigure}{0.32\linewidth}
\centering
\includegraphics[width=1\linewidth]{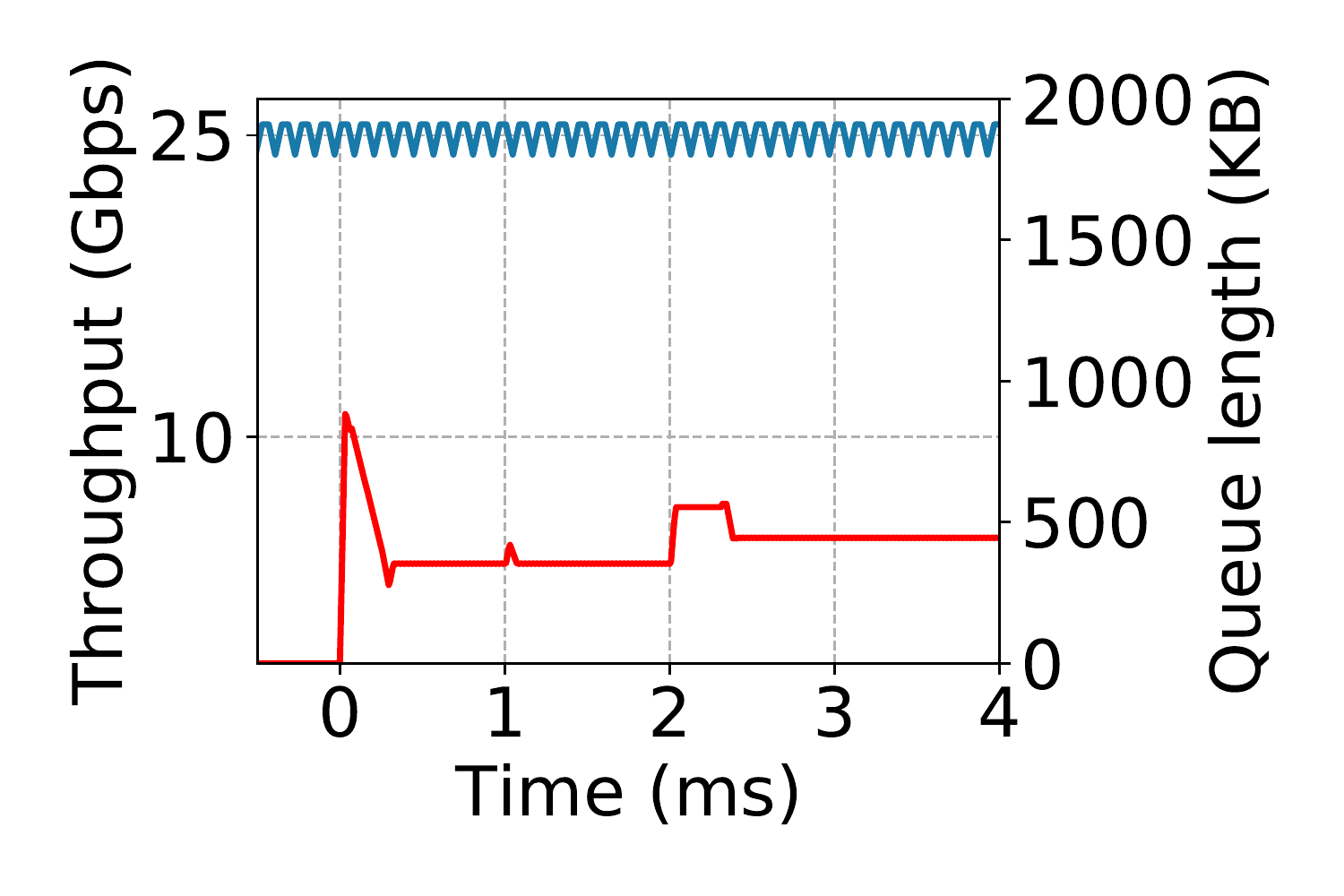}
\subcaption{Over-Commitment: $2$}
\end{subfigure}
\begin{subfigure}{0.32\linewidth}
\centering
\includegraphics[width=1\linewidth]{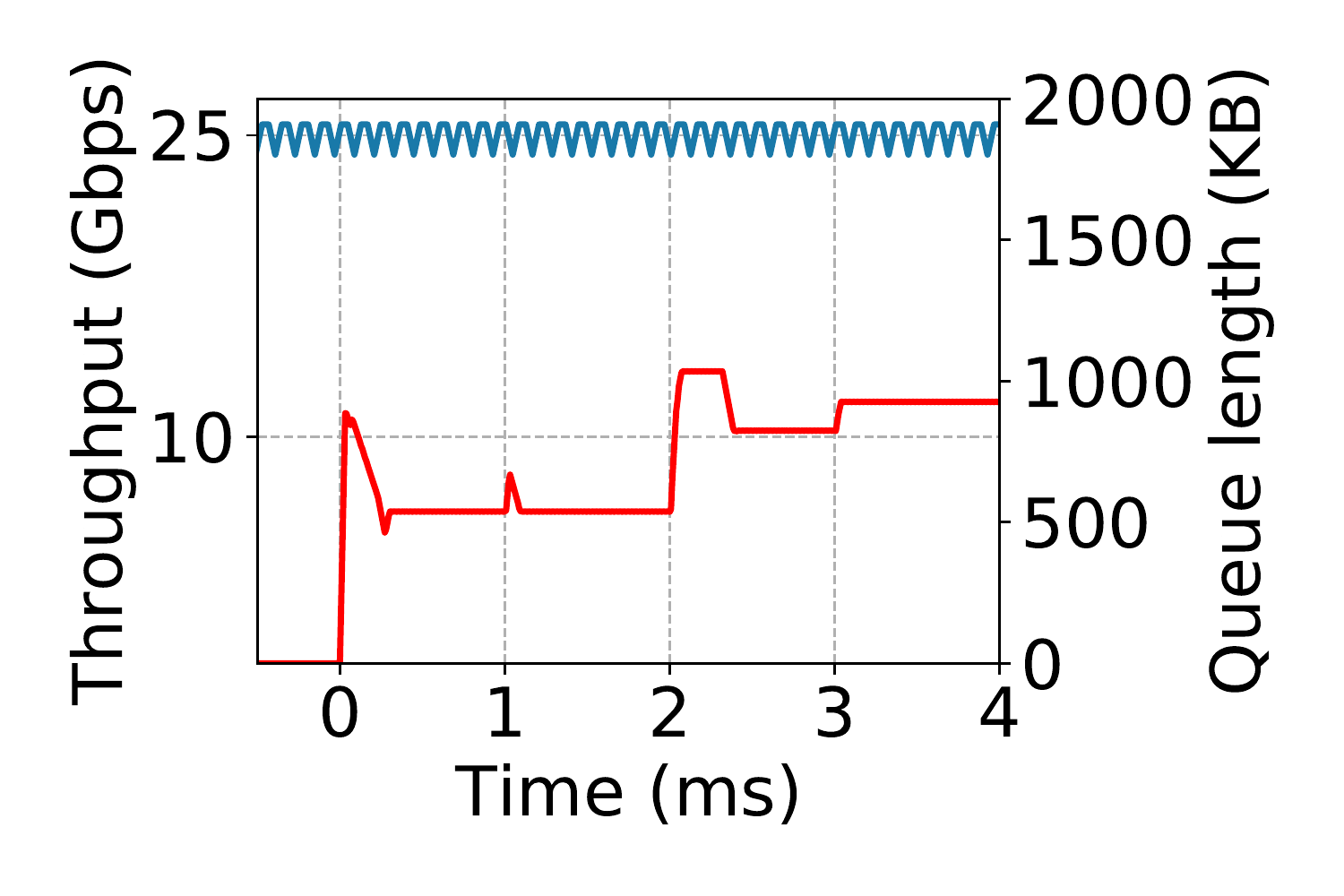}
\subcaption{Over-Commitment: $3$}
\end{subfigure}
\begin{subfigure}{0.32\linewidth}
\centering
\includegraphics[width=1\linewidth]{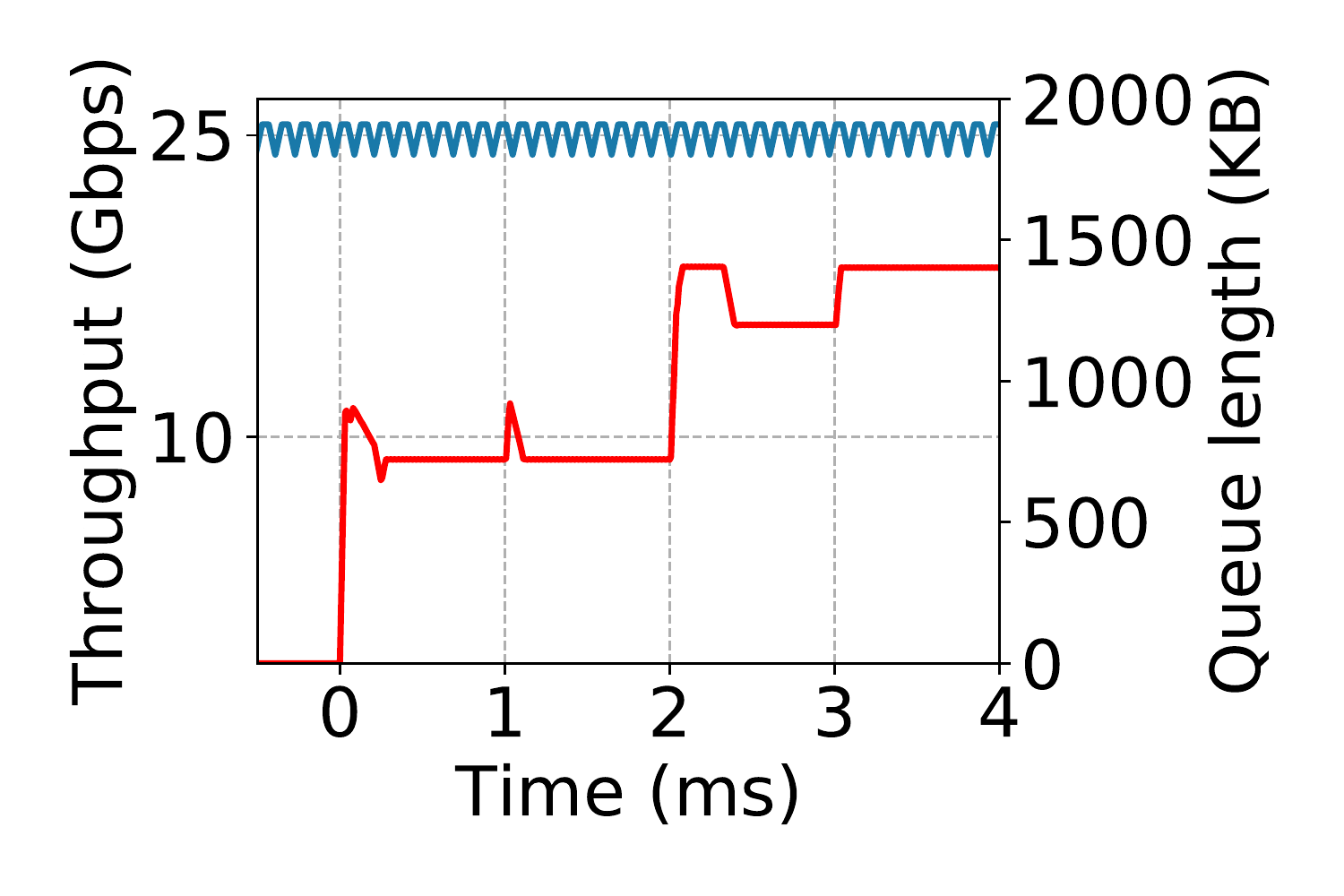}
\subcaption{Over-Commitment: $4$}
\end{subfigure}
\begin{subfigure}{0.32\linewidth}
\centering
\includegraphics[width=1\linewidth]{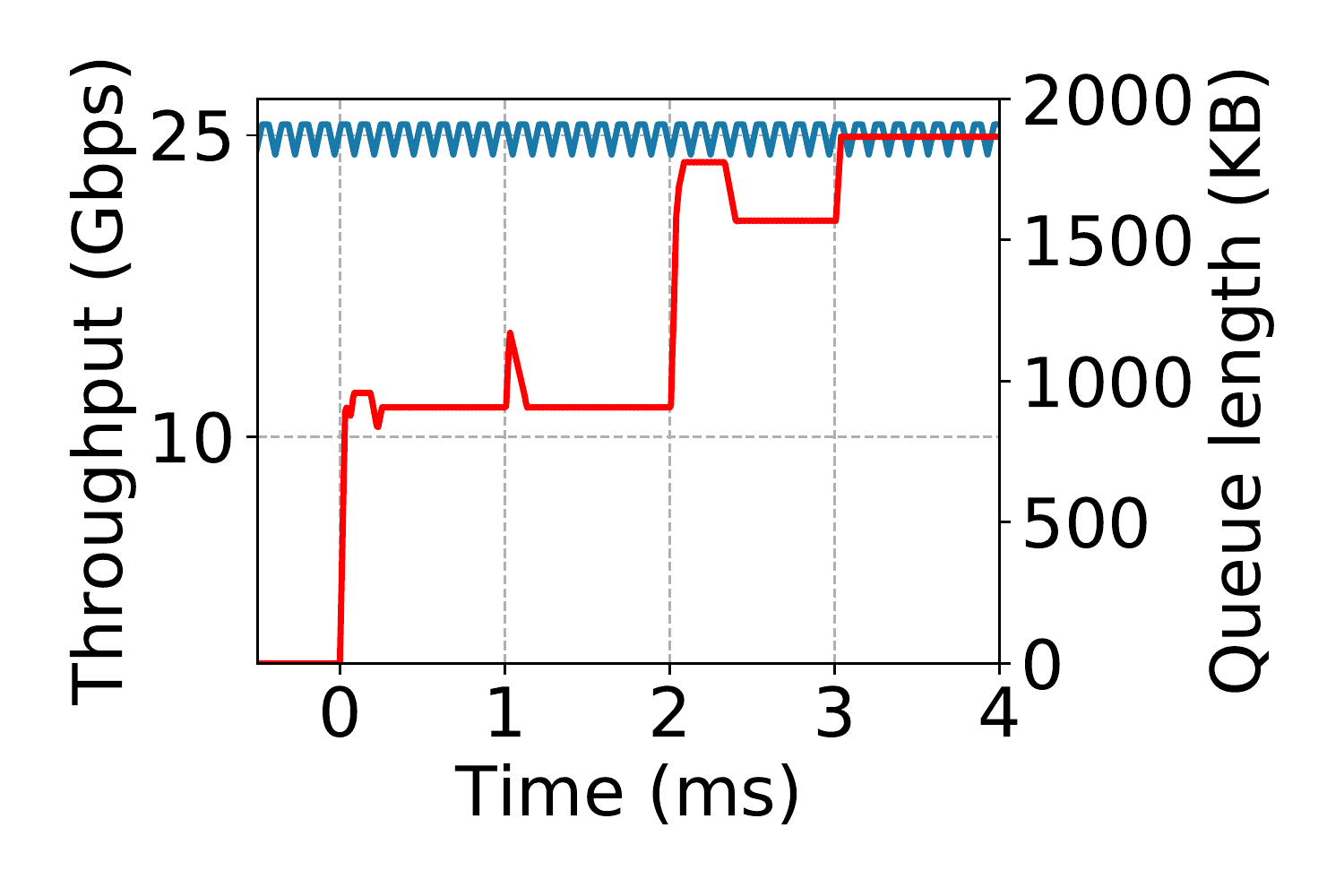}
\subcaption{Over-Commitment: $5$}
\end{subfigure}
\begin{subfigure}{0.32\linewidth}
\centering
\includegraphics[width=1\linewidth]{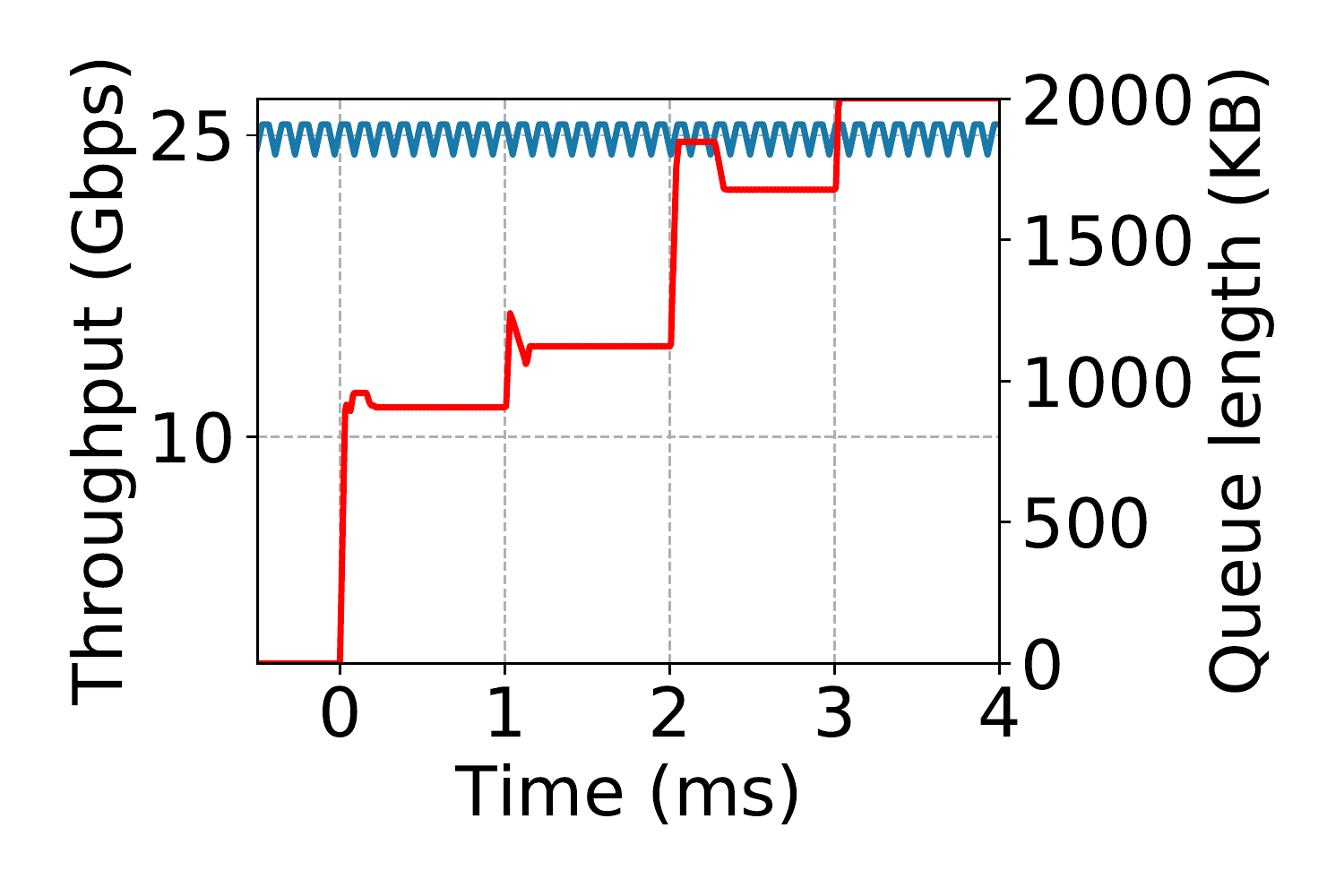}
\subcaption{Over-Commitment: $6$}
\end{subfigure}
\caption{HOMA's reaction to $10:1$ incast at different over-commitment levels. }
\label{fig:homa-10-incast}
\end{figure*}

\label{LastPage}

\end{document}